\documentclass[10pt,aps,prc,amsmath,amssymb,twocolumn,floatfix,nofootinbib,showpacs,superscriptaddress]{revtex4-1}

\usepackage[utf8]{inputenc}

\usepackage{graphicx,amsmath,amssymb,bm,mathtools}
\usepackage{amsfonts}
\usepackage{verbatim}
\usepackage{float}
\usepackage{amsbsy}

\usepackage{fancyhdr}

\usepackage{color}
\usepackage[labelfont={normalsize},subrefformat=parens,caption=false]{subfig}

\graphicspath{{./figures/}}

\newcommand{\beq}{\begin{equation}}
\newcommand{\eeq}{\end{equation}}
\newcommand{\bml}{\begin{multline}}
\newcommand{\eml}{\end{multline}}

\newcommand{\bseq}{\begin{subequations}}
\newcommand{\eseq}{\end{subequations}}

\newcommand{\fm}{\, \text{fm}}
\newcommand{\fmi}{\, \text{fm}^{-1}}

\newcommand{\eft}{$\chi$EFT}
\newcommand{\kf}{\, \textit{k}_\textrm{{F}}}

\newcommand{\mpi}{m_{\pi}}
\newcommand{\fpi}{F_{\pi}}
\newcommand{\ga}{g_A}
\newcommand{\lchi}{\Lambda_{\chi}}
\newcommand{\spinvec}{\boldsymbol{\sigma}}
\newcommand{\isovec}{\boldsymbol{\tau}}
\newcommand{\spinone}{\boldsymbol{\sigma}_1}
\newcommand{\spintwo}{\boldsymbol{\sigma}_2}
\newcommand{\isoone}{\boldsymbol{\tau}_1}
\newcommand{\isotwo}{\boldsymbol{\tau}_2}

\newcommand{\NLTN}{\text{MSNL}}
\newcommand{\LTN}{\text{MSL}}
\newcommand{\ratio}{R^{3N}_{\text{SO}}}

\newcommand{\la}{\langle}
\newcommand{\ra}{\rangle}

\newcommand{\NNLO}{\ensuremath{{\rm N}{}^2{\rm LO}}}

\newcommand{\kvec}{\mathbf{k}}
\newcommand{\kvecp}{\mathbf{k'}}
\newcommand{\qvec}{\mathbf{q}}
\newcommand{\qvecp}{\mathbf{q'}}
\newcommand{\jvec}{\mathbf{j}}
\newcommand{\jvecp}{\mathbf{j'}}
\newcommand{\pvec}{\mathbf{p}}

\newcommand{\rvec}{\mathbf{r}}

\newcommand{\rhat}{\mathbf{\hat{r}}}
\newcommand{\qhat}{\mathbf{\hat{q}}}

\newcommand{\csunits}{\, \text{MeV}^{-2}}
\newcommand{\ciunits}{\, \text{GeV}^{-1}}
\newcommand{\cdunits}{}
\newcommand{\ceunits}{}

\newcommand{\bi}{\begin{itemize}}
\newcommand{\ei}{\end{itemize}}
\newcommand{\I}{\item}
\newcommand{\be}{\begin{enumerate}}
\newcommand{\ee}{\end{enumerate}}
\newcommand{\bc}{\begin{center}}
\newcommand{\ec}{\end{center}}

\newcommand{\NNcut}{\Lambda_{\text{NN}}}
\newcommand{\TNcut}{\Lambda_{\text{3N}}}

\newcommand{\appropto}{\mathrel{\vcenter{
  \offinterlineskip\halign{\hfil$##$\cr
    \propto\cr\noalign{\kern2pt}\sim\cr\noalign{\kern-2pt}}}}}

\newcommand\numberthis{\addtocounter{equation}{1}\tag{\theequation}}

\begin{document}

\title{Regulator Artifacts in Uniform Matter for Chiral Interactions}

\author{A. Dyhdalo}
\email{dyhdalo.2@osu.edu}
\affiliation{Department of Physics, The Ohio State University, Columbus, OH 43210, USA}

\author{R.J. Furnstahl}
\email{furnstahl.1@osu.edu}
\affiliation{Department of Physics, The Ohio State University, Columbus, OH 43210, USA}

\author{K. Hebeler}
\email{kai.hebeler@physik.tu-darmstadt.de}
\affiliation{Institut f\"ur Kernphysik, Technische Universit\"at Darmstadt, 64289 Darmstadt, Germany}
\affiliation{ExtreMe Matter Institute EMMI, GSI Helmholtzzentrum f\"ur Schwerionenforschung GmbH, 64291 Darmstadt, Germany}

\author{I. Tews}
\email{itews@uw.edu}
\affiliation{Institut f\"ur Kernphysik, Technische Universit\"at Darmstadt, 64289 Darmstadt, Germany}
\affiliation{ExtreMe Matter Institute EMMI, GSI Helmholtzzentrum f\"ur Schwerionenforschung GmbH, 64291 Darmstadt, Germany}
\affiliation{Institute for Nuclear Theory, University of Washington, Seattle, WA 98195-1550}

\date{\today}

\begin{abstract}
Regulator functions applied to two- and three-nucleon forces are a necessary
ingredient in many-body calculations based on chiral effective field theory interactions.
	These interactions have been developed recently with a variety of different cutoff forms, including regulating both the momentum transfer (local) and the relative momentum (nonlocal).  
	While in principle any regulator that suppresses high momentum modes can be employed, in practice artifacts are inevitable in current power counting schemes.
Artifacts from particular regulators
may cause significant distortions of the physics or may affect 
many-body convergence rates, so understanding their nature is important.
Here we characterize the differences between cutoff effects using uniform matter at
Hartree-Fock and second-order in the interaction as a testbed.  
This provides a clean laboratory to isolate phase-space effects of various
regulators on both two- and three-nucleon interactions.
We test the normal-ordering approximation for three-nucleon forces in nuclear matter and find that the relative size of the residual 3N contributions is sensitive to the employed regularization scheme.
\end{abstract}

\maketitle

%%%%%%%%%%%%%%%%%%%%%%%%%%%%%%%%%%%%%%%%%%%%%%%%%%%%%%%%%%%%%%%%%%%%%%%%%%
%%%%%%%%%%%%%%%%%%%%%%%%%%%%%%%%%%%%%%%%%%%%%%%%%%%%%%%%%%%%%%%%%%%%%%%%%%

\section{Introduction}

Chiral Effective Field Theory (\eft)~\cite{Epelbaum:2008ga,Machleidt:20111} has become the method
of choice for input Hamiltonians and other operators
needed for \textit{ab initio} calculations of 
few- and many-body nuclear systems \cite{Roth:2011ar,
Roth:2011vt,
Lahde:2013uqa,
Maris:2008ax,
Navratil:2010jn,
Bogner:2011kp,
Hagen:2010gd,
Hergert:2013uja,
Hergert:2015awm,
Carbone:2013rca,
Fiorilla:2011sr,
Hebeler:2010xb,
PhysRevC.89.014319,
PhysRevLett.110.032504,
Hagen:2013nca,
Holt:2014hma,
PhysRevC.91.051301,
doi:10.1146/annurev-nucl-102313-025446,
PhysRevC.93.011302,
Cipollone:2013zma,
Barrett:2013nh,
nphys3529}. 
\eft~respects the low-energy symmetries of QCD and promises to be 
  model-independent, systematically improvable in an order-by-order expansion, 
  and have controlled uncertainties from omitted terms.
\eft~is not uniquely specified and there are different competing implementations. 
Hereafter, when we use the term \eft, we are referring specifically to the 
Weinberg power counting scheme with no explicit
$\Delta$-isobar~\cite{Epelbaum:2008ga,Machleidt:20111}. 
   
As with any quantum field theory, the presence of loops 
requires the introduction of a regularization scheme and scale. 
Nonperturbativeness of the nucleon-nucleon (NN) system, as manifested by the shallow deuteron 
bound state and large singlet S-wave scattering length, implies the need to 
resum certain classes of diagrams. 
For the power counting prescription introduced by 
Weinberg~\cite{Weinberg:1990rz,Weinberg:1991um}, 
the NN potential is truncated at a specified order in the chiral expansion
and then iterated, e.g., in the Lippmann-Schwinger equation. 
An analogous procedure is used for many-body forces, e.g., three-nucleon (3N) forces, which are constructed in the chiral expansion and iterated, e.g.,
in the Faddeev equations~\cite{VanKolck:1994yi}.
Ultraviolet (UV) divergences arise
both in the construction of the nuclear potential and in its iteration. 
For the latter, cutoff regularization is applied in all current applications
of \eft.

In implementing the cutoff regularization we specify a function, 
called a regulator, that suppresses the nuclear potentials above
a regularization scale $\Lambda$, called the cutoff. 
The regulator is treated as
an \textit{intrinsic} part of the potential and not a separate entity 
associated only with divergent loops.
Regulators by construction separate unresolved UV
	physics from explicit infrared (IR) physics, whereupon the UV physics is 
	implicitly incorporated via the Lagrangian low-energy constants (LECs).
We require that
the regulator be sufficiently smooth (i.e., not a step function), so that it can be 
used in basis transformations, but this leaves much freedom in the functional
form.

The inclusion of long-range pions in the iteration
for Weinberg power counting means that \eft\
is not fully renormalized order by order~\cite{Kaplan:1996xu}. That is,
there remains a residual cutoff dependence in the theory
at each order. 
The residual scale and scheme dependences are what we call ``regulator artifacts'' (note that regulator artifacts also include regularization dependencies due to breaking symmetries e.g., Lorentz invariance). 
To achieve full model-independence in an EFT, the \emph{predictions} of the theory must 
demonstrate an insensitivity to the choice of regulator and cutoff scale. 
But in contrast to other field theories (e.g., QED), the physics in \eft~does not vary
logarithmically but much more rapidly with the cutoff.
Thus, special attention must be paid to the scheme and scale
being adopted. 
The present work seeks to 
make the impact of these choices and associated regulator artifacts
more transparent. 

	As many-body methods have become increasingly accurate, the focus has shifted back to the chiral Hamiltonian.
	Better understanding of renormalization in Weinberg power counting and being able to quantify uncertainties will be crucial to future precision tests of \eft. 
	Below, we highlight various issues involving regulators arising in current applications. 
\bi
	\I There needs to be adequate suppression of the
	short-range parts of the long-range (pion) potentials.
	Regularization of the highly singular structure in
	two-pion-exchange (TPE) diagrams demonstrates some of these
	subtleties~\cite{Epelbaum:2004fk}, e.g., spurious bound states if the cutoff is chosen too high.
	The functional form of the regulator is also found to impact artifacts in the form of residual cutoff dependence~\cite{Epelbaum:2014efa}.
	\I In addition to cutting off UV physics, regulators 
	should avoid distorting the long-range 
	(IR) parts of the nuclear potentials~\cite{Epelbaum:2014efa} as these parts 
	of the force are assumed to 
	be rigorously connected to QCD through chiral symmetry. 
	\I Some many-body methods, such as 
	Auxiliary Field Diffusion Monte Carlo (AFDMC) and Green's Function Monte Carlo
	(GFMC), need local potentials and local 
	regulators
	to avoid large statistical uncertainties~\cite{Pieper:2001mp,Pieper:2004qh,Gezerlis:2013ipa}.
	\I Regulators can impact the convergence of many-body methods at finite density. 
	A common many-body approximation used with 3N forces is to
	normal-order them with respect to a finite density reference 
	state~\cite{Hagen:2007ew}.
	This leads to density-dependent 
	0-, 1-, and 2-body terms plus a residual 3-body part. 
	The residual contribution is usually assumed to be small 
	(in some cases there has been a numerical check) and discarded
	for computational efficiency (e.g., see Ref.~\cite{Roth:2011vt}).  
\ei
The regulator choice has effects on each of these issues.

 To assess the regulator
dependence in \eft, we  propose studying these interactions
\emph{perturbatively} in a  uniform system. Applying many-body
perturbation theory (MBPT) is  particularly clean and simple in this case, 
and allows the effects of the regulator to be isolated
without worrying about complications such as finite size effects. 
We confine ourselves to the regulator's impact on the Hartree-Fock (HF)
and second-order energy to demonstrate effects for the IR and UV parts of the interaction. 
We also restrict our attention in this paper to the LO NN and $\NNLO$ 3N
interaction terms derived in \eft~at, respectively, order $\nu = 0$
and $\nu = 2$ in the chiral expansion.
These are sufficiently rich for the present investigation.
We assume
natural sizes for all LEC coefficients and do not fit the forces to experimental data (e.g., phase shifts).

With one exception at 3N second-order, we work with pure neutron matter (PNM),
which is more perturbative and simpler to analyze 
than symmetric nuclear matter (SNM). 
In doing so, we build on recent
results by Tews et al.\ in Ref.~\cite{PhysRevC.93.024305}, 
where it was found that the  HF energy in PNM for the $\NNLO$
3N forces has a large dependence on the choice of  the regulator function. 
We emphasize that we do \emph{not} resolve here the question of 
how regulator artifacts are absorbed by the implicit renormalization that occurs 
when constructing realistic interactions; 
our intent is to describe the origin of these artifacts and stimulate further investigations.

When studying the effects of the regulators on the energy, we make extensive use of 
decomposing the NN/3N contributions into their direct and exchange components. 
While the individual pieces in this decomposition are not physical, it is useful
to isolate effects of the regulator on the corresponding different sectors of the potential. 
As an example, certain parts of the $\NNLO$ 3N forces (the $c_4, c_D, c_E$ terms)
vanish in a system of only neutrons~\cite{Hebeler:2009iv}. 
However, the vanishing for the $c_D, c_E$ components
is presupposed on a complete
cancellation between the different 3N
antisymmetric components. Some regulator
choices alter this cancellation by regulating
direct and exchange terms differently, resulting in
non-zero $c_D, c_E$ contributions even in a pure neutron system%
\footnote{The $c_4$ term vanishes 
due to its isospin structure and not due to an antisymmetric cancellation.
As a result the $c_4$ term is always zero in PNM.}.

In all cases, our strategy is to analyze the effects of introducing
the regulator by considering the interaction phase space,
which provides the dominant influence on  
the energy integrand at a given order in MBPT (and the other parts of the
integrand are readily approximated).
Except for the simple case of NN HF, the analytic reduction of MBPT
integrals is quite cumbersome and the resulting expressions not enlightening. 
Instead, we propose analyzing
momentum-space histograms that are Monte Carlo samplings of the relevant momenta. 
These histograms denote where the
primary strength is located in the energy integrands. 
How they are constructed will be explained in
Section~\ref{sec:results} below.

The plan of the paper is as follows.
In Section~\ref{sec:forces_review} we review the basic chiral interactions
at LO and $\NNLO$ for the NN and 3N forces, respectively, and 
define a range of regulators that have been chosen for calculations
in each sector. 
In Section~\ref{sec:results} we analyze the energy
contributions at first and second order in MBPT using different combinations of forces and regulators. 
Section~\ref{sec:outlook} then concludes with a summary and future
issues that need to be examined.

%%%%%%%%%%%%%%%%%%%%%%%%%%%%%%%%%%%%%%%%%%%%%%%%%%%%%%%%%%%%%%%%%%%%%%%%%%
%%%%%%%%%%%%%%%%%%%%%%%%%%%%%%%%%%%%%%%%%%%%%%%%%%%%%%%%%%%%%%%%%%%%%%%%%%

\section{NN/3N Chiral Forces and Regulators}
\label{sec:forces_review}

\subsection{LO NN Forces}

\begin{figure}[tbh]
	\includegraphics[scale = 0.55]{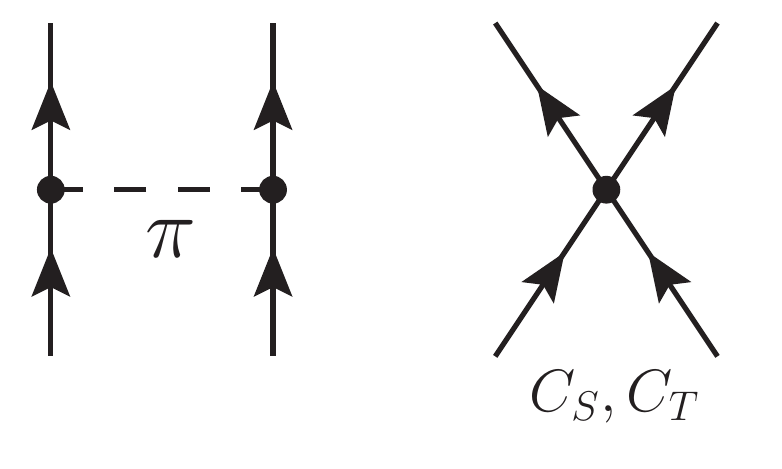}
	\caption{The leading order chiral NN forces~\cite{Epelbaum:2008ga,Machleidt:20111}.}
	\label{fig:NN_forces}
 \end{figure}

The NN forces at leading-order (LO) in the chiral  expansion~\cite{Epelbaum:2008ga,Machleidt:20111}
are sufficiently general for our first look at regulator artifacts in that they contain 
both long-range and short-range pieces.
At lowest order, there are two independent contact terms with
LECs $C_S, C_T$  and a static one-pion exchange
(OPE) diagram (see Fig.~\ref{fig:NN_forces}), so the potential
in momentum space can be written:
\bseq
  \beq
    V^{\text{NN}}_{\text{LO}} = V^{\text{NN}}_{\text{c}} + V^{\text{NN}}_{\pi} \; ,
  \eeq
where
  \beq
   V^{\text{NN}}_{\text{c}} = C_S + C_T (\spinone \cdot
   \spintwo) \; , 
  \eeq
  \beq
    V^{\text{NN}}_{\pi}(\qvec) = - \frac{\ga^2}{4\fpi^2}
    \frac{(\spinone
    \cdot
    \qvec) \; (\spintwo \cdot \qvec)}
    {q^2 + \mpi^2} \isoone \cdot
    \isotwo \; .
    \label{eq:LO_NN_forces}
  \eeq
\eseq
In terms of incoming (outgoing) single-particle momenta
$\pvec_1,\pvec_2$ $(\pvec'_{1},\pvec'_{2})$,
the momentum transfer $\qvec$ and the relative
momentum  $\kvec$ (for later use) are
\beq
 \qvec \equiv \pvec_1 - \pvec_{1}' = \pvec'_{2} - \pvec_2 \;,
 \quad 
 \kvec \equiv \frac{\pvec_1 - \pvec_2}{2} \;.
\label{eq:rel/tran_mom}
\eeq  
For all calculations in this paper, the axial coupling constant $\ga = 1.267$ is used along with $C_S = 1.0~\csunits$. 

Because nucleons are fermions, our potentials need to be antisymmetric under particle exchange. To this end, we define the antisymmetrizer $A_{12}$ 
\beq
	A_{12} \equiv  (1 - P_{12}),
\eeq
where $P_{12}$ is the exchange operator for particles $1$ and $2$. At HF and 2nd order, expressions with an even number of exchange operators are dubbed ``direct" diagrams while expressions with an odd number of exchange operators are called ``exchange" diagrams.

The static OPE potential can also be separated in momentum space 
into two different terms, long-range (LR) and short-range (SR)\footnote{The terminology long-range and short-range is somewhat a misnomer here.  It is used for convenience to distinguish the contact part of the OPE potential. The long-range part of the OPE still has `short-range' components, i.e., a $1/r^3$ term in the tensor.},
\bseq
\beq
V^{\text{NN}}_{\pi}(\qvec) =
V^{\text{NN}}_{\pi, \text{LR}}(\qvec)
+
V^{\text{NN}}_{\pi, \text{SR}}(\qvec)
 \;,
\eeq
where
\beq
 V^{\text{NN}}_{\pi, \text{LR}}(\qvec) = - \frac{\ga^2}{12\fpi^2}
    \left[ \frac{q^2 \; S_{12}(\qhat)}
    {q^2 + \mpi^2} 
	- \frac{\mpi^2 \spinone \cdot \spintwo}{q^2 + \mpi^2}    
    \right] \isoone \cdot
    \isotwo \; ,
    \label{eq:OPE_no_delta}
\eeq
\beq
V^{\text{NN}}_{\pi, \text{SR}}(\qvec) =
 -\frac{\ga^2}{4\fpi^2} 
 \frac{\spinone \cdot \spintwo}{3}
 \isoone \cdot \isotwo
 \label{eq:OPE_contact_mom}
 \;.
\eeq
The tensor operator $S_{12}(\qhat)$ defined as, 
\beq
       S_{12}(\qhat) \equiv 3(\spinone \cdot \qhat)
       (\spintwo 
       \cdot \qhat) - \spinone \cdot
	   \spintwo \;,
\eeq
\eseq
where $\qhat$ denotes the momentum transfer
unit vector and $q \equiv |\qvec|$.
The above separation corresponds to subtracting off the short-range contact part of the OPE potential.

By taking the Fourier transform of $V_{\pi}(\qvec)$ in \eqref{eq:LO_NN_forces}, we can 
express the OPE potential in coordinate space:
   \bseq
	 \beq
	 V^{\text{NN}}_{\pi}(\rvec)
	 = V^{\text{NN}}_{\pi, \text{LR}}(\rvec)
	 +
	 V^{\text{NN}}_{\pi, \text{SR}}(\rvec)
	 \;,
	 \eeq   
where   
     \begin{multline}
     V^{\text{NN}}_{\pi, \text{LR}}(\rvec) = \frac{\ga^2}{4\fpi^2} \frac{\mpi^2}{12\pi}
      \frac{e^{-\mpi r}}{r} \\ \null\times\bigg[\left(1 + \frac{3}{\mpi r} + 
     \frac{3}{(\mpi r)^2} \right) S_{12}(\rhat)
     +  \spinone \cdot
	 \spintwo \bigg] \isoone \cdot \isotwo \; ,
	 \label{eq:coor_opep}
     \end{multline}
     \beq
	 V^{\text{NN}}_{\pi, \text{SR}}(\rvec) =     
     - \frac{\ga^2}{12\fpi^2} \delta^3(\rvec) 
	 \spinone \cdot
	 \spintwo \; 
	 \isoone 
	 \cdot
	 \isotwo \;.
	 \label{eq:OPE_contact_pos}
     \eeq
   \eseq
Here $r$ denotes the magnitude  of the relative distance and $\rhat$ is its
unit vector. As before, the potential
can be separated into a short-range contact part 
along with long-range central and tensor
contributions.

In the following, we work exclusively with the long-range part
of the OPE potential. That is to say, by OPE we are referring \emph{only}
to $V^{\text{NN}}_{\pi, \text{LR}}(\qvec)$ in \eqref{eq:OPE_no_delta} for momentum space and $V^{\text{NN}}_{\pi, \text{LR}}(\rvec)$ in \eqref{eq:coor_opep} for coordinate space. 
Including the contact part of OPE is superfluous for our purposes as its behavior under regularization is the same as for the $C_S,C_T$ terms.
	Furthermore,
absorbing the OPE delta function into the leading-order contact avoids mixing contact regularization effects with the
remaining central and tensor parts of the OPE potential. 
Explicitly separating out the delta function from the OPE potential is standard practice for potentials regulated in coordinate space.

%%%%%%%%%%%%%%%%%%%%%%%%%%%%%%%%%%%%%%%%%%%%%%%%%%
%
%%%%%%%%%%%%%%%%%%%%%%%%%%%%%%%%%%%%%%%%%%%%%%%%%%

For energies to be finite at second-order in
MBPT, a regularization scheme must be introduced. For a general local
NN potential, there will only be one independent momentum
that needs to be regulated after taking momentum conservation into 
consideration. Regulators in general can either be local or nonlocal. By definition, 
local regulators (and potentials) are functions purely of the relative distance $\rvec$ 
in coordinate space or the momentum transfer $\qvec$ in momentum space. 
Nonlocal regulators (and potentials) have additional dependencies other than just $\rvec$ or $\qvec$. 

\begin{figure}[tbh]
\includegraphics[width=0.8\columnwidth]{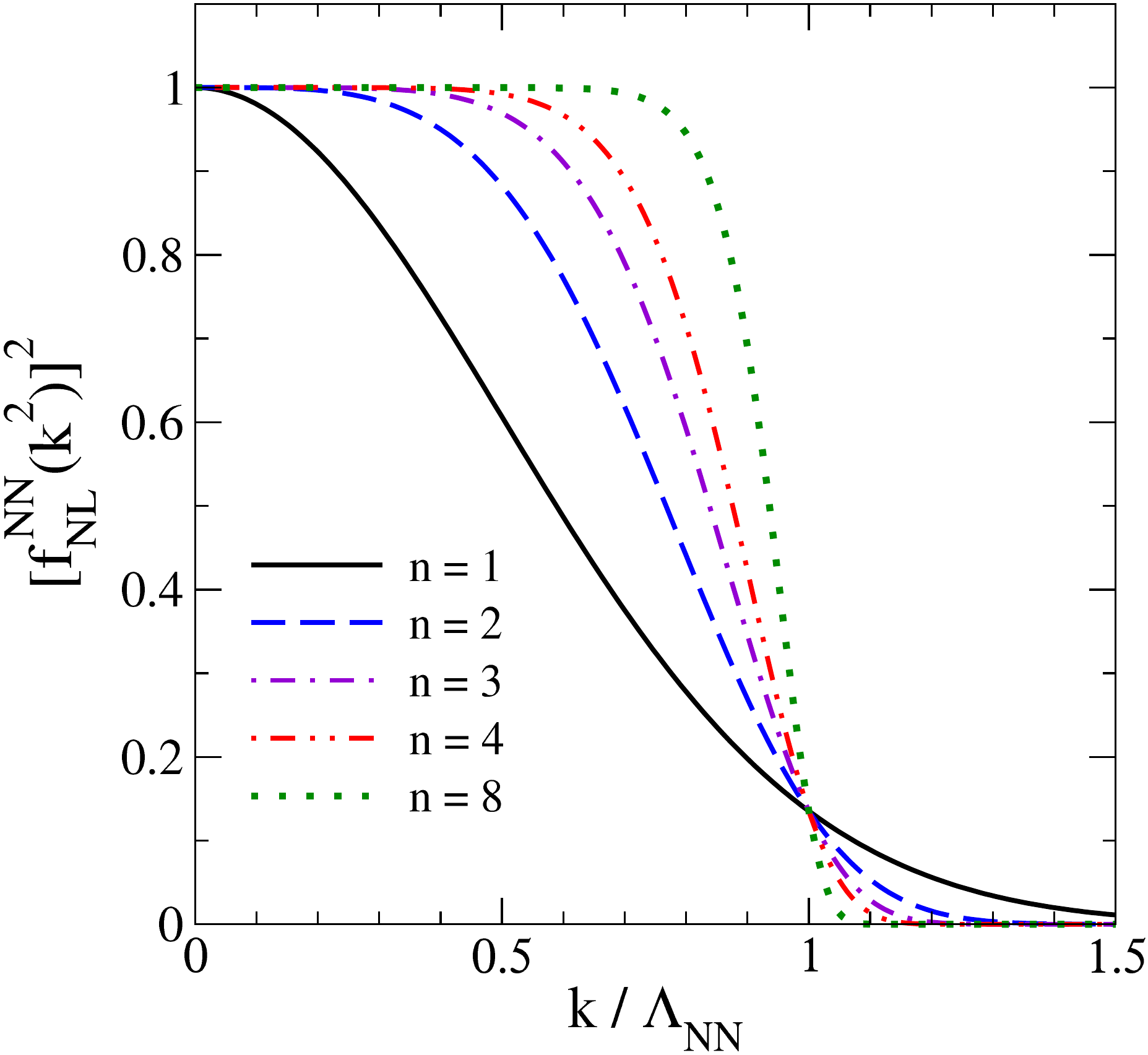}~
\caption{Examples of different choices of $n$, from 1 to 8, in the
nonlocal regulator exponential of Eq.~\eqref{eq:NN_reg_nonlocal}. The x-axis is the magnitude of the
relative momentum scaled by the cutoff and the y-axis is the
regulator function squared.}
\label{fig:NN_NL_reg_choices}
\end{figure}

One popular choice
is a nonlocal regulator, which we call momentum space nonlocal (MSNL), defined to exponentially regulate the relative
momentum magnitude $k$~\cite{Entem:2003ft, Hebeler:2009iv,Epelbaum:2004fk},
\beq
	f_{\text{MSNL}}^{\text{NN}}(k^2) = \exp\left[- 
	        \left( k^2 / \NNcut^2 \right)^n  \right] \;,
  \label{eq:NN_reg_nonlocal}
\eeq
where $\NNcut$ is the NN cutoff in momentum
space and $n$ is a fixed integer. For current NN calculations, typical 
values include $n=1-3$ and $\NNcut = 450-600~\text{MeV}$~\cite{Epelbaum:2014efa,Hebeler:2009iv,Kruger:2013kua}.
The relative momentum magnitudes both before and after the 
interaction are regulated to satisfy hermiticity, so the potential 
assumes the form:
\beq
	V^{\text{NN}}(\kvec,\kvec') \implies
	f_{\text{MSNL}}^{\text{NN}} (k^2) \; 
	  V^{\text{NN}}(\kvec,\kvec') \; 
	f_{\text{MSNL}}^{\text{NN}} (k'^2) \; ,
	\label{eq:NL_regulated_potential}
\eeq
where $\kvec\,(\kvec')$ denotes the relative momentum before (after) the interaction. 
These regulators are symmetric under individual nucleon permutation so
that direct and  exchange pieces of the antisymmetric  potential are
regulated identically. Under a partial wave decomposition of the
potential, all waves are also cut off in the same way.
Figure~\ref{fig:NN_NL_reg_choices} shows the effect of different values of $n$
(e.g., on a diagonal potential in Eq.~\eqref{eq:NL_regulated_potential}), which can
be compared to the $n\rightarrow\infty$ limit of a step function at $k=\NNcut$.

A different approach is to
use a local regulator, which we call momentum space local (MSL), that depends on the momentum transfer magnitude $q$, as in 
Ref.~\cite{Lepage:1997cs},
\beq
	f_\text{MSL}^{\text{NN}} (q^2) 
	= \exp\left[ - \left( q^2 / \NNcut^2 \right)^n  \right] \; ,
\label{eq:NN_reg_local}
\eeq
such that,
\beq
  V^{\text{NN}}(\qvec) \implies 
  V^{\text{NN}}(\qvec) 
  f_\text{MSL}^{\text{NN}} (q^2)	 
  \; ,
\eeq
where we have written the NN
potential in local form as
a pure function of $\qvec$.
As the regulator in \eqref{eq:NN_reg_local} is not symmetric under
single-particle  permutation, the direct and exchange parts of the
potential are not regulated in the same way. Likewise, different partial
waves will experience different cutoff artifacts.

An alternative local  approach is to regulate
in coordinate space on the magnitude of the relative distance $r$ with
some coordinate space cutoff $R_0$. 
Depending on the exponent $n$, these regulators may have oscillatory behavior when transformed to momentum space and display different
 behavior from local momentum space regulators.
For the coordinate-space regulated OPE, these different regularization schemes have the least
effect in high partial waves because one is cutting off short-distance (small $r$) parts of the potential.
A fully local choice used in some quantum Monte Carlo calculations, which we label CSL,
is to use~\cite{Gezerlis:2014zia},
\beq
	f_{\text{CSL}}^{\text{NN}} (r^2) = 
	\left(1 - \exp\left[ - \left(r^2 / R_0^2 \right)^n \right] \right) \; ,
	\label{eq:CSL}
\eeq
to regulate the long-range part of the OPE potential,
\beq
   V^{\text{NN}}_{\pi, \text{LR}}(\rvec) \implies V^{\text{NN}}_{\pi, \text{LR}}(\rvec) 
   \; f_{\text{CSL}}^{\text{NN}} (r^2)
   \;,
\eeq
which cuts off the short distance (small r) parts
of the OPE potential~\cite{Epelbaum:2014efa}. 
The short-range contacts and short-range OPE are regulated by replacing the
Dirac delta function with a smeared delta
function~\cite{Gezerlis:2013ipa,Gezerlis:2014zia},
\beq
	\delta(\rvec) \to \delta_{R_0}(\rvec) = \alpha_n e^{-(r^2/R_0^2)^n}
	\;,
\label{eq:smeared_delta}
\eeq
where $\alpha_n$ is a normalization coefficient, chosen such that
\beq
	\int d^3r \; \delta_{R_0}(\rvec) = 1 \; .
\eeq

It is also possible to mix local and nonlocal forms. 
One semi-local choice developed by Epelbaum, Krebs, and Mei{\ss}ner, 
which we label EKM, is to use~\cite{Epelbaum:2014efa}
\beq	f_{\text{EKM}}^{\text{NN}} (r^2) = 
	\left(1 - \exp\left[ - \left(r^2 / R_0^2 \right) \right] \right)^n \; ,
	\label{eq:ekm}
\eeq
for the long-range OPE potential,
\beq
V^{\text{NN}}_{\pi, \text{LR}}(\rvec) \implies V^{\text{NN}}_{\pi, \text{LR}}(\rvec) \; f_{\text{EKM}}^{\text{NN}} (r^2) \; ,
\eeq
and use \eqref{eq:NN_reg_nonlocal} on the 
short-range contacts (and short-range OPE). The EKM long-range regularization
is sufficient to make
the previously used spectral function regularization
of the highly singular TPE potential unnecessary for $n\geq 4$~\cite{Epelbaum:2014efa}. 
Current NN implementations use 
$R_0 = 0.8-1.2~\fm$ as typical cutoffs~\cite{PhysRevLett.115.122301}.

We summarize the different NN regulator
schemes used in this paper in
Table~\ref{tab:reg_table}.

%%%%%%%%%%%%%%%%%%%%%%%%%%%%%%%%%%%%%%%%%%%
%
%%%%%%%%%%%%%%%%%%%%%%%%%%%%%%%%%%%%%%%%%%%

\begingroup
\squeezetable
\begin{table}
 \caption{\label{tab:reg_table}
Summary table for various regulator combinations in the NN sector defined in the
text, with equation 
references to the regulators in parenthesis. OPE refers to \eqref{eq:OPE_no_delta} 
in momentum space and \eqref{eq:coor_opep} in coordinate space. 
Contacts refers to both the $C_S,C_T$ terms along with the OPE contact in
 \eqref{eq:OPE_contact_mom} and \eqref{eq:OPE_contact_pos}.}
 \begin{ruledtabular}
 \begin{tabular}{c || c | c | c }
  Scheme & Type & OPE & Contacts \\
 \hline
 MSNL & nonlocal & nonlocal \eqref{eq:NN_reg_nonlocal} & nonlocal \eqref{eq:NN_reg_nonlocal}   \\
 
 MSL  & local & local \eqref{eq:NN_reg_local} & local \eqref{eq:NN_reg_local}\\

 EKM  & semi-local & local \eqref{eq:ekm} & nonlocal \eqref{eq:NN_reg_nonlocal} \\
 
 CSL  & local & local \eqref{eq:CSL} & local \eqref{eq:smeared_delta} \\
 \end{tabular}
 \end{ruledtabular}
\end{table}
\endgroup

\subsection{\NNLO~3N Forces}

\begin{figure}[tbh]
	\includegraphics[scale = 0.70]{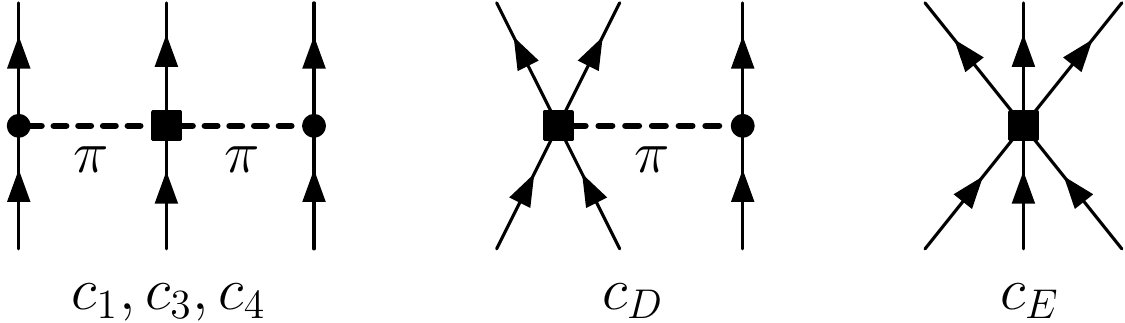}
	\caption{The \NNLO~chiral 3N forces~\cite{Epelbaum:2008ga,Machleidt:20111}. 
 }
	\label{fig:3N_forces}
 \end{figure}	  

The \NNLO~3N forces
\cite{VanKolck:1994yi,Epelbaum:2002vt}
(see also \cite{Hammer:2012id})
in the $\Delta$-less 
\eft~consist
of a long-range TPE with $c_i$ coefficients determined from
$\pi N$ scattering, a single-pion exchange with a short-range contact $c_D$, and a pure
contact $c_E$ term (see Fig.~\ref{fig:3N_forces}):
\bseq
  \beq
  	V_{\NNLO}^{\text{3N}} = V^{\text{3N}}_{2\pi} + V^{\text{3N}}_{\text{D}}
  	+ V^{\text{3N}}_{\text{E}} \; ,
  \eeq
  \beq
  	V^{\text{3N}}_{2\pi} = \frac{1}{2} \left(\frac{\ga}{2\fpi}\right)^2
  	\sum_{i \neq j \neq k} 
  	  \frac{\left(\spinvec_i \cdot \qvec_i \right) 
  	        \left(\spinvec_j \cdot \qvec_j \right)}
  	       {\left(q_i^2 + \mpi^2 \right) 
  	        \left(q_j^2 + \mpi^2 \right)} 
  	 F^{\alpha \beta}_{ijk} \tau^{\alpha}_i \tau^{\beta}_j 
  	 \; ,
  	 \label{eq:TPE}
  \eeq
  \begin{multline}
  	F^{\alpha \beta}_{ijk} = \delta^{\alpha \beta} 
  	  \left[
  	     - \frac{4 c_1 \mpi^2}{\fpi^2} 
  	     + \frac{2c_3}{\fpi^2} \qvec_i \cdot \qvec_j 
  	  \right]
  	\\
  	\null + 
  	\sum_{\gamma} \frac{c_4}{\fpi^2} 
  	   \epsilon^{\alpha \beta \gamma} \tau^{\gamma}_k \spinvec_k 
  	   \cdot \left(\qvec_i \times \qvec_j \right) \; ,
    	   \label{eq:ci_terms}
  \end{multline}
  \beq
  	V^{\text{3N}}_{\text{D}} = -\frac{\ga}{8 \fpi^4} \frac{c_D}{\lchi}
  	  \sum_{i \neq j \neq k}
  	  \frac{\left(\spinvec_j \cdot \qvec_j  \right) 
  	        \left(\spinvec_i \cdot \qvec_j \right)}
  	       {q_j^2 + \mpi^2}
  	       \isovec_i \cdot \isovec_j 
  	   \; ,
  	   \label{eq:3N_OPE}
  \eeq
  \beq
  	V^{\text{3N}}_{\text{E}} = \frac{c_E}{2 \fpi^4 \lchi}
  	\sum_{j \neq k} \left(\isovec_j \cdot \isovec_k \right) \; ,
  \eeq
  \label{eq:LO_3N_forces}
\eseq

\noindent
where the subscripts $i,j,k$ are particle indices.%
\footnote{Note that the $\Lambda_{\chi}$ appearing in the 3N potentials
is distinct from $\NNcut$
and $\TNcut$. $\Lambda_{\chi}$ 
denotes the estimated breakdown scale
of \eft~while $\NNcut$ and 
$\TNcut$ come purely from regulating the EFT.} For all calculations in this paper, $c_i = 1.0~\ciunits$, $c_D = 1.0$, $c_E = 1.0$.

As in the 2-body sector we define an antisymmetrizer $A_{123}$ to ensure that our 3N potential is antisymmetric under particle exchange,
\beq
	A_{123} \equiv 
	(1 - P_{12})(1 - P_{13} - P_{23})  \; .
	\label{eq:3N_antisym}
\eeq
Depending on the number of exchange operators in our energy expressions, we have ``direct", ``single-exchange", and ``double-exchange" diagrams.
	
%%%%%%%%%%%%%%%%%%%%%%%%%%%%%%%%%%%%%%%%%%%%%%%%%%
%
%%%%%%%%%%%%%%%%%%%%%%%%%%%%%%%%%%%%%%%%%%%%%%%%%%	

For calculations to be finite past first order in perturbation theory,
we again need to introduce a regularization scheme for our 3-body
potentials. For a local 3N potential, there will in general be
2 independent momenta after momentum conservation.
One commonly used choice is a nonlocal regulator completely
symmetric in the single-particle momenta, 
\begin{multline}
	f_{\NLTN}^{\text{3N}}
	 (\mathbf{p}_1,\mathbf{p}_2,\mathbf{p}_3) =
	 \\
	\exp \bigg[ - \bigg(\frac{p_1^2 + p_2^2 + p_3^2- \mathbf{p}_1
	\cdot \mathbf{p}_2 - \mathbf{p}_1 \cdot \mathbf{p}_3 	
	-
	\mathbf{p}_2 \cdot \mathbf{p}_3}{3 \TNcut^2} \bigg)^n
	\bigg] \; ,
	\\
	\label{eq:3N_nonlocal}
\end{multline}
which we call $\NLTN$.
Like its 2-body nonlocal counterpart, this regulator retains its functional
form under permutation of the nucleon indices and thus regulates each
antisymmetric piece of the 3-body potential in the same way. 
The nonlocal regulator can be equivalently
written in terms of the magnitudes of the 3-body Jacobi momenta, 
\beq
	f_{\NLTN}^{\text{3N}}
	 (k^2, j^2) = \exp \bigg[ - \bigg(\frac{k^2 + \frac{3}{4} j^2}{\TNcut^2} \bigg)^n \bigg] \; ,
	\label{eq:3N_nonlocal_alt}
\eeq
where we define the Jacobi momenta $\jvec, \kvec$ with respect to the $1,2$ particle subsystem,
\beq
	\jvec = \frac{2}{3} \left({\bf p}_3
	- \frac{{\bf p}_2 + {\bf p}_1}{2} \right) \; ,
	\qquad
	\kvec = \frac{{\bf p}_2 - {\bf p}_1}{2} \; .
	\label{eq:3N_Jacobi}
\eeq
To satisfy hermiticity, again we regulate on both the incoming and outgoing Jacobi momenta,
\begin{align*}
	&V^{\text{3N}} (\jvecp, \kvecp ; \jvec, \kvec) 
	\implies  \\
	f_{\NLTN}^{\text{3N}}
	 (k'^2, j'^2) \;
	 &V^{\text{3N}} (\jvecp, \kvecp ; \jvec, \kvec) \;
	 f_{\NLTN}^{\text{3N}}
	 (k^2, j^2) \; .
	\numberthis 
\end{align*}
Common choices for the 3N $\NLTN$ regulator 
include $n = 2-3$ and $\TNcut = 400-600~\text{MeV}$ ~\cite{Epelbaum:2002vt,
Bogner:2009un,PhysRevC.89.014319}.
Usually $\TNcut$ is chosen to be equal to $\NNcut$, but the necessity for this
has not been established.

Another choice, which we dub $\LTN$, is the Navratil local regulator defined as
\cite{Navratil:2007zn}, 
\beq
	f_{\LTN}^{\text{3N}}
	(q_i^2) = \exp \left[- \left( \frac{\mathbf{q}_i^2} 
	{\TNcut^2} \right)^n \right]
	\;,
	\label{eq:3N_local}
\eeq
where e.g., $\mathbf{q}_1 = \mathbf{p}_1 - \mathbf{p}_{1}'$ is the momentum
transfer in terms of individual nucleon three-momenta with 
$\mathbf{p}_1~(\mathbf{p}_{1}')$ being the momenta before (after) the interaction. 
The 3N potential expressed in local form after regularization becomes,
\beq
	V^{\text{3N}} ({\bf q}_i, {\bf q}_j) 
	\implies 
	f_{\LTN}^{\text{3N}}	(q_i^2) \;
	V^{\text{3N}} ({\bf q}_i, {\bf q}_j) \;
	f_{\LTN}^{\text{3N}} (q_j^2) \; ,
	\label{eq:3N_reg_potential_local}	
\eeq
where the subscripts $i,j$ refer to momentum transfers between different single-particle momenta.
Like the local momentum space regulator in \eqref{eq:NN_reg_local}, the Navratil local 
regulator is not symmetric under individual nucleon permutations.
As such, the different parts of the fully antisymmetric 3N potential are all regulated differently. 
This also results in ambiguities in deciding how to regulate different
parts of the long-range 3N forces depending on if 
the regulator momentum labels $i,j$ match the 
spin-isospin labels in the 3N potential~\cite{PhysRevC.93.024305, Navratil:2007zn, PhysRevC.85.024003}. 
Taking the potential  
$V^{\text{3N}}_{D}$
with LEC $c_D$ as an example,
we denote below two 
different regularization structures following Ref.~\cite{Navratil:2007zn},
\bseq
\begin{align}
&f_{\LTN}^{\text{3N}}	(q_i^2) \;
\frac{\left(\spinvec_j \cdot \qvec_j  \right) 
  	        \left(\spinvec_i \cdot \qvec_j \right)}
  	       {q_j^2 + \mpi^2}
  	       \;
  	       \isovec_i \cdot \isovec_j \;
f_{\LTN}^{\text{3N}}	(q_j^2) \; , 
\label{eq:3N_local_choice_a}
\\
&f_{\LTN}^{\text{3N}}	(q_k^2) \;
\frac{\left(\spinvec_j \cdot \qvec_j  \right) 
  	        \left(\spinvec_i \cdot \qvec_j \right)}
  	       {q_j^2 + \mpi^2}
		\;  	      
  	       \isovec_i \cdot \isovec_j
\;
f_{\LTN}^{\text{3N}}	(q_j^2) \; .
\label{eq:3N_local_choice_b}
\end{align}
\eseq
The momentum transfer labels in the regulators match the spin-isospin indices 
in \eqref{eq:3N_local_choice_a}, whereas only one index is matched in \eqref{eq:3N_local_choice_b}. 
In this paper, for the purposes of calculation, we adopt the convention of 
Eq.~\eqref{eq:3N_local_choice_b}.

\section{Results}
\label{sec:results}

To explore the regulator dependence
in \eft, we study the uniform system (infinite, homogeneous, isotropic matter) 
in MBPT. The uniform system has the desirable feature that certain non-perturbative aspects 
of nuclear systems in free-space, e.g., the fine-tuning of the 
NN S-waves, are rapidly damped at finite density~\cite{Bogner:2005sn}. 
In this paper, with the exception of 3N second-order, we work 
exclusively with PNM up to the first two orders in MBPT. 
This is because PNM is simpler and more perturbative than SNM and serves as
a testbed without the complications of including isospin. 
In test cases, we have found similar trends in these two
limiting systems.

In the following, we look first at the NN forces at HF and second-order 
in MBPT, then we examine 3N forces in the same sequence. 
Examining both the HF and the second-order energy allows the probing of different 
parts of the nuclear potentials with a regulator scheme.
The HF energy has the feature of being computable without a regulator and serves 
as a touchstone for examining scheme/scale dependence. As all HF momenta are 
on-shell, regulator effects here are described as IR effects. 
The second-order energy is divergent in the absence of regularization, 
hence artifacts from the regulator here are called UV effects.

\subsection{NN Forces at HF}

For a 2-body interaction, the  HF energy per particle in terms of
single-particle momenta is given by
\begin{multline}
	\frac{E_{\text{HF}}^{\text{NN}}}{N} = \frac{1}{2\rho} 
	  \sum_{\sigma_1, \sigma_2} \sum_{\tau_1, \tau_2} 
	  \int \frac{d^3p_1}{(2\pi^3)}
	  \int \frac{d^3p_2}{(2\pi)^3}
	\\
	  \null\times
	   n({\bf p}_1) n({\bf p}_2) 
	   \la 12 | A_{12} V^{\text{NN}}_{\text{LO}} | 12 \ra
	\;,
	\label{eq:NN_HF_energy}
\end{multline}
where
\bseq
 \beq
	|1 \ra \equiv |{\bf p}_1 \sigma_1 \tau_1 \ra, 
 \eeq
 \beq
    n({\bf p}_1) \equiv
	\Theta (\kf - |{\bf p}_1|),
 \eeq
\eseq
$\rho$ is the nucleon number density, 
$\sigma_i$ ($\tau_i$) is the spin (isospin) operator for the $i$th particle, and  $V^{\text{NN}}_{\text{LO}}$ are the LO chiral
NN forces with a particular regularization scheme.

\begin{figure*}[tbh]
	\includegraphics[width=0.33\textwidth]{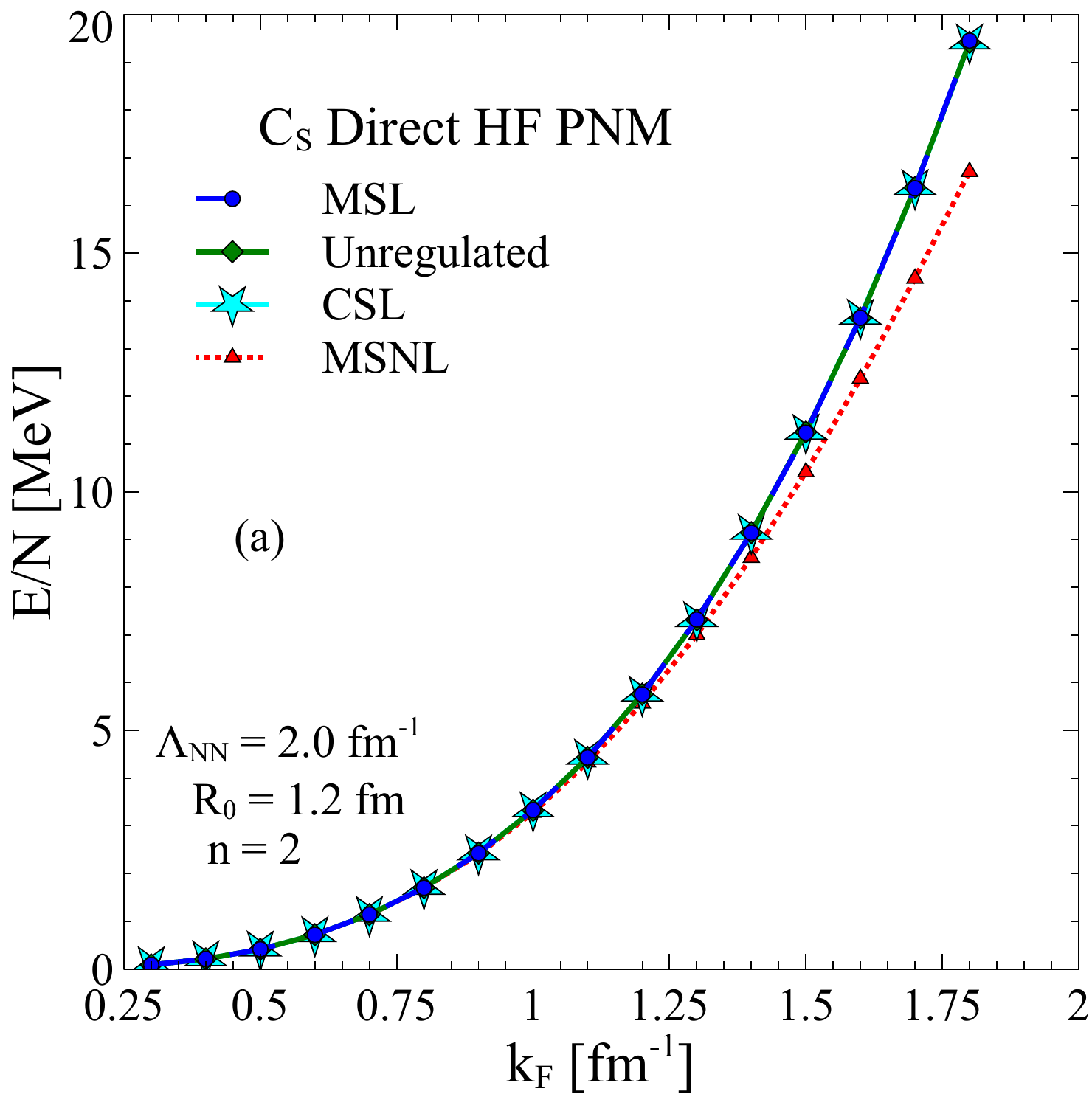}~%
	\includegraphics[width=0.33\textwidth]{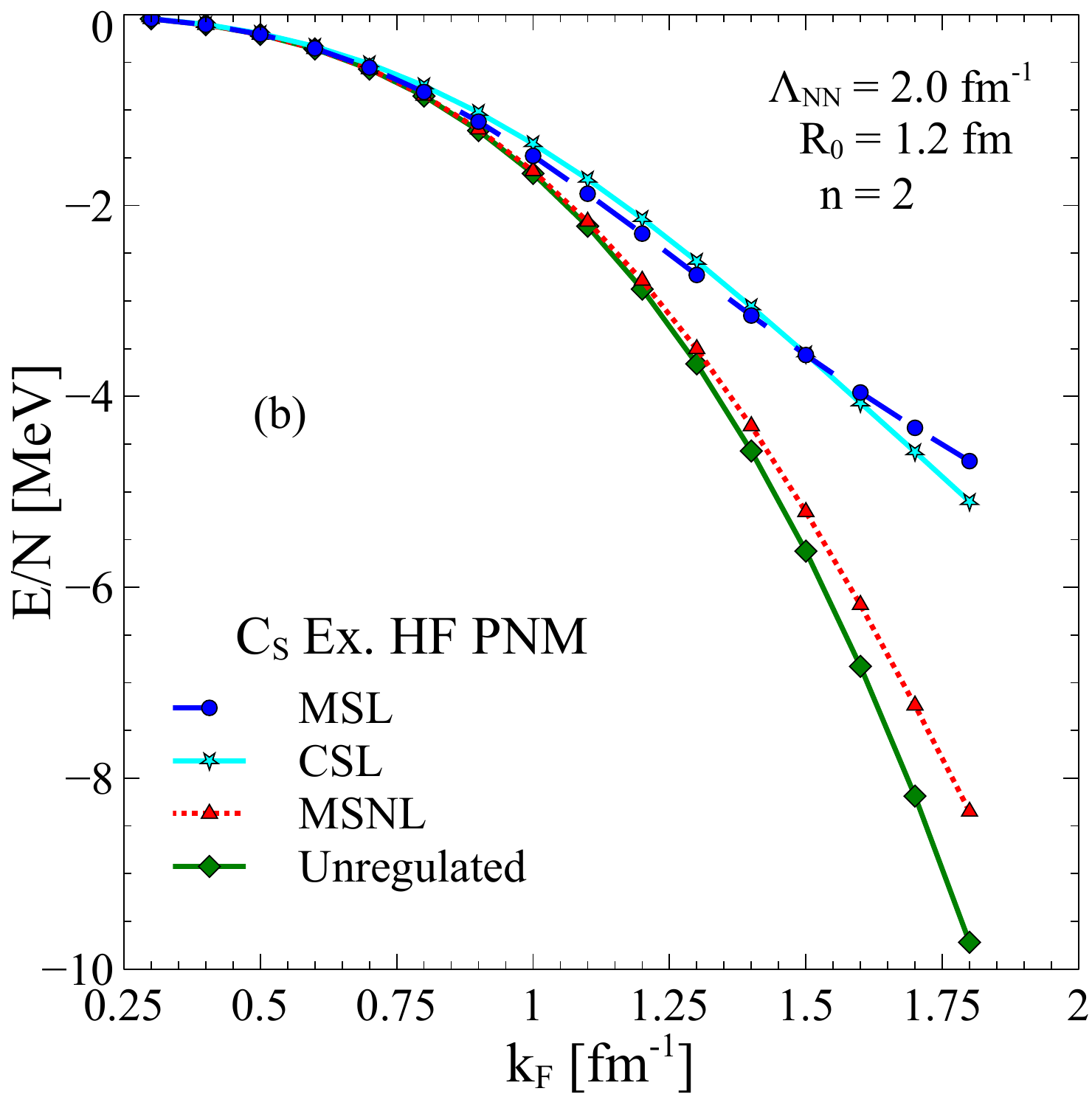}~%
	\includegraphics[width=0.33\textwidth]{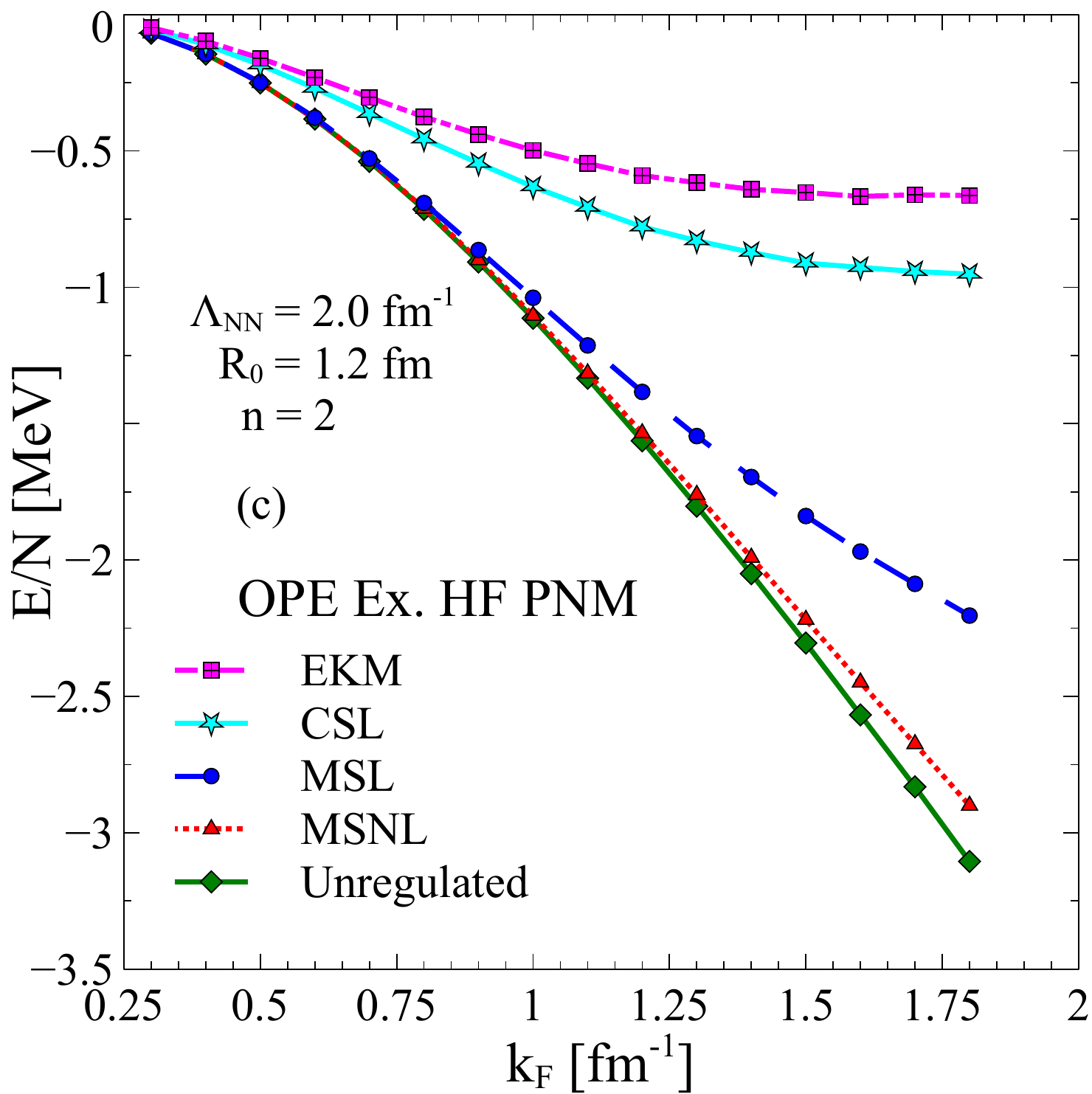}
	\caption{Neutron matter calculations of the HF energy per particle for the direct (a)
	and exchange (b) terms for
	$C_S$ and the OPE exchange term (c) using the regularization schemes in Table~\ref{tab:reg_table}.
	 The $C_T$ calculation has similar behavior to the $C_S$ exchange term. 
	The EKM scheme uses the same regularization as the MSNL scheme for the contact terms. 
	The trends in SNM (not shown) are comparable to those in PNM.
	The calculations use $n = 2$, $\NNcut = 2.0~\fmi$, and $R_{0} = 1.2~\fm$.}
	\label{fig:NN_HF_energy}
\end{figure*}

Evaluating the HF energy using the different regulator schemes in 
Table~\ref{tab:reg_table} yields
the curves in Fig.~\ref{fig:NN_HF_energy} for the $C_S$ and OPE terms.
At this stage we have already separated the direct and exchange parts of the potential, 
$\la 12 | V^{\text{NN}}_{\text{LO}} | 12 \ra $ and
$\la 12 | V^{\text{NN}}_{\text{LO}} | 21 \ra$ respectively, to illustrate differences in regulator behavior on energy calculations. 
Note that there is no direct OPE energy as spin-isospin dependent interactions at HF vanish when performing spin-isospin traces.
Calculations are presented here for soft cutoffs of $\NNcut = 2.0~\fmi$ and $R_0 = 1.2~\fm$ to better highlight regulator artifacts at high density.  
Performing calculations at a more common $\NNcut = 2.5~\fmi$ or $R_0 = 0.9~\fm$ does not alter our qualitative discussion below (see supplemental material). 
The regulator situation at HF is particularly simple and our analysis in this section serves as a proof of principle of how our various diagnostic tools can
explain the systematics of the energy.

The unregulated direct $C_S$ HF energy in Fig.~\ref{fig:NN_HF_energy}(a) is
exactly reproduced for the MSL and CSL regulator schemes because $\qvec = 0$ for the direct diagram. In contrast, the MSNL result is suppressed. 
The exchange $C_S$ HF energy shows a different hierarchy where, in order of absolute magnitude, one finds MSNL $>$ CSL $\sim$ MSL.

\begin{figure*}[tbh]
	\includegraphics[width=0.35\textwidth]{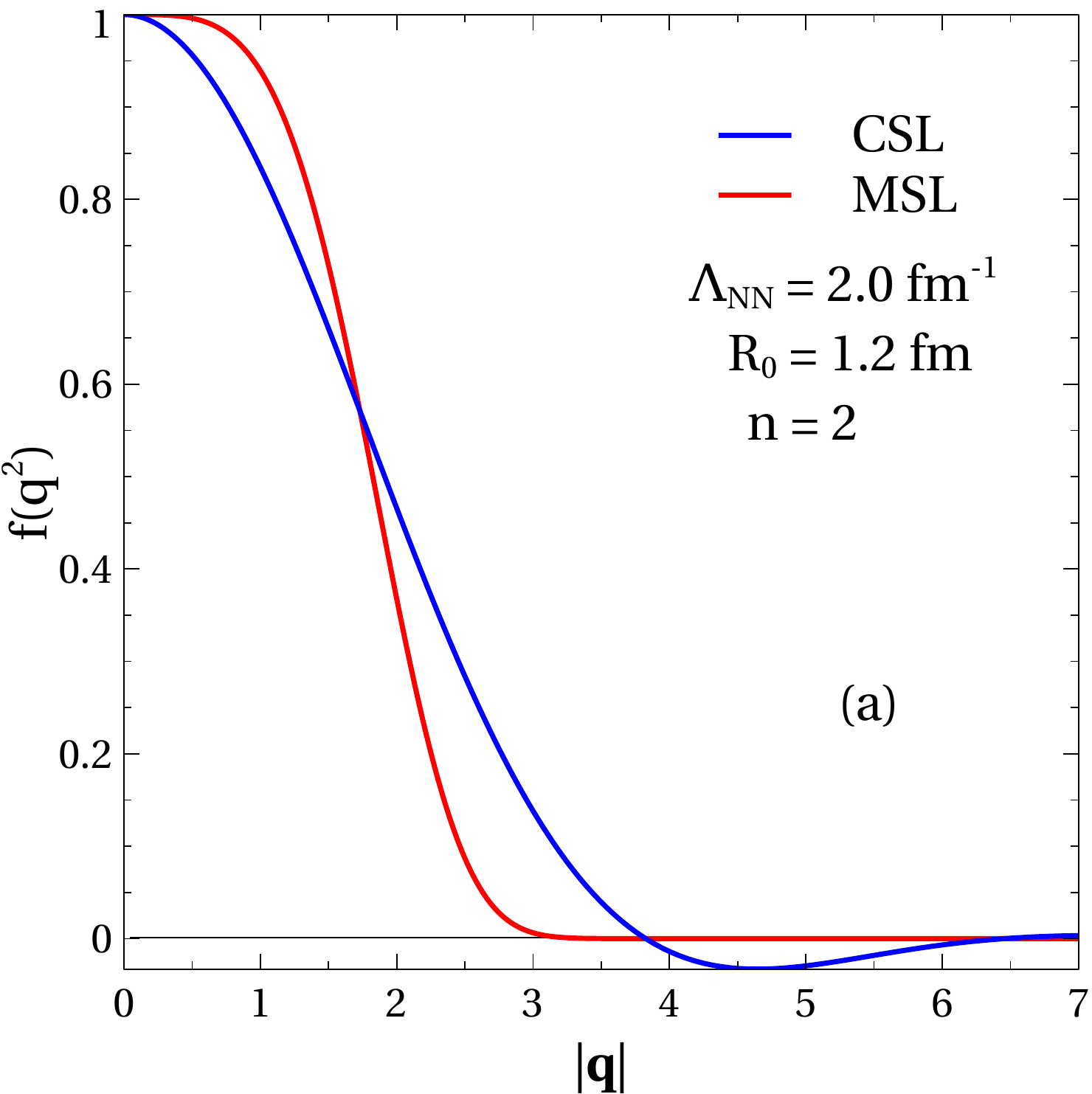}~
\includegraphics[width=0.35\textwidth]{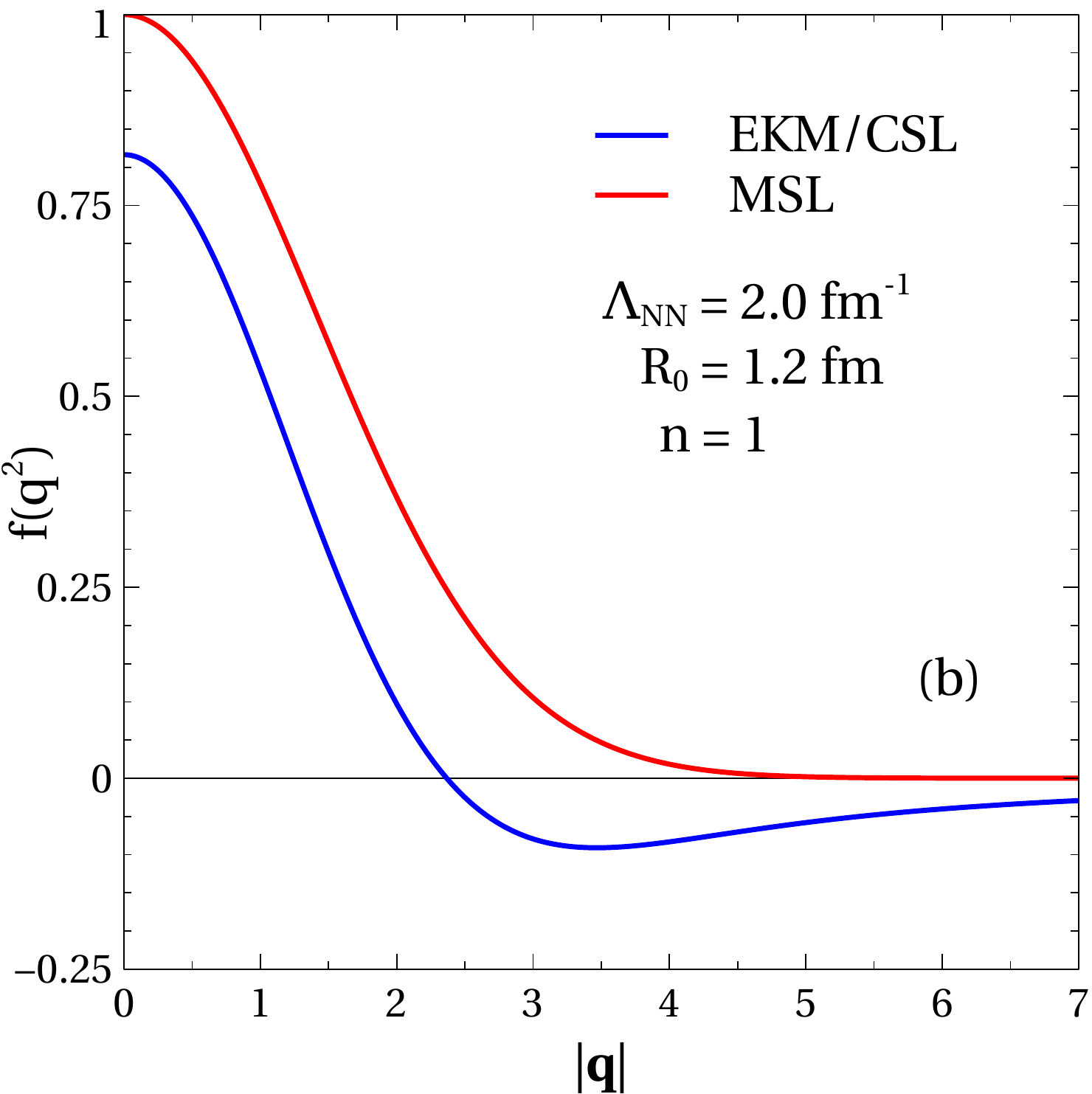}~

	\caption{Plot (a) shows the MSL regulator in \eqref{eq:NN_reg_local} and CSL 
	contact regulator in \eqref{eq:smeared_delta} for $n = 2$, $R_{0} = 1.2~\fm$, $\NNcut = 2.0~\fmi$.
	Plot (b) shows the MSL regulator in \eqref{eq:NN_reg_local} and CSL/EKM OPE regulator in \eqref{eq:CSL} or \eqref{eq:ekm} for $n=1$, $R_{0} = 1.2~\fm$, $\NNcut = 2.0~\fmi$. The regulator function is plotted as a function of the momentum transfer magnitude $|\qvec|$. 
	The $S_{12}$ operator, which vanishes for HF, was set to zero before performing the Fourier transform for EKM/CSL.	
	Hence the OPE EKM/CSL regulator functional form as plotted is valid for HF only.}
	\label{fig:CSL_LOCAL_COMPARE}
\end{figure*}

The exchange $C_S$ HF energies in Fig.~\ref{fig:NN_HF_energy}(b) imply that
the CSL contact regulator in \eqref{eq:smeared_delta} has similar 
behavior to the MSL regulator in \eqref{eq:NN_reg_local} for the 
cutoffs $\NNcut = 2.0~\fmi$ and $R_{0} = 1.2~\fm$.
In the special
case of $n = 1$ with a no-derivative contact,
a straightforward Fourier transform connects these two regulators, i.e.,
\begin{align*}
	\mathcal{F}\left[V^{\text{NN}}_{C_S} 
	\; f_\text{MSL}^{\text{NN}} (q^2)\right] &= C_{S} \int 
	\frac{d^3q}{(2\pi)^3} e^{i \qvec 
	\cdot \rvec} \; e^{-q^2 / \NNcut^2} \\
	&= C_S \alpha_1 \; e^{-r^2 /R_0^2}
	\label{eq:fourier_trans_local}
	\numberthis
\end{align*}
where $R_0 = 
	2 / \NNcut$ and $\alpha_1 = (R_0^3 
	\; \pi^{3/2})^{-1}$. 
	Only for this special case will the regulators be directly related. 
At larger $n$, the relations become more complicated hypergeometric functions and the correspondence between $\NNcut$ and $R_{0}$ is no longer clean. 
We plot the choice of $n = 2$ for contact CSL and MSL in Fig.~\ref{fig:CSL_LOCAL_COMPARE}(a) to illustrate this different behavior. 
The oscillatory nature of the Fourier transformed regulator implies that no simple redefinition of $\NNcut$ or $R_0$ will completely equate the two regulators.

\begin{figure}[tbh]
\includegraphics[width=0.9\columnwidth]{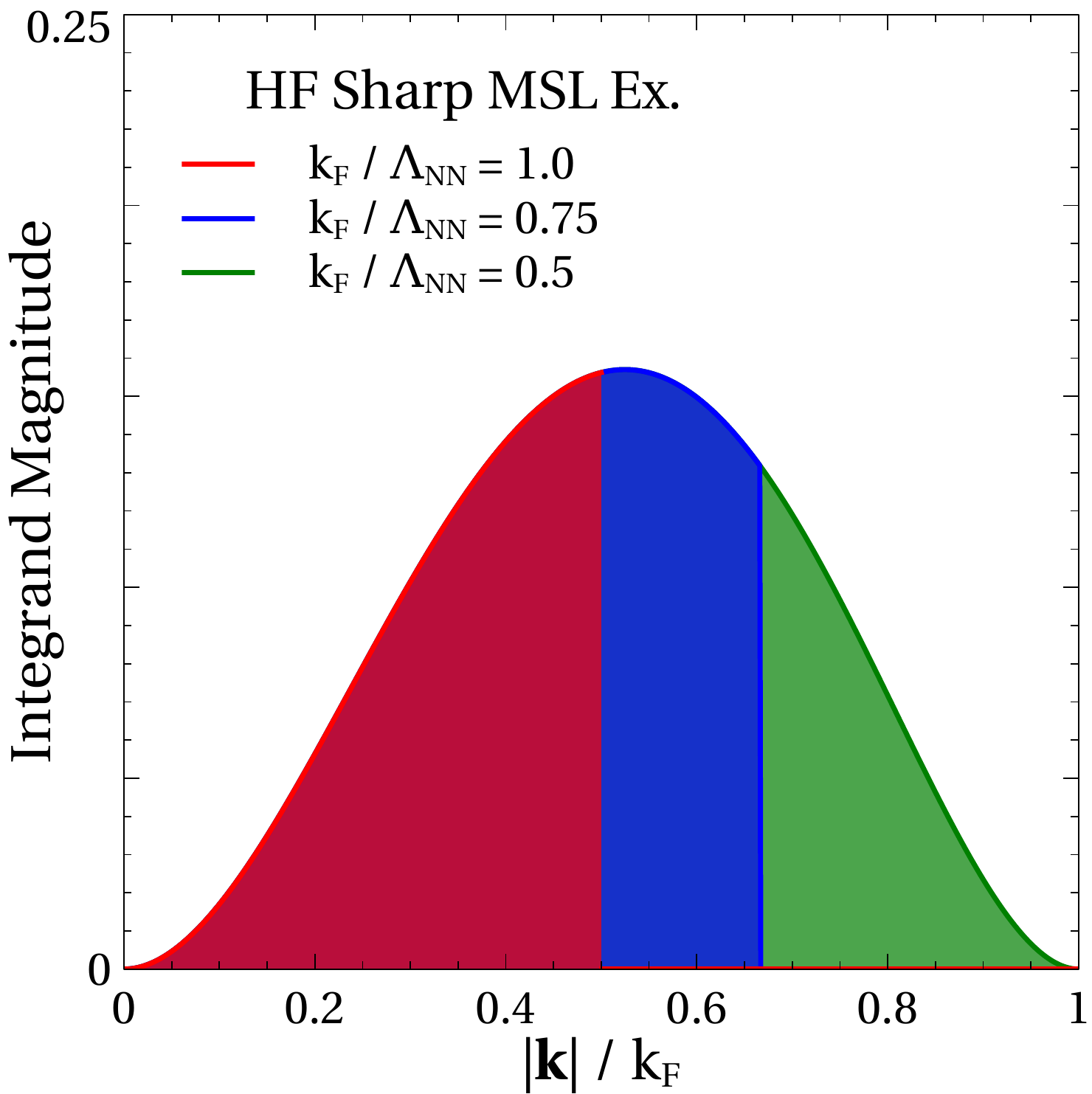}~
	\caption{Phase space of the Hartree-Fock
	exchange term regulated with the sharp MSL regulator of \eqref{eq:sharp_local}. The 
	magnitude of the dimensionless integrand in~(\ref{eq:HF_phasespace})
	is plotted as a function of $k / \kf$. The colored
	regions indicate the phase space
	included for different values of $\kf / 
	\NNcut$.}
	\label{fig:NN_HF_phase_space_local}
\end{figure}

For the exchange OPE, the hierarchy in energy values in Fig.~\ref{fig:NN_HF_energy}(c), 
in decreasing order of magnitude, is MSNL $>$ MSL $>$ CSL $>$ EKM.
The significant deviation
in the MSL, EKM, and CSL OPE energies 
compared to unregulated HF can be traced to the regulation of the 
small $r$ parts of the OPE potential. The energy 
density of uniform nuclear matter is dominated
by the low partial waves (e.g., S,P,D waves).
The MSL, CSL, and EKM regulators, in \eqref{eq:NN_reg_local}, \eqref{eq:CSL}, and \eqref{eq:ekm} respectively,
by construction cut off the potential at small $r$ and will thus affect these low partial waves to a greater extent than the MSNL scheme.

The energy trends in the graphs of Fig.~\ref{fig:NN_HF_energy} are directly linked to 
the interaction phase space, as we now demonstrate.
This is most apparent for a sharp regulator, for which 
five of the six integrals in \eqref{eq:NN_HF_energy}
can be done analytically  for pure S-wave or contact potentials. Dropping
prefactors, we find the phase space is proportional to the
dimensionless integrand~\cite{Furnstahl:1999ix,FETTER71}, 
\beq
%  \text{PS} \propto 
%    \int_0^1 
     \left(\frac{k}{\kf}\right)^2
     \left( 2 - 3\frac{k}{\kf} + \left(\frac{k}{\kf}\right)^3 \right) f_{\text{reg}}
    \;,  
  \label{eq:HF_phasespace}
\eeq
where $f_{\text{reg}}$ refers generically to any regularization scheme. We have also suppressed the overall dependence on $\kf$ and the
potential. Making the MSNL regulator in
\eqref{eq:NN_reg_nonlocal} sharp results in,
\beq
f_{\text{MSNL}}^{\text{NN}}(k^2) 
\xrightarrow[n\to\infty]{} \; \Theta
\left(\frac{\NNcut - k}{\kf}\right)
\label{eq:sharp_nonlocal}
\eeq
while making the MSL regulator in \eqref{eq:NN_reg_local} sharp
gives  different results for direct and exchange terms due to
regulating in the momentum transfer,
\begin{align*}
f_\text{MSL}^{\text{NN}} (q^2) \xrightarrow[n\to\infty]{} \quad &{\bf 1}	\quad &&\text{Direct}, \\
f_\text{MSL}^{\text{NN}} (q^2) \xrightarrow[n\to\infty]{} \quad &\Theta\left(\frac{\NNcut - 2k}{\kf}\right) \quad &&\text{Exchange} 
\; .
\numberthis
\label{eq:sharp_local}
\end{align*}

Therefore, for a sharp cutoff chosen above the Fermi  surface $\kf$,
the HF phase space will be unaltered by the MSNL regulator. In contrast, the
exchange term regulated in the sharp MSL scheme gets cut off as soon as
$\kf / \NNcut > 0.5$ (i.e., the effective
cutoff in the MSL exchange case is half that in the
MSNL case). This is shown in
Fig.~\ref{fig:NN_HF_phase_space_local}, where the colored region
indicates the integration region for different values of $\kf / \NNcut$. 
For example, for $\kf / \NNcut = 1.0$ all the phase space above $k / \kf = 0.5$ has been completely removed by the sharp MSL regulator while for $\kf / \NNcut = 0.5$, the full phase space is still extant.
As a result, in regions where the Fermi momentum is small compared with the cutoff,
we expect little deviation between
the unregulated, MSNL, and MSL HF energy.

\begin{figure*}[t]

	\includegraphics[scale=0.435]{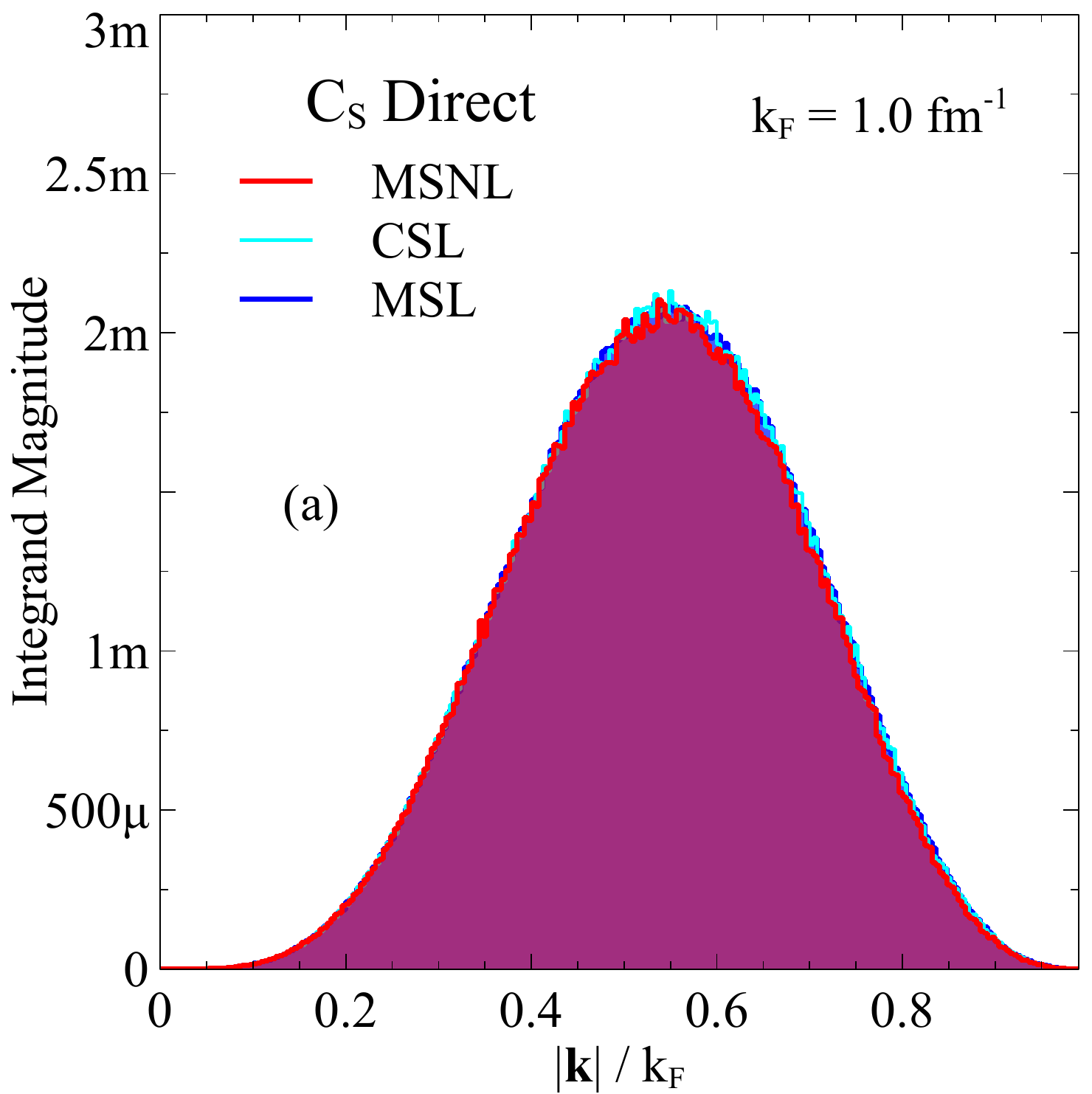}
	\hspace{-.45in}
	\includegraphics[scale=0.435]{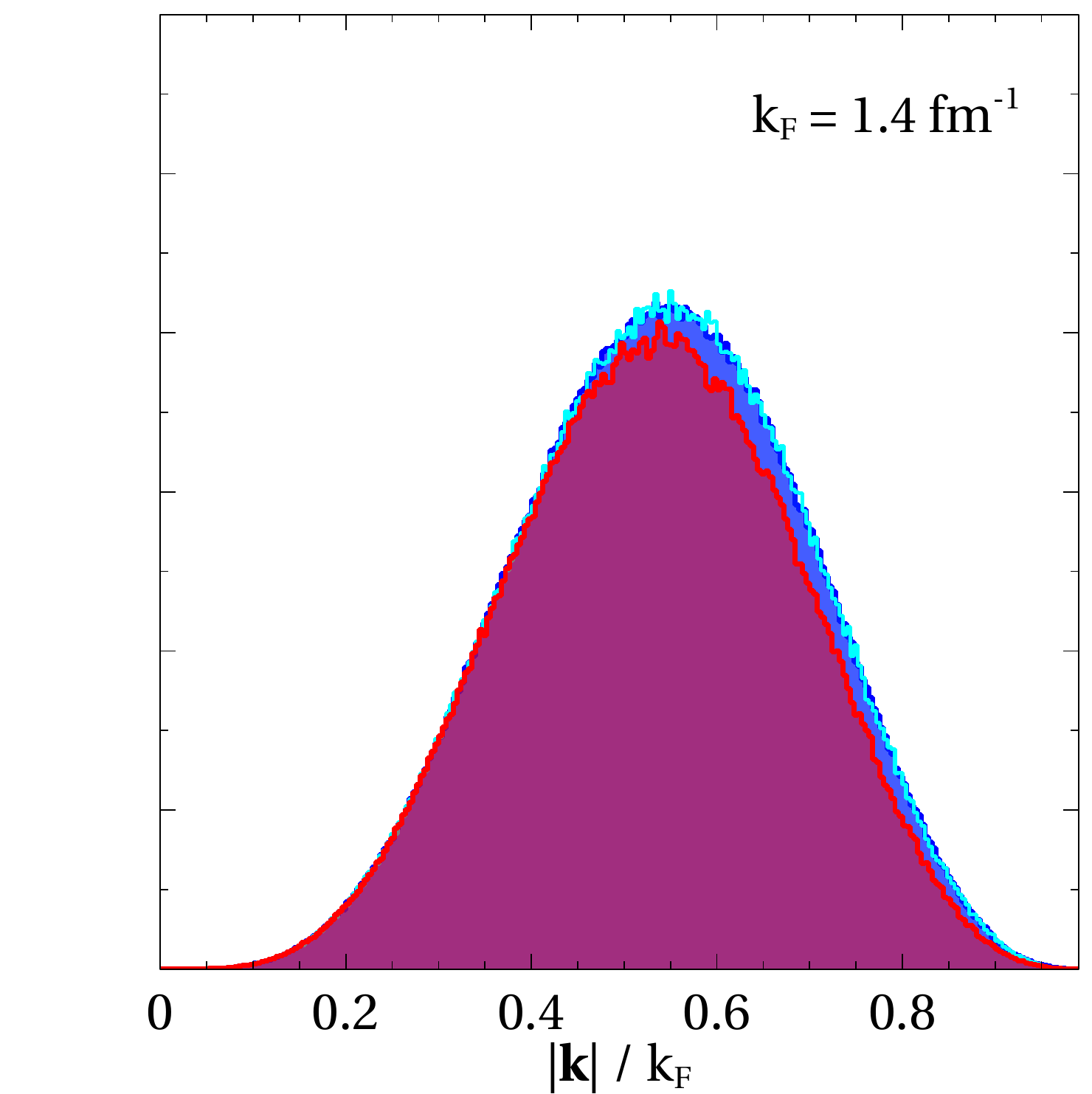}
	\hspace{-.45in}
	\includegraphics[scale=0.435]{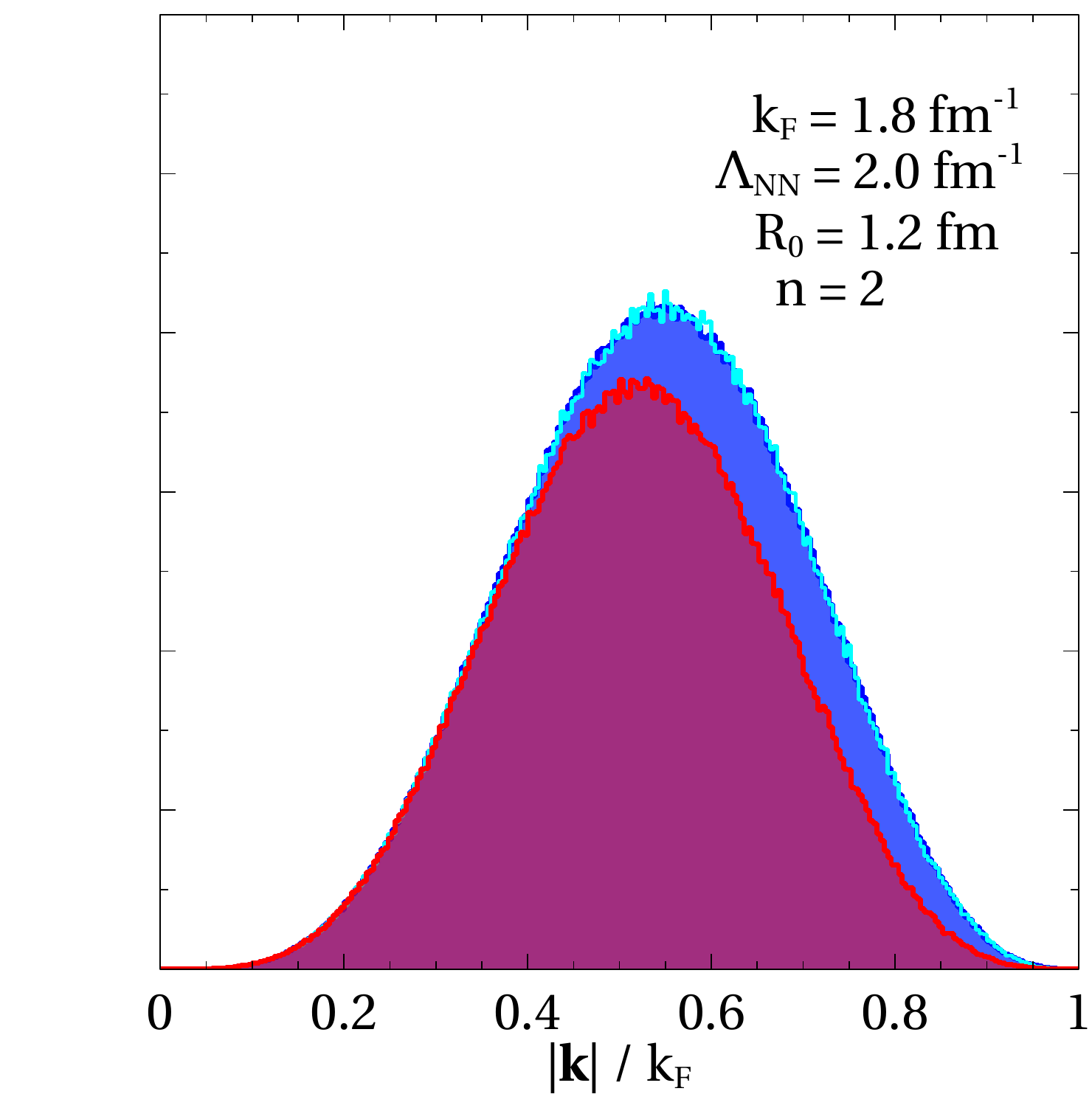}
	
	\medskip

	\includegraphics[scale=0.435]{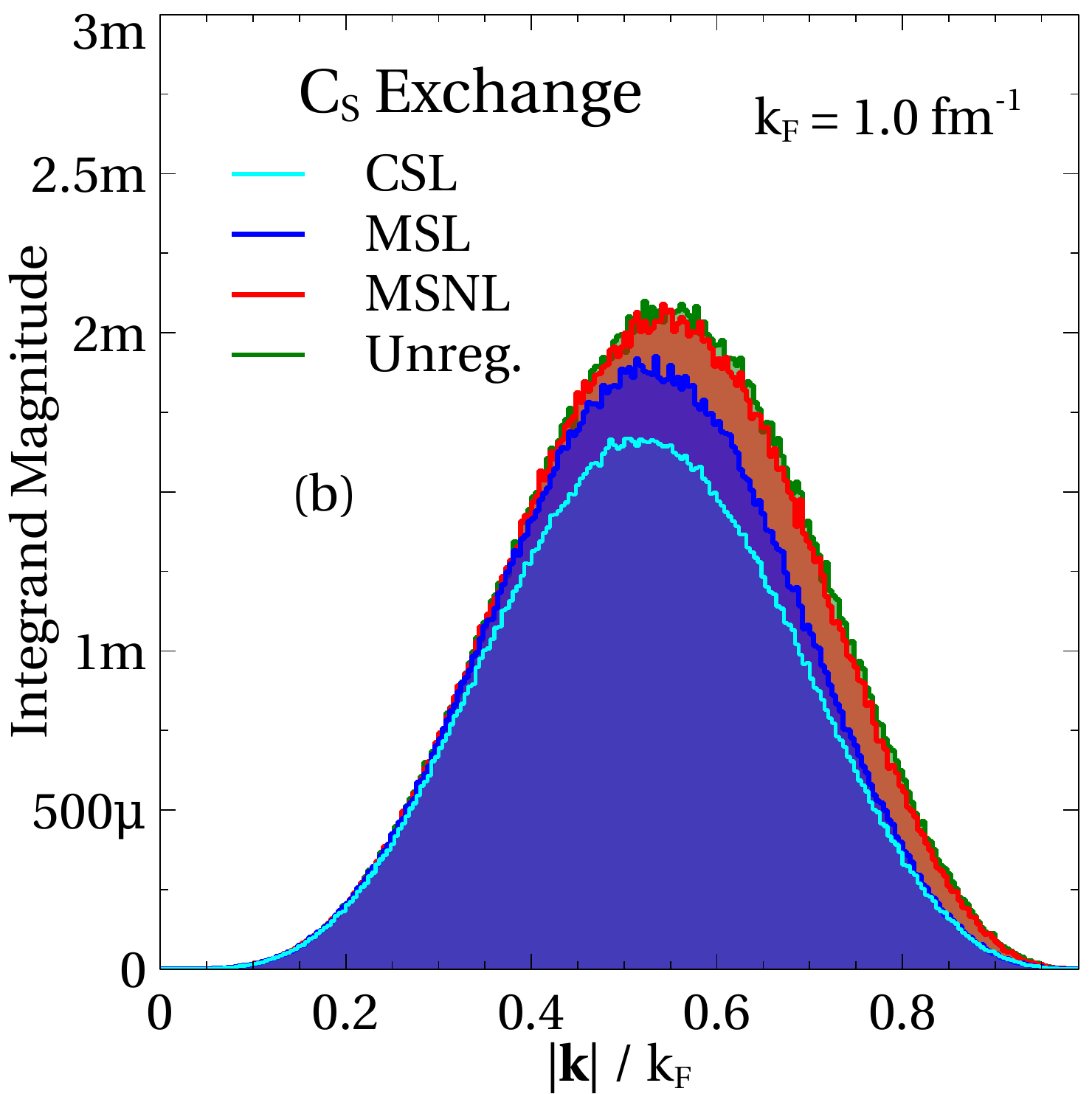}
	\hspace{-.45in}
	\includegraphics[scale=0.435]{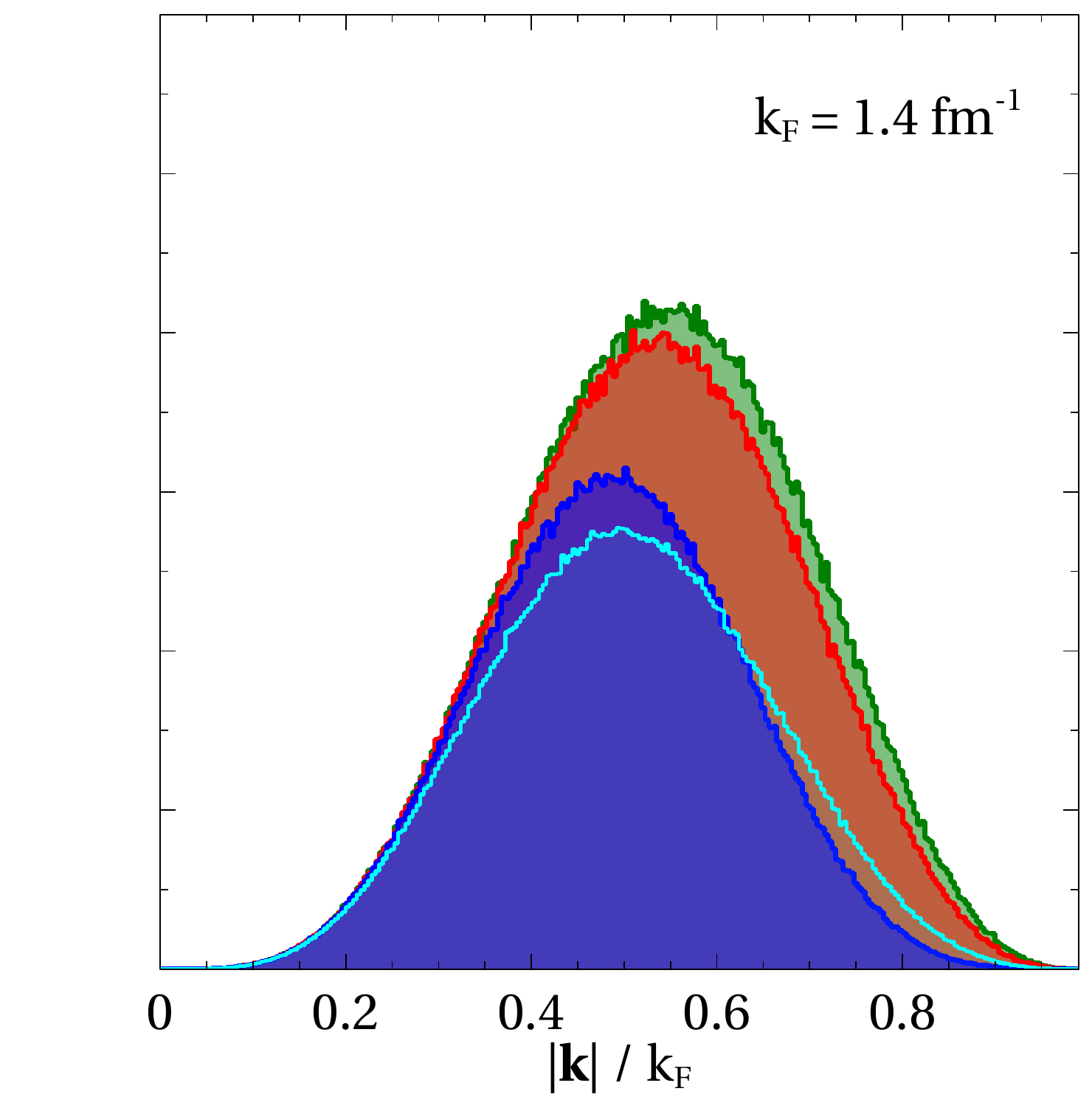}
	\hspace{-.45in}
	\includegraphics[scale=0.435]{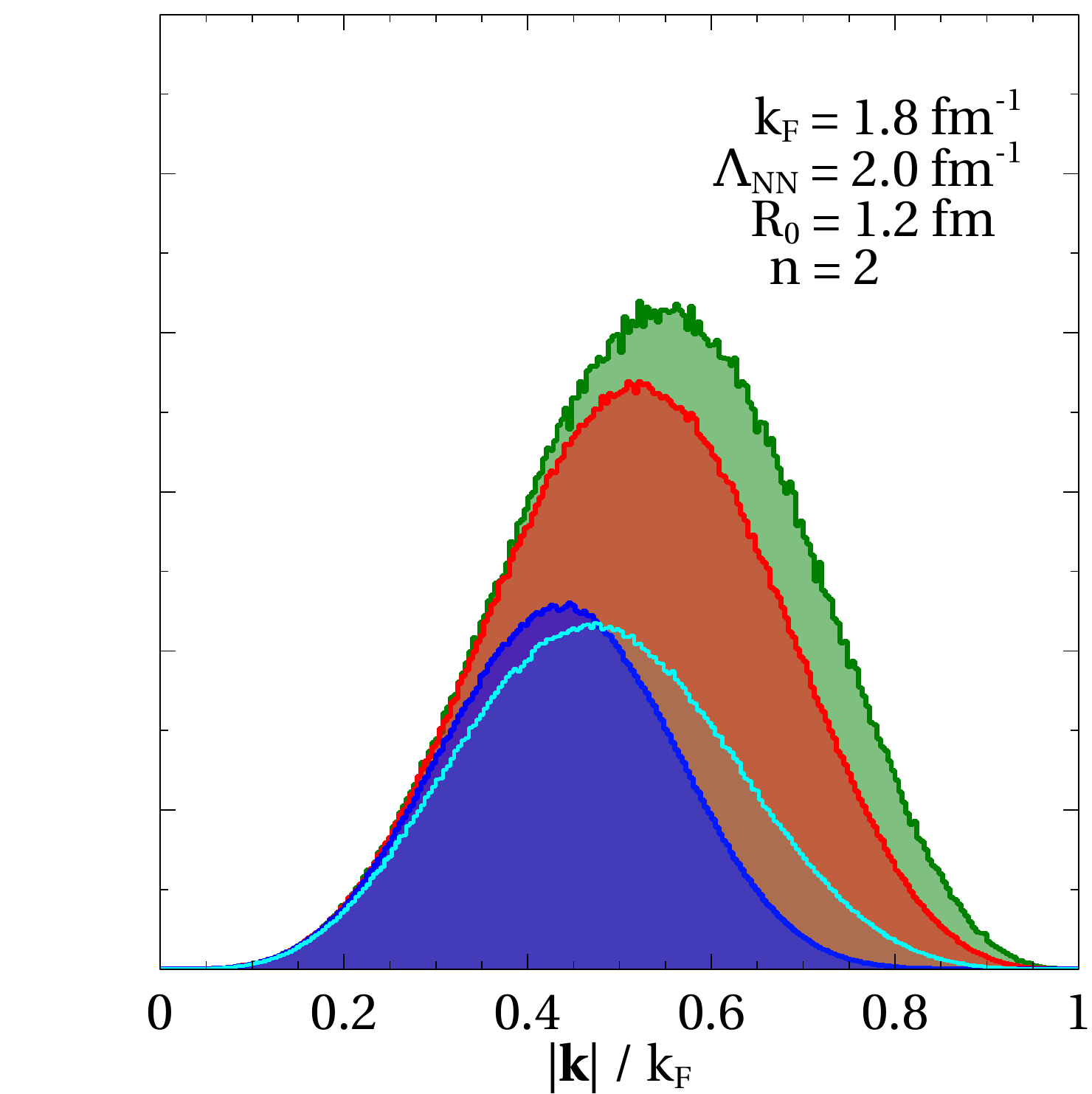}
	
	\medskip
	\includegraphics[scale=0.435]{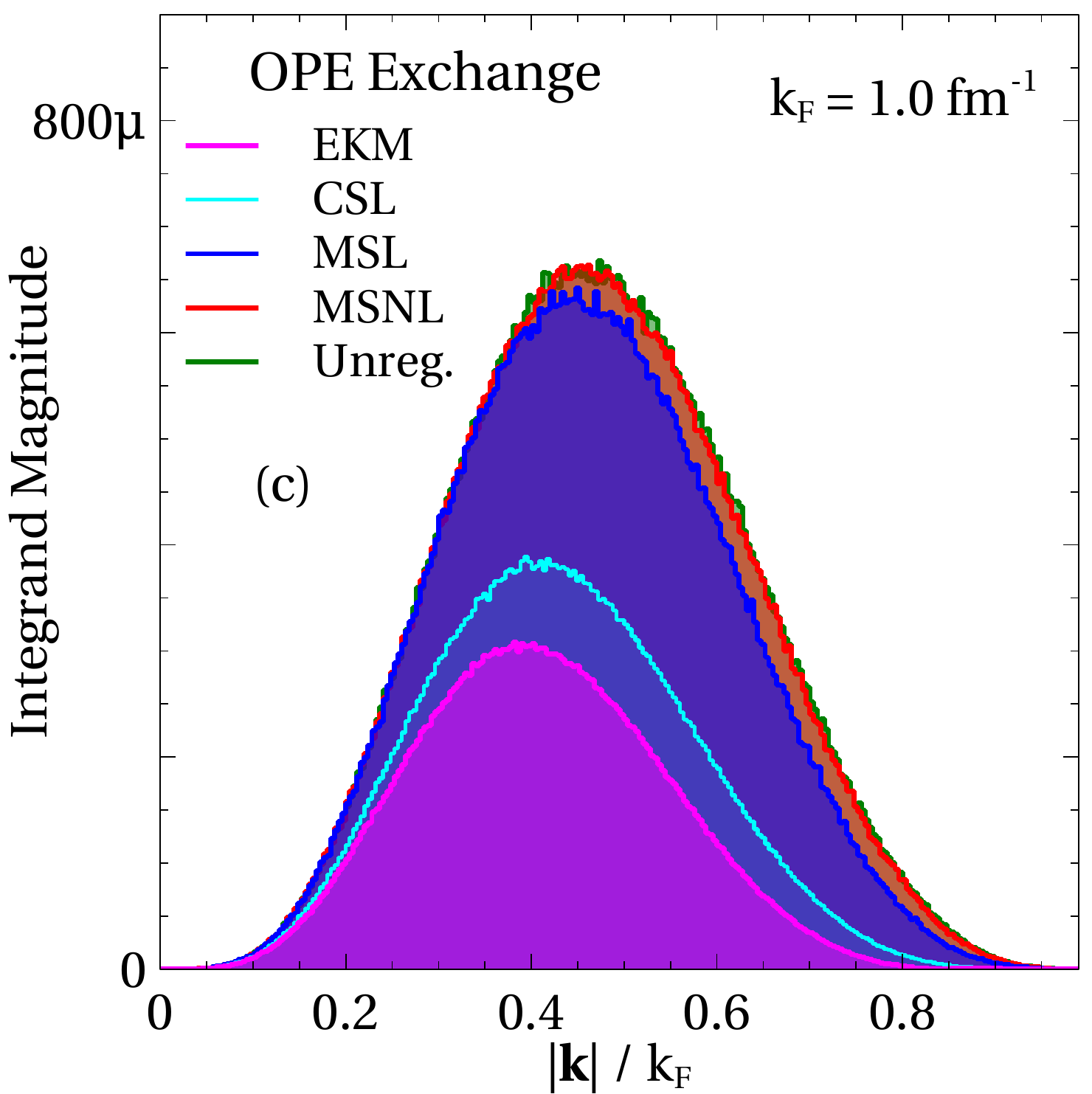}
	\hspace{-.45in}
	\includegraphics[scale=0.435]{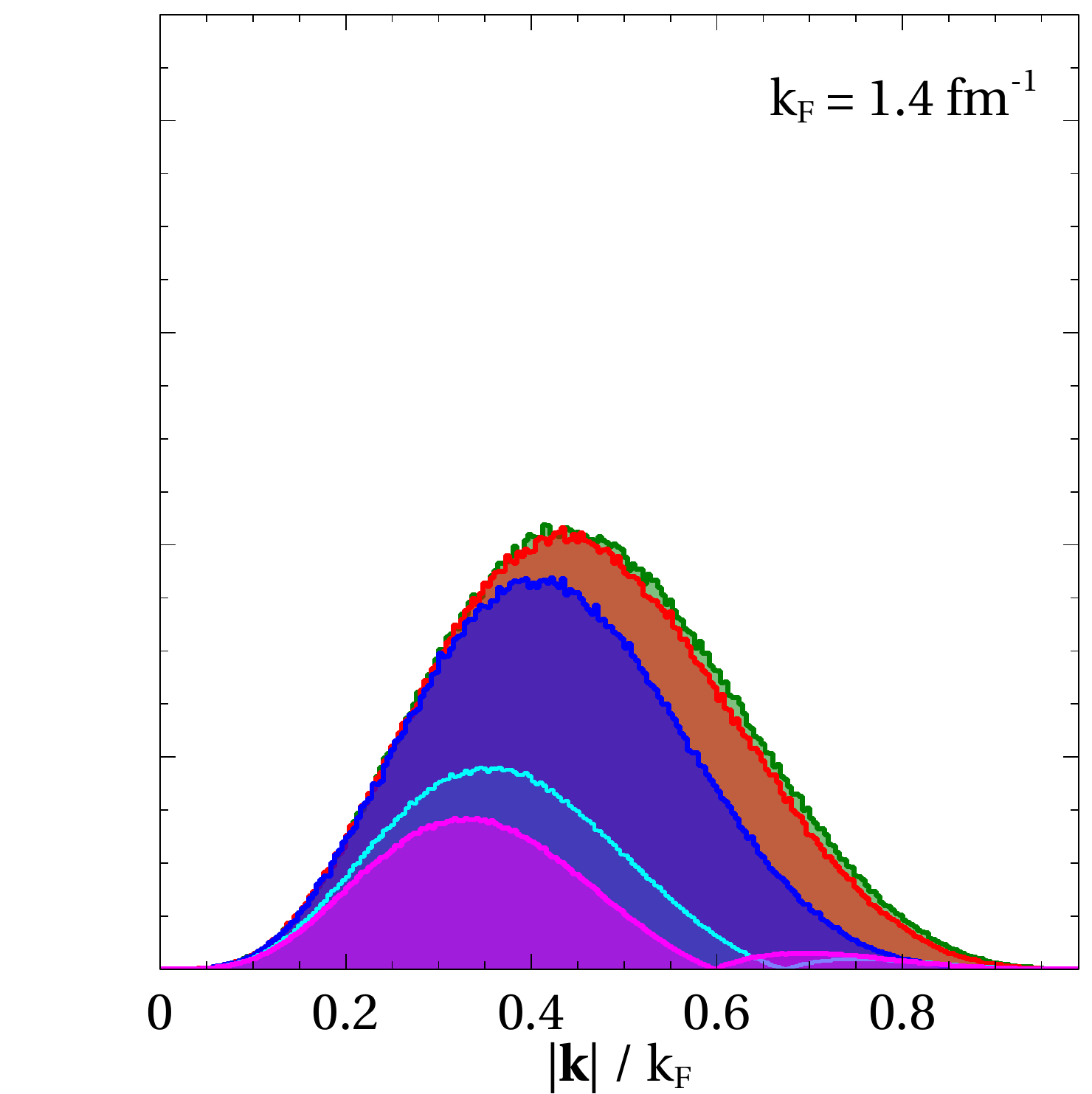}
	\hspace{-.45in}
	\includegraphics[scale=0.435]{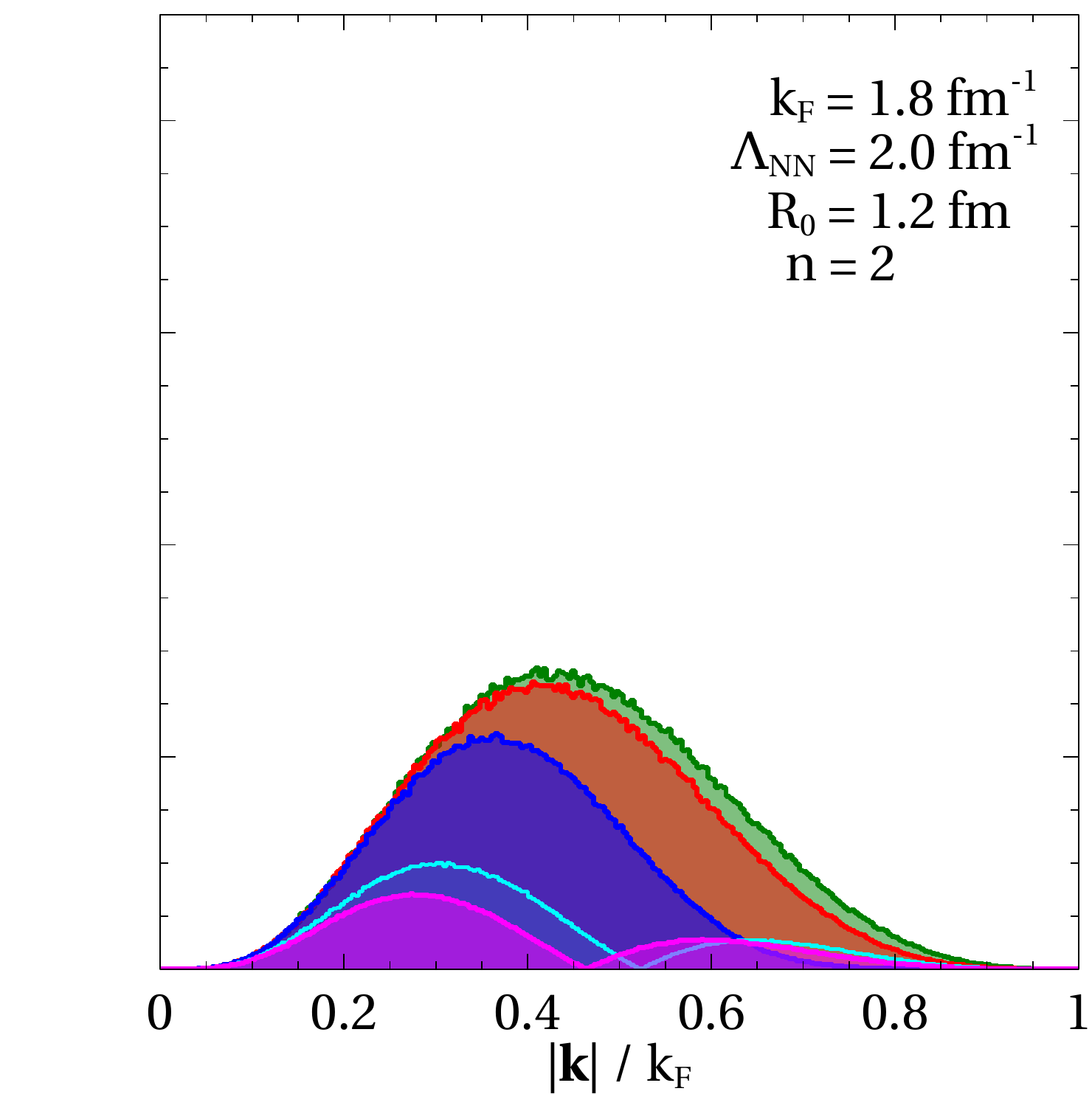}
	
	\caption{(color online) Momentum histograms of the 2-body HF integrand
	for the LO NN forces at $\kf = 1.0$, $1.4$, and $1.8~\fmi$. The integrand magnitude $I_1$ in \eqref{eq:NN_HF_histogram_equation} for the direct $C_S$ (a), exchange $C_S$ (b), and exchange OPE (c) is plotted as a function of the dimensionless $k / \kf$. $\mu$ and $m$ on the vertical axis denote $10^{-6}$ and $10^{-3}$ respectively.
	All graphs are evaluated at $\NNcut = 2.0~\fmi$, $R_{0} = 1.2~\fm$, with $n = 2$.}
	\label{fig:NN_HF_histograms}
\end{figure*}

	Although at HF the phase space in \eqref{eq:HF_phasespace} can be analytically derived, the situation is considerably more complicated at second-order and in the 3-body sector.
	In anticipation of this, we develop a new way to visualize the regulator phase space occlusion using a diagnostic based on Monte Carlo sampling. 
	To understand regulator effects and the hierarchy of energy values, we propose creating plots of the HF integrand in \eqref{eq:NN_HF_energy} and plotting it against the relative momentum magnitude as is done in Fig.~\ref{fig:NN_HF_histograms}.
	These histogram plots will be the main analysis tool for regulator effects on the potential both at HF and at higher orders in perturbation theory.
	They are created by randomly generating single-particle momenta ${\bf p}_1, {\bf p}_2$ by Monte Carlo sampling and then calculating the scaled HF energy integrand $I_1$, 
\begin{align*}
	I_1 = 
	|f_{\text{reg}}| \;
	\frac{k^2}{\kf^2} \; 
	\frac{P^2}{\kf^2} \;
	n({\bf P}/2 + {\bf k}) \; 
	n({\bf P}/2 - {\bf k}) \;
	\\
	\times \begin{dcases}
	1, & \text{Contact}
	\\
	\frac{\mpi^2}{q^2 + \mpi^2}	
	\;, & \text{OPE}
	\end{dcases}
	\numberthis
	\label{eq:NN_HF_histogram_equation}	
\end{align*}
where $f_{\text{reg}}$ refers to a regularization scheme in Table~\ref{tab:reg_table}, ${\bf P}$ is the total momentum, ${\bf P} = {\bf p}_1 + {\bf p}_2$, and the integrand is weighted by 
a contact or OPE interaction%
\footnote{The term in \eqref{eq:OPE_no_delta} proportional to the $S_{12}$ operator is zero at HF after preforming spin traces.}. 
The value of the integrand $I_1$ is then binned for the corresponding relative momentum magnitude $k$ (normalized by $\kf$) and the process is repeated.
After a sufficiently high number of momenta are generated, the final plot is normalized by the total number of iterations.
The scaling of the momentum magnitudes $k$ and $P$ by $\kf$ is done here for convenience. 

	These histograms can be interpreted as the phase space available to the system at HF in MBPT now weighted by momenta and the interaction $V^{\text{NN}}$. 
	The interaction weighting is included to demonstrate how different interactions weight different parts of the phase space and how this interplays with differing regularization schemes.

\begin{figure*}[t!]

	\includegraphics[scale=0.435]{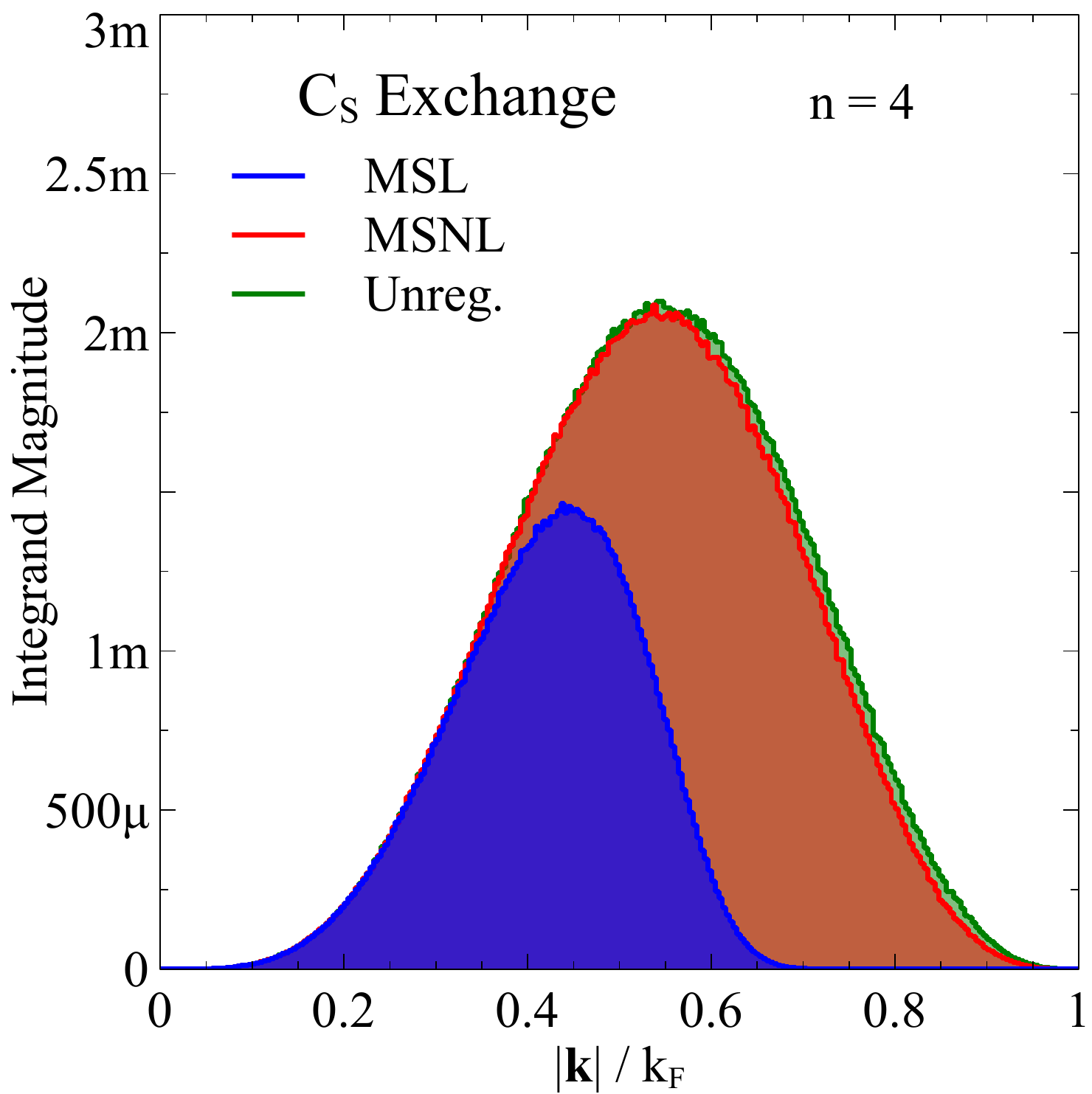}
	\hspace{-.45in}
	\includegraphics[scale=0.435]{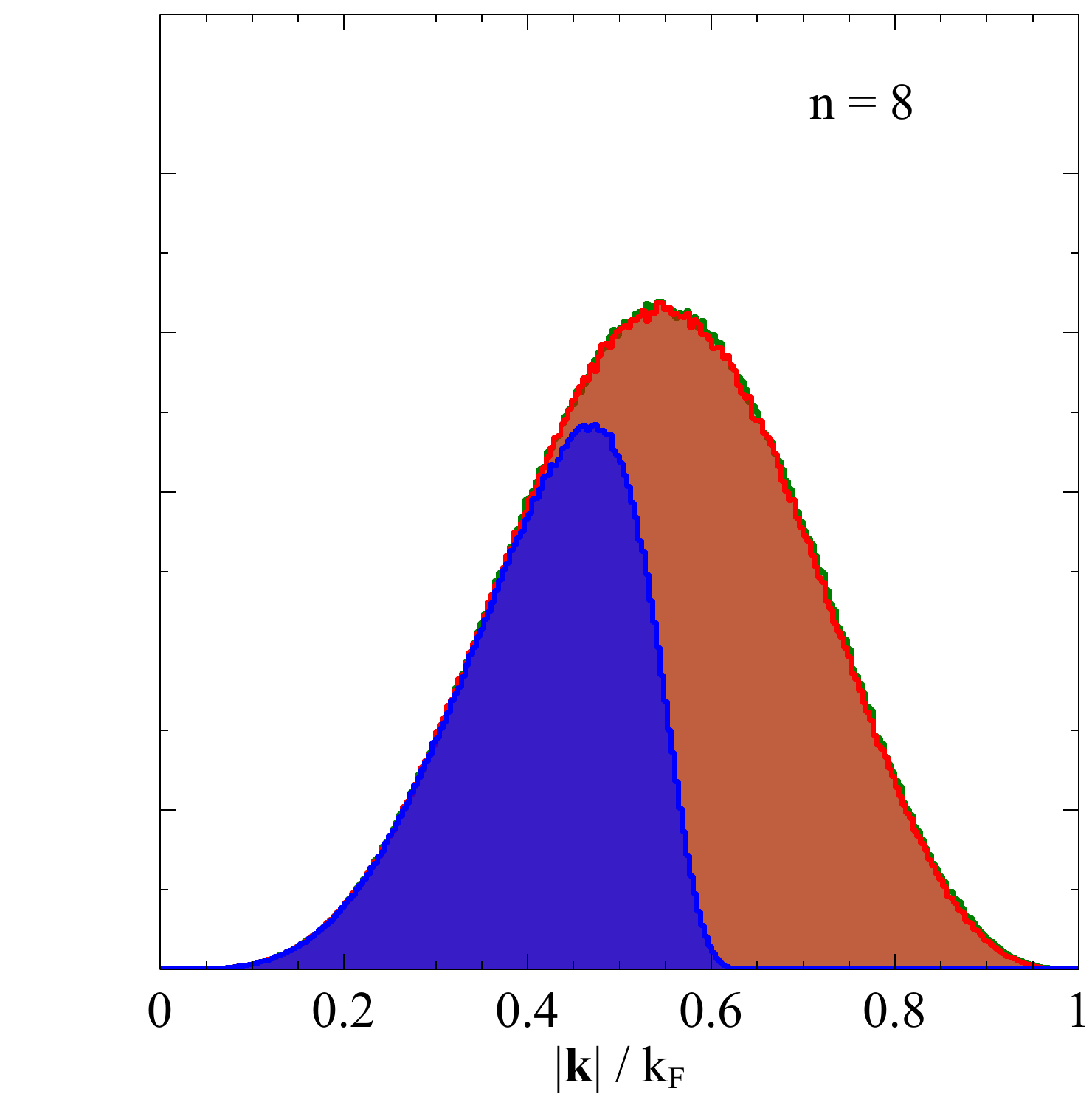}
	\hspace{-.45in}
	\includegraphics[scale=0.435]{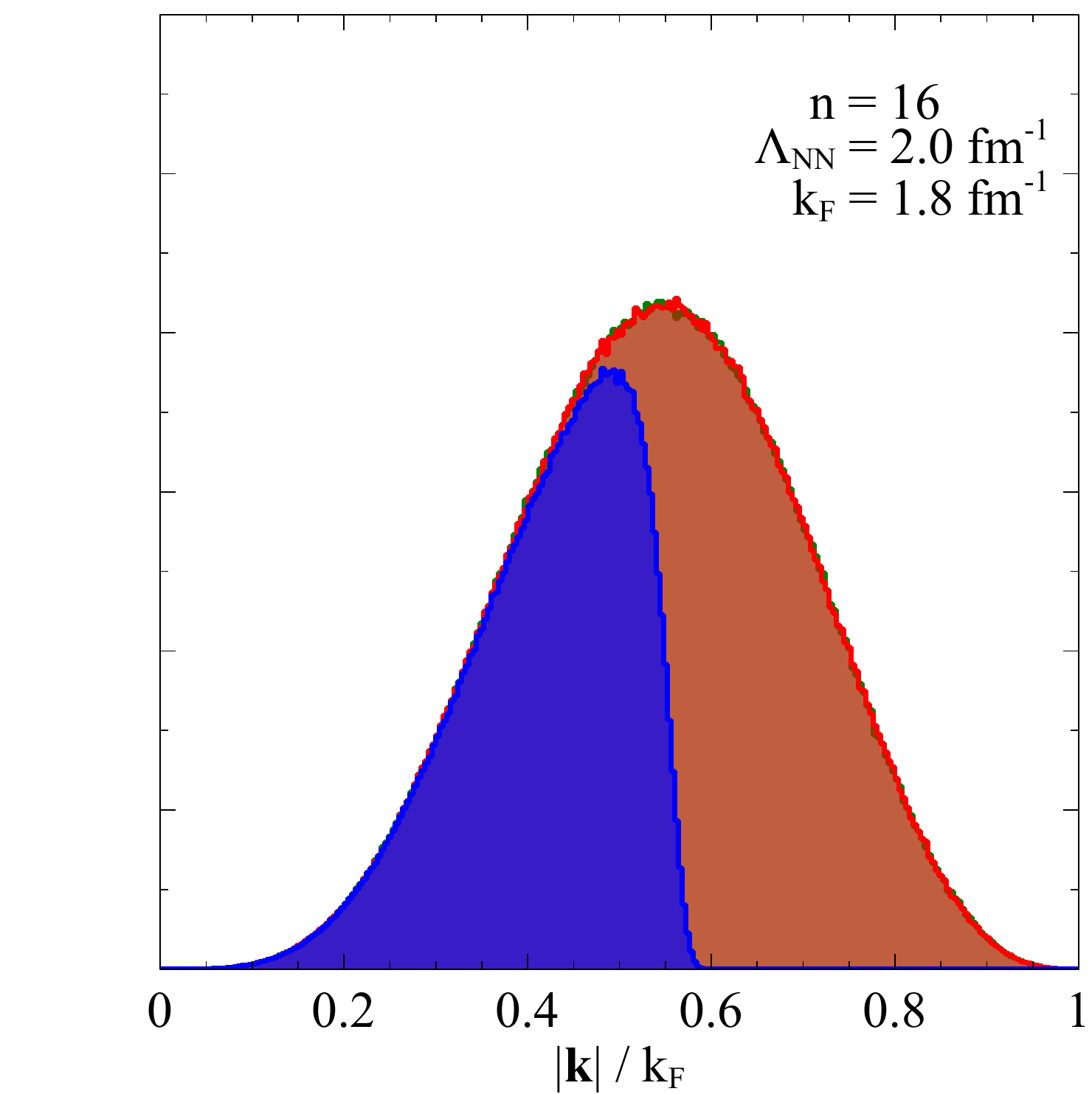}

	\caption{(color online) Momentum histograms of the 2-body HF integrand
	for the exchange $C_S$ term at $n = 4$, $8$, $16$. The integrand magnitude $I_1$ in \eqref{eq:NN_HF_histogram_equation} is plotted as a function of $k/\kf$.
	As the exponent $n$ in the regulator is increased, the full phase space is recovered for the MSNL regulator, while the MSL regulator approaches a theta function, in agreement with \eqref{eq:sharp_nonlocal} and \eqref{eq:sharp_local}.
	All graphs are evaluated at $\kf = 1.8~\fmi$, $\NNcut = 2.0~\fmi$.}
	\label{fig:NN_HF_histograms_N_vary}
\end{figure*}

\begin{figure}[tbh]
   \includegraphics[width=0.85\columnwidth]{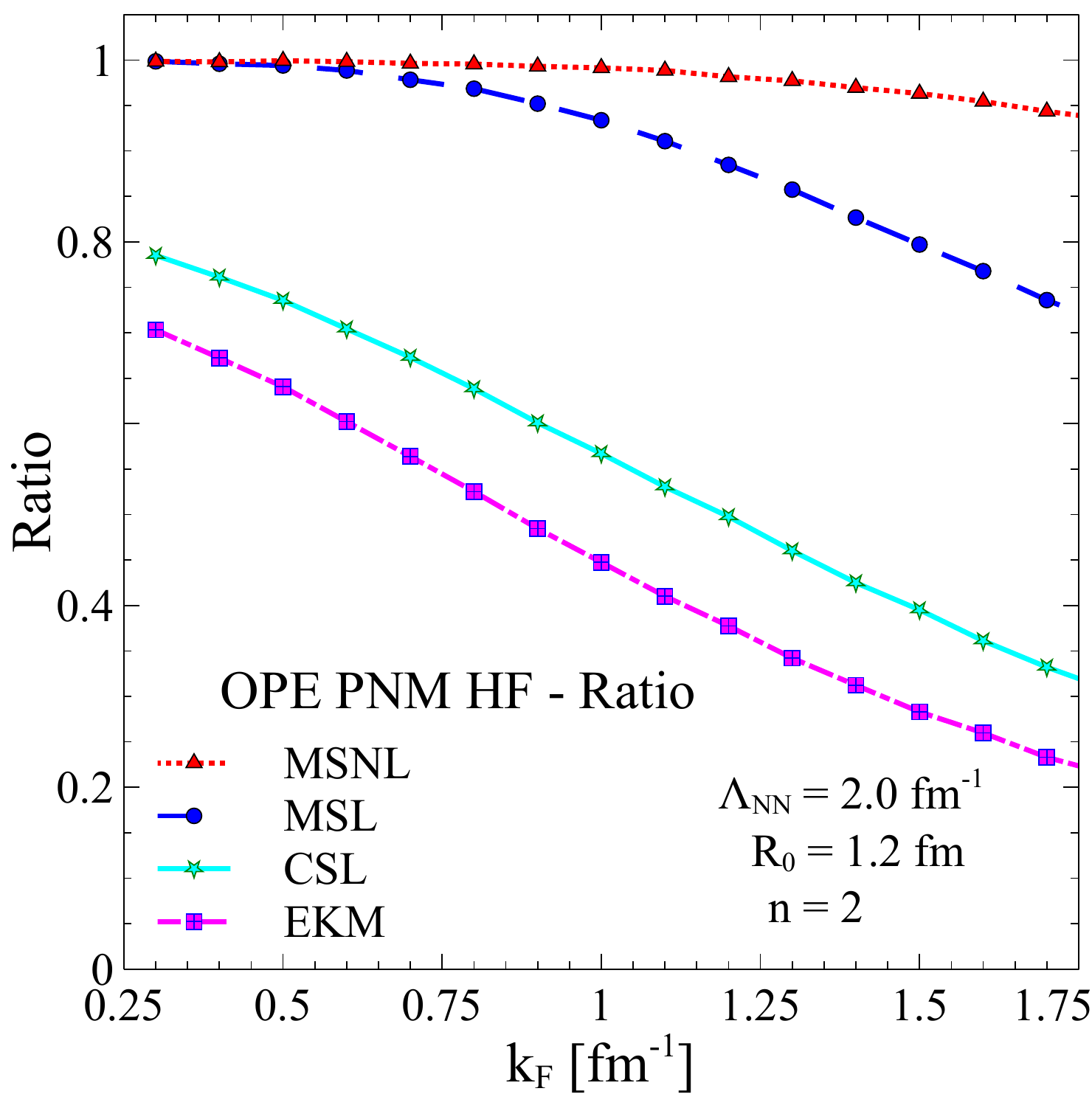}
	\caption{
	Ratio of the regulated HF OPE energy in PNM to the unregulated one,	 as a function of $\kf$.
	Unlike the 
	momentum space 
	regulators, the long-range CSL
	and EKM regulators do \emph{not}
	reproduce the full HF energy in the
	low-density limit. 
	The calculations use $n = 2$, $R_0 = 1.2~\fm$, and $\NNcut = 2.0~\fmi$.}
	\label{fig:COOR_DISTORTION}
\end{figure}

We use these plots for three key purposes:
\begin{enumerate}
	\item to show that the hierarchy in computed MBPT energy values matches the volumes of the weighted phase space,
	\item to illuminate where in the phase space different regulators act, i.e., where
	the contribution to the energy integral becomes small,
	\item to demonstrate how different interactions interplay with the regularization schemes.
\end{enumerate}

\noindent
Addressing these points in order, we first note that the volume of weighted phase space tracked for different regulator choices in Fig.~\ref{fig:NN_HF_histograms} exactly matches the hierarchy in energy values of Fig.~\ref{fig:NN_HF_energy}. 
For example, the $C_S$ direct energy in Fig.~\ref{fig:NN_HF_energy}(a) is unaltered for
the CSL and MSL regulator schemes while the MSNL scheme shows an increasing 
suppression for increasing $\kf$. 
Looking at Fig.~\ref{fig:NN_HF_histograms}(a), the direct $C_S$ histograms show an 
increasing loss of phase space at large $|\kvec|/\kf$ for the MSNL scheme as $\kf$ 
increases while the MSL and CSL phase spaces are unaltered.
A corresponding matching of weighted phase space volume to energy calculations exists 
for the exchange $C_S$ and OPE terms as well.
Increasing the integer $n$ in the MSNL and MSL regulators (i.e., making the regulators sharper) for 
the $C_S$ exchange term at a fixed density results in the plots in 
Fig.~\ref{fig:NN_HF_histograms_N_vary}. 
As $n$ increases, one recovers the full space for the MSNL scheme and the sharp 
cutoff limit for the MSL scheme in agreement with \eqref{eq:sharp_nonlocal} 
and \eqref{eq:sharp_local}.

Secondly, we see that the primary regions that get suppressed in the weighted phase
 space, for both the contact and OPE plots, are regions of large $|\kvec|/ \kf$. 
 This is expected given the form of the MSNL regulator 
 in \eqref{eq:NN_reg_nonlocal}, 
 that all local regulators will suppress large $\qvec$, 
 and that there is a simple relation between $\qvec$ and $\kvec$ at HF,
\begin{align*}
\qvec &= 0	\quad &&\text{Direct}, \\
\qvec &= 2\kvec \quad &&\text{Exchange} 
\; .
\numberthis
\label{eq:q_k_relation_HF}
\end{align*}
As such, how the phase space is cut off for this class of regulators is mostly
universal at HF. 
Note also the interesting behavior in Fig.~\ref{fig:NN_HF_histograms}(c) in the OPE case in that both the CSL and EKM phase 
spaces go to zero at some value of $|\kvec|/\kf$ at larger densities ($\kf = 1.8~\fmi$) and then increase 
again. This reflects the oscillatory nature of the Fourier-transformed regulator 
(see Fig.~\ref{fig:CSL_LOCAL_COMPARE}(b)). 
Fig.~\ref{fig:CSL_LOCAL_COMPARE}(b) also reveals that the EKM/CSL regulator function 
does not approach $1$ for $|\qvec| \to 0$. This can be seen in the modification of the phase space in Fig.~\ref{fig:NN_HF_histograms}(c) at low $k / \kf$. 
	As a consequence, the ratio of the regulated to unregulated HF OPE energy,
plotted in Fig.~\ref{fig:COOR_DISTORTION}, does not go to 1 at 
low $\kf$ for the EKM/CSL regulators.

\begin{figure*}[tbh]
	\includegraphics[width=0.41\textwidth]{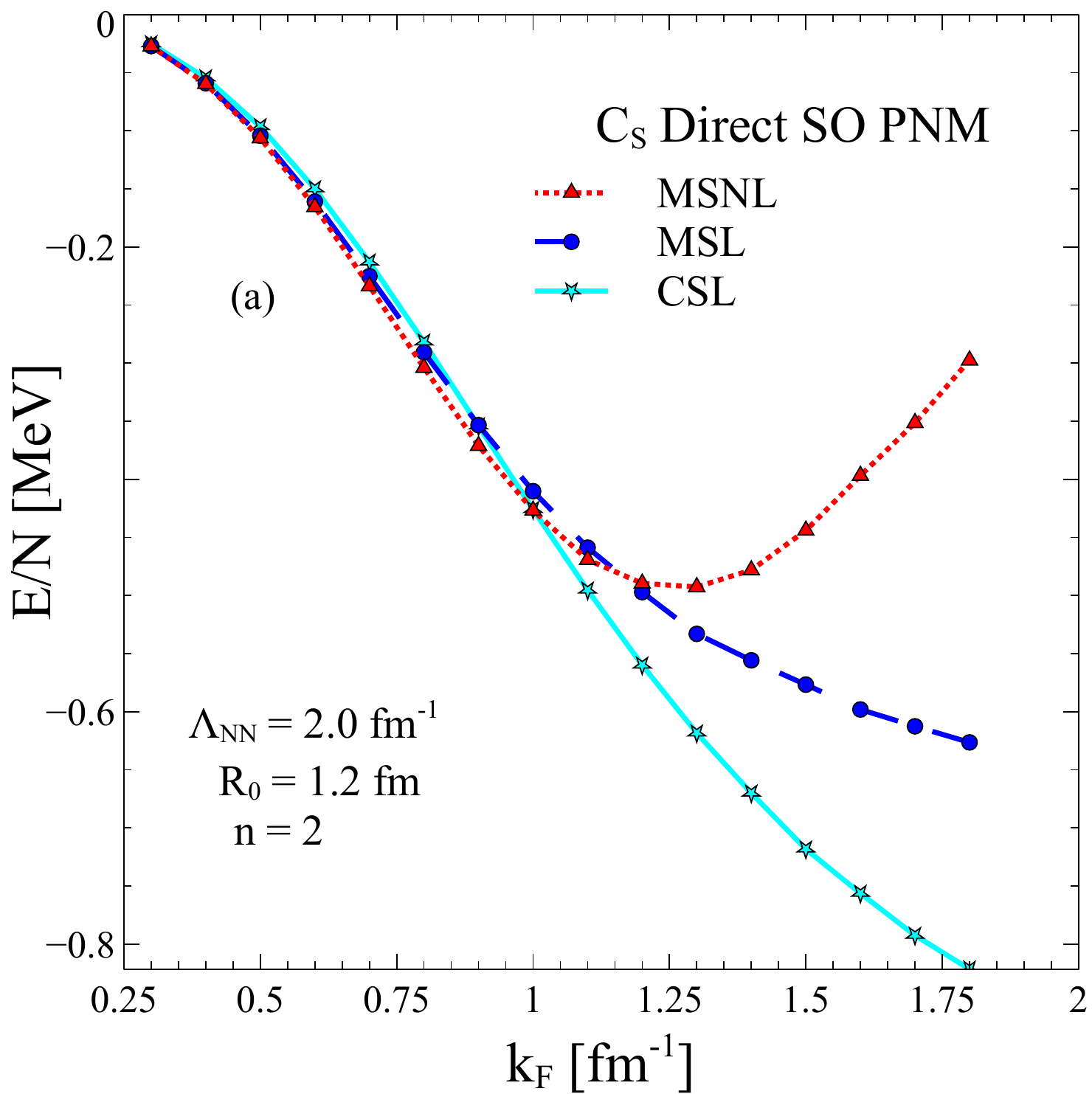}~~~
	\includegraphics[width=0.41\textwidth]{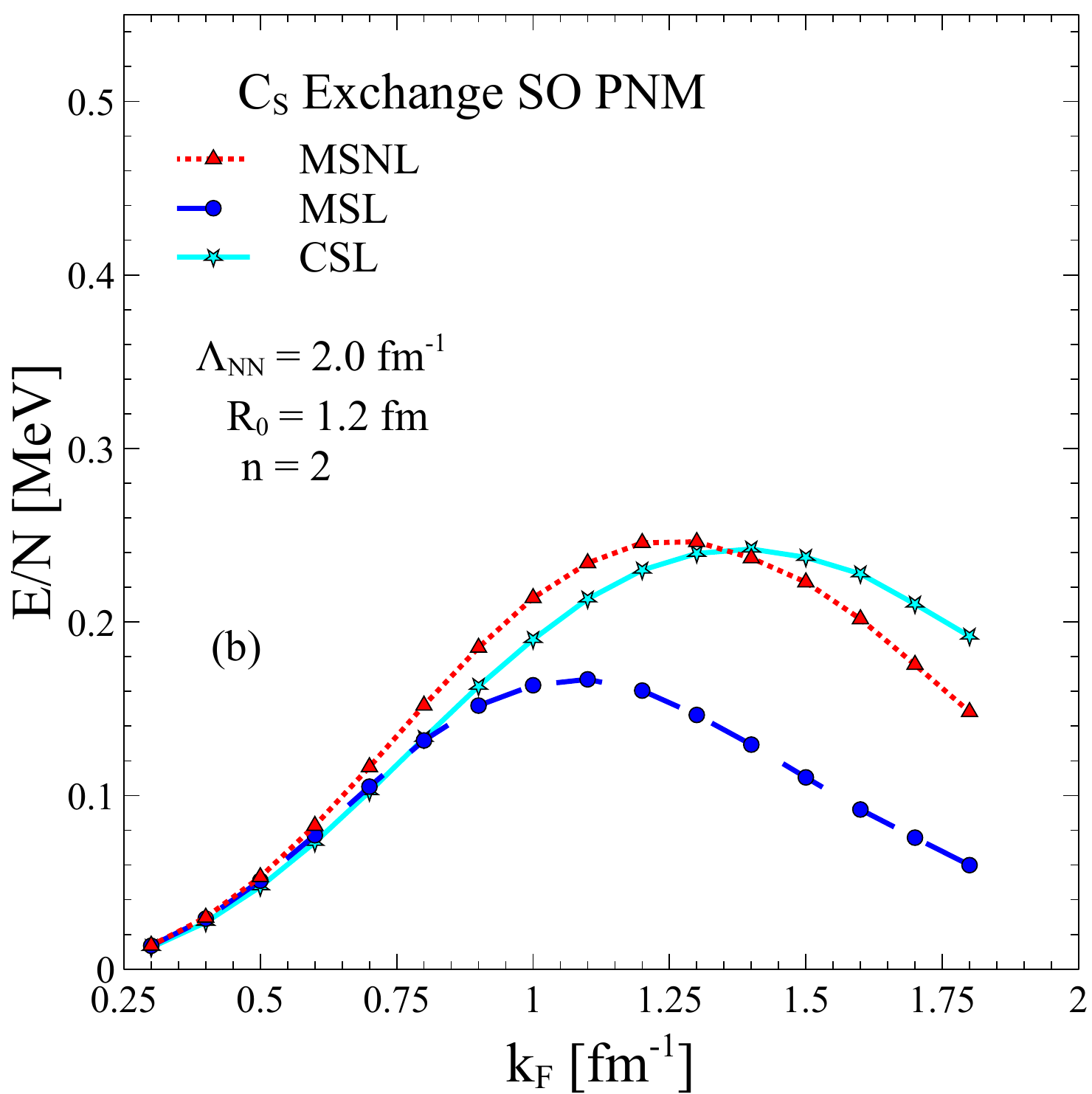}
	\medskip
	
	\includegraphics[width=0.41\textwidth]{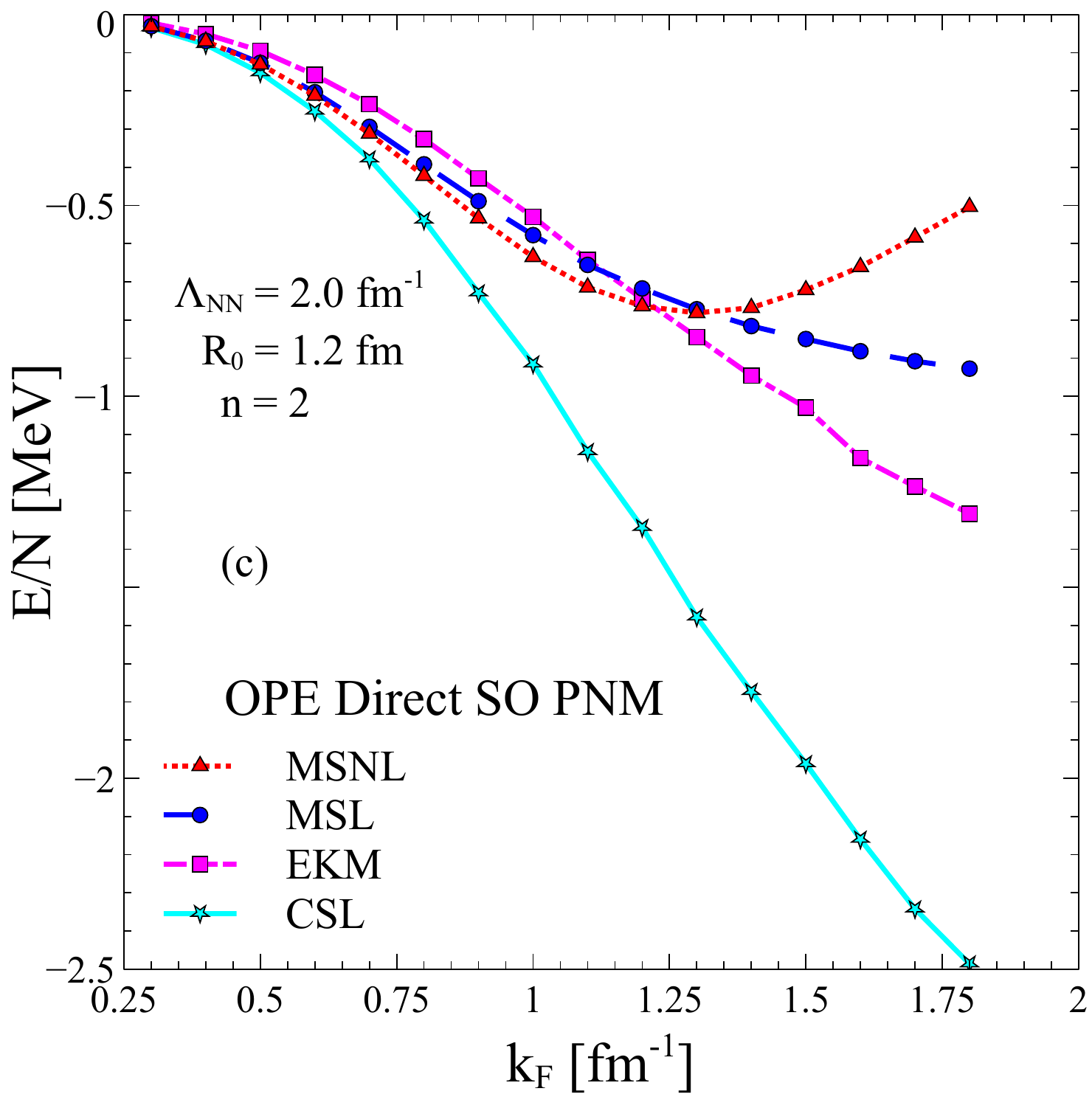}~~~
	\includegraphics[width=0.41\textwidth]{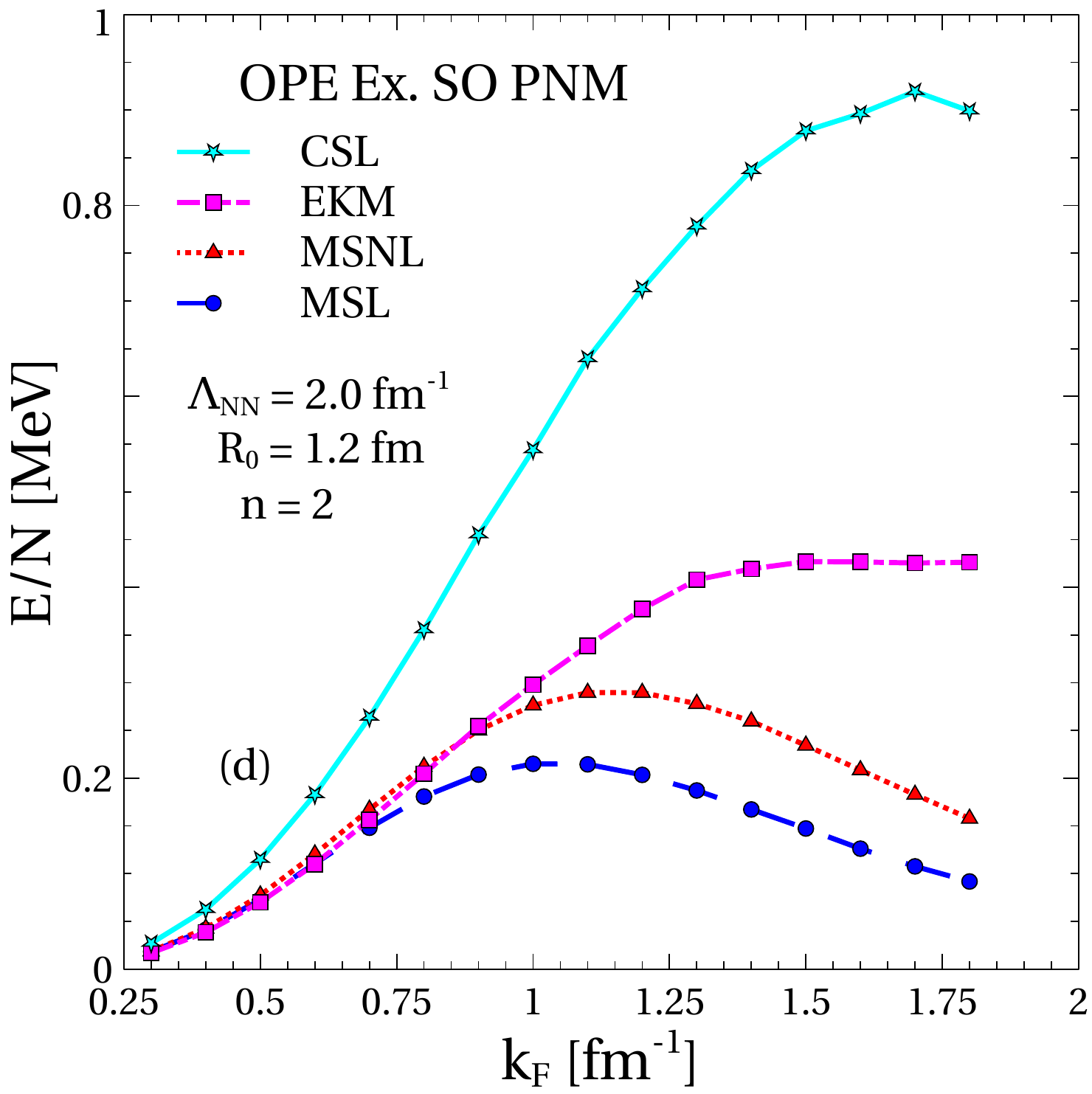}
	
	\caption{Neutron matter calculations of the second-order energy for the $C_S$--$C_S$ direct (a)
	and exchange (b) terms along with the OPE--OPE direct (c) and exchange (d) terms using different regulators.	
	 The calculations are done at $\NNcut = 2.0~\fmi$, $R_0 = 1.2~\fm$, and $n = 2$.}
	\label{fig:NN_SO_energy}
\end{figure*}	

Thirdly, we can compare the weighted phase space distribution to see the effect 
of the different interactions, contact vs.\ OPE\footnote{Note that in scaling the momentum magnitudes $k$ and $P$ by $\kf$ in $I_1$, larger $\kf$ will tend to shrink the distribution when weighting by the OPE interaction.}. 
The $C_S$ and OPE plots are very similar to one another suggesting that the regulators are primarily determining the distribution.
	The key difference between the two is the shifting of the maximum OPE phase space distribution towards smaller $\kvec$ (cf.\ the distribution peak in $C_S$ and OPE in Fig.~\ref{fig:NN_HF_histograms} (b) and (c)). 
	This shifting of the peak of the OPE phase space distribution results in less suppression for the regulated energy values, as can be seen in comparing the OPE and $C_S$ exchange energies in Fig.~\ref{fig:NN_HF_energy} (b) and (c).

%%%%%%%%%%%%%%%%%%%%%%%%%%%%%%%%%%%%%%%%%%%%%%%%%%
%
%%%%%%%%%%%%%%%%%%%%%%%%%%%%%%%%%%%%%%%%%%%%%%%%%%

\begin{figure*}[tbh]
	\includegraphics[width=0.32\textwidth]{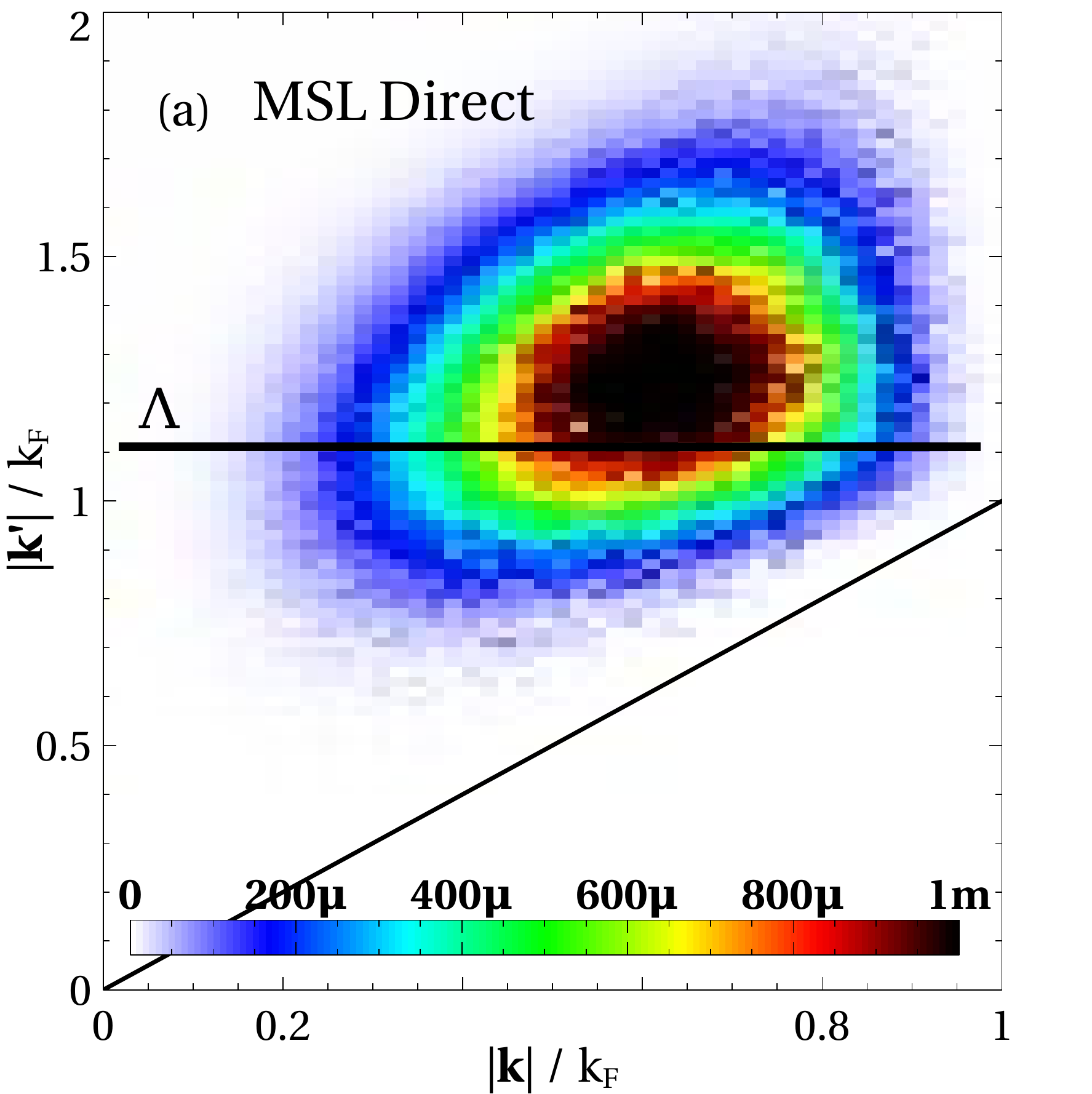}
	\includegraphics[width=0.32\textwidth]{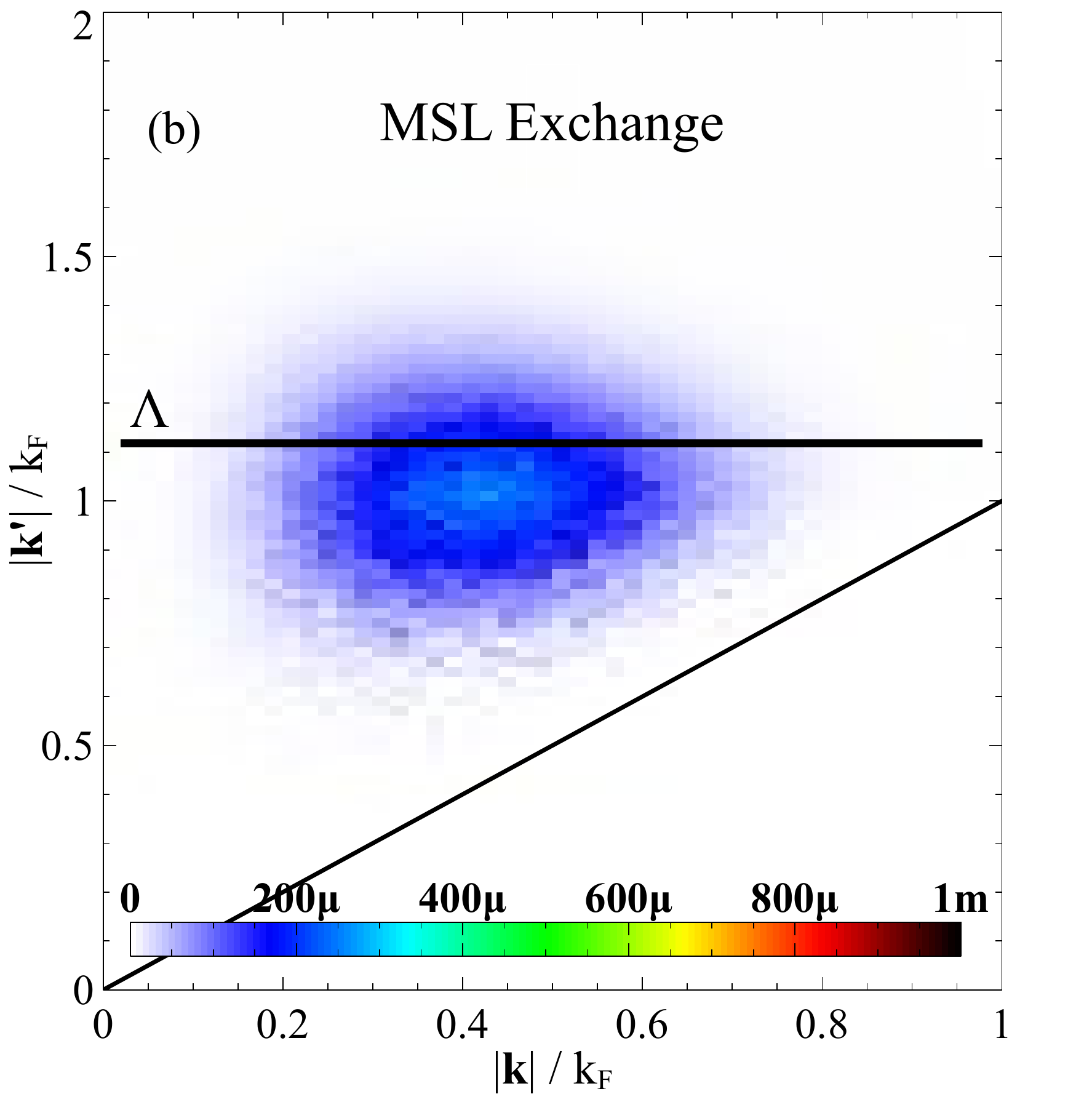}

	%\medskip

	\includegraphics[width=0.32\textwidth]{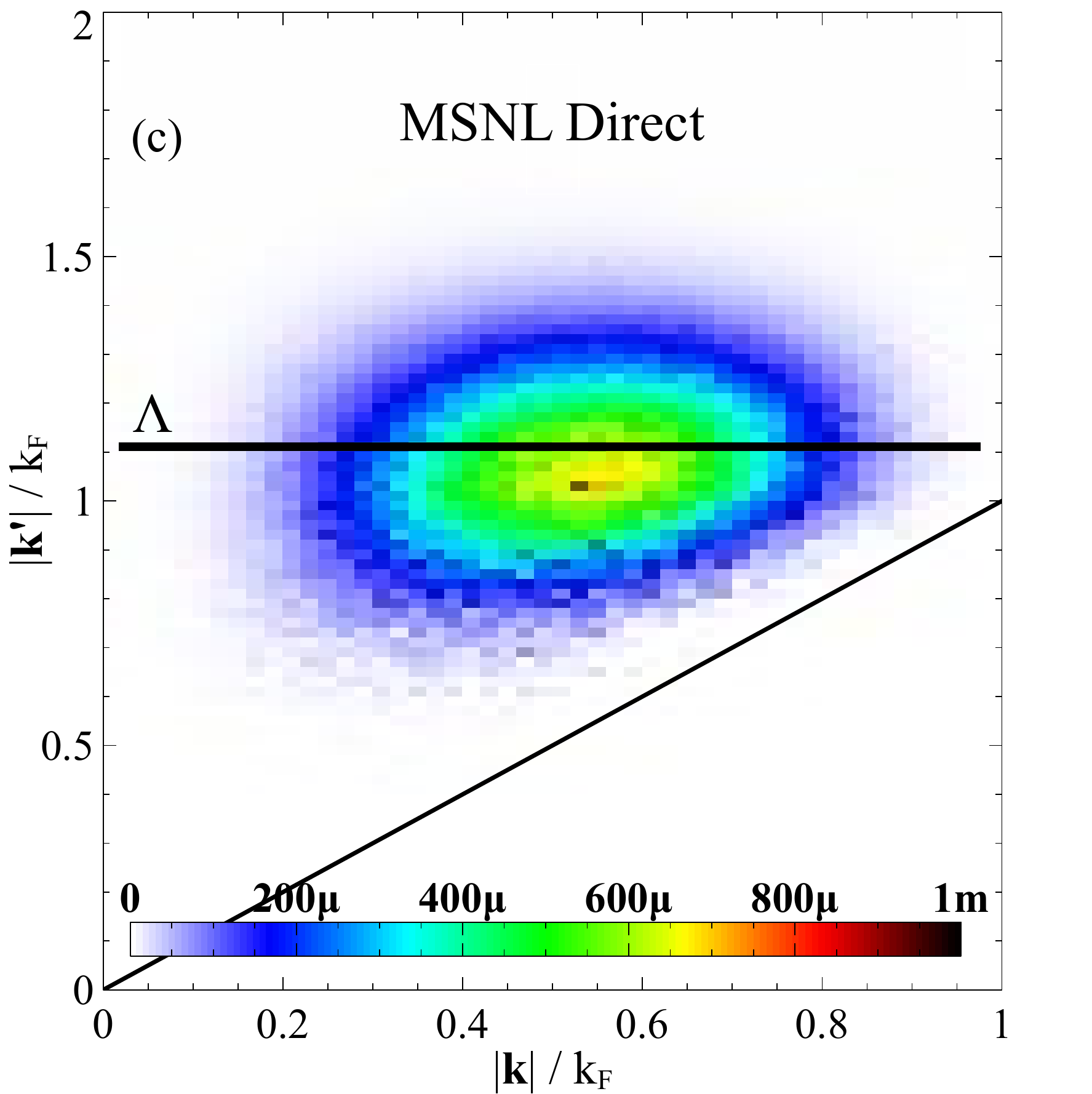}
	\includegraphics[width=0.32\textwidth]{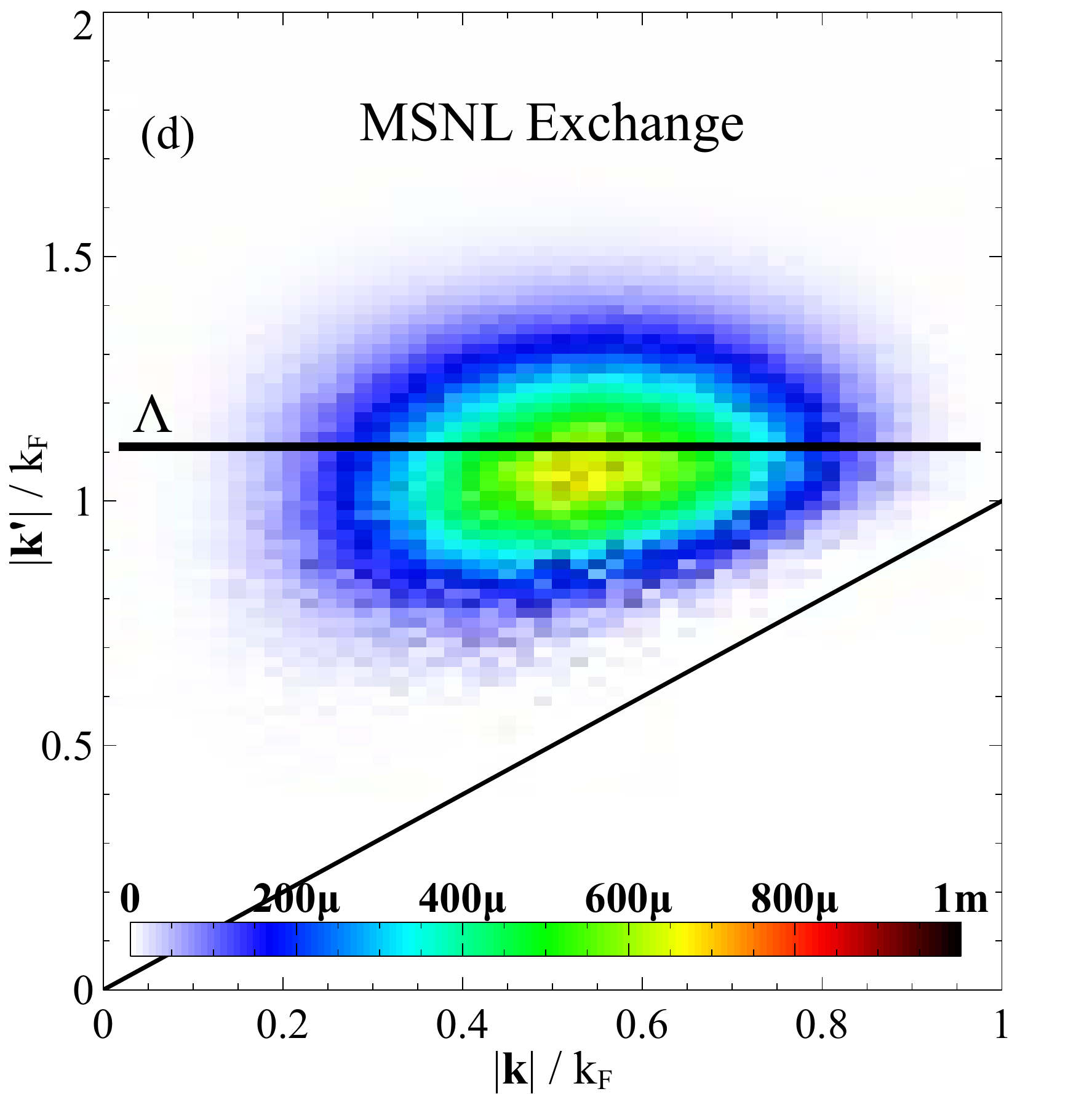}

	\caption{(color online) Momentum histograms for the integrand magnitude $I_{2}$ in \eqref{eq:NN_2nd_order_regs_only} at $\kf =
	1.8~\fmi$, $n = 2$, and $\NNcut = 2.0~\fmi$.
	The y-axis gives
	the particle relative momentum while the 
	x-axis gives the hole relative momentum.
	Colors indicate
	the integrand magnitude for a particular
	$k,k'$ pair. The horizontal black line indicates the cutoff $\NNcut$ while the sloping black line separates out the inaccessible region due to Pauli blocking. $\mu$ and $m$ in the color bar stand for $10^{-6}$ and $10^{-3}$ respectively.}
	\label{fig:NN_SO_histograms_REGS}
\end{figure*}

\subsection{NN Forces at Second-Order}
\label{sec:NN_SO}

For a 2-body interaction at second-order in MBPT, the energy per particle
in terms of single-particle  momenta is,
\begin{align*}
	\frac{E^{\text{NN}}_{\text{SO}}}{N} =&
	\frac{1}{4 \rho} 
	\bigg[ 
	\prod_{i=1}^{4} 
	\sum_{\sigma_i}
	\sum_{\tau_i}
	\int \frac{d^3p_i}
	{(2\pi)^3}
	\bigg] n({\bf p}_1) n({\bf p}_2)
	\bar{n}({\bf p}_3) 
	\bar{n}({\bf p}_4)
	\\
	& \null\times \frac{\la 12 | A_{12} 
	V^{\text{NN}}_{\text{LO}}
	|34 \ra \la 34 | A_{12} 
	V^{\text{NN}}_{\text{LO}}
	|12 \ra} 
	{\varepsilon_{{\bf p}_1}
	+ \varepsilon_{{\bf p}_2}
	- \varepsilon_{{\bf p}_3}
	- \varepsilon_{{\bf p}_4}} 
	\\
	& \null \times (2\pi)^3
	\delta^3 ({\bf p}_1 
	+ {\bf p}_2
	- {\bf p}_3
	- {\bf p}_4) \; ,
	\label{eq:NN_SO_energy}
	\numberthis
\end{align*}
where
\bseq
	\beq
	\bar{n}({\bf p}_i) \equiv 
	\Theta(|{\bf p}_i| - \kf) \; ,
	\eeq
and	
	\beq
	\varepsilon_{{\bf p}_i} = 
	\frac{\hbar^2 p_i^2}{2m} \; .
	\eeq
\eseq
For simplicity we use a free spectrum, but we do not expect a different choice
to change our discussion.
It is also useful to define a new relative momentum,
\beq
	\kvecp = 
	\frac{{\bf p}_3 - {\bf p}_4}{2} \; ,
	\label{eq:NN_particle_relative_mom}
\eeq
where ${\bf p}_3,{\bf p}_4$ correspond
to single-particle momenta with magnitudes above
the Fermi momentum $\kf$. 

The momentum transfer for a particular matrix element is defined differently depending 
on which part of the antisymmetrizer $A_{12}$ acts in the matrix element:
\begin{align*}
{1 \implies \la 12 | V | 34 \ra :}&  \quad& \qvec &= {\bf p}_1 - {\bf p}_3 = \kvec - \kvecp \; , \\
{P_{12} \implies \la 21 | V | 34 \ra :}&  \quad& \qvecp &= {\bf p}_1 - {\bf p}_4 = \kvec + \kvecp  \; .
\numberthis
\label{eq:q_2nd_order}
\end{align*}
using the relative momenta definitions in \eqref{eq:rel/tran_mom} and \eqref{eq:NN_particle_relative_mom}. 
	As such, the second-order direct term will have only $\qvec$ (or $\qvecp$) dependence while the exchange term will have both $\qvec$ and $\qvecp$ dependence due to the different particle order in the two matrix elements.
	Both $\kvec$ and $\kvecp$ are independent momenta implying that it is not generally possible for both $\qvec$ and $\qvecp$ to simultaneously have small magnitudes.
	Therefore, we expect that local regulators, which act to cut off large momentum transfers, will have suppressed energies (and phase spaces) for exchange terms relative to the direct terms.

\begin{figure*}
\includegraphics[width=0.32\textwidth]{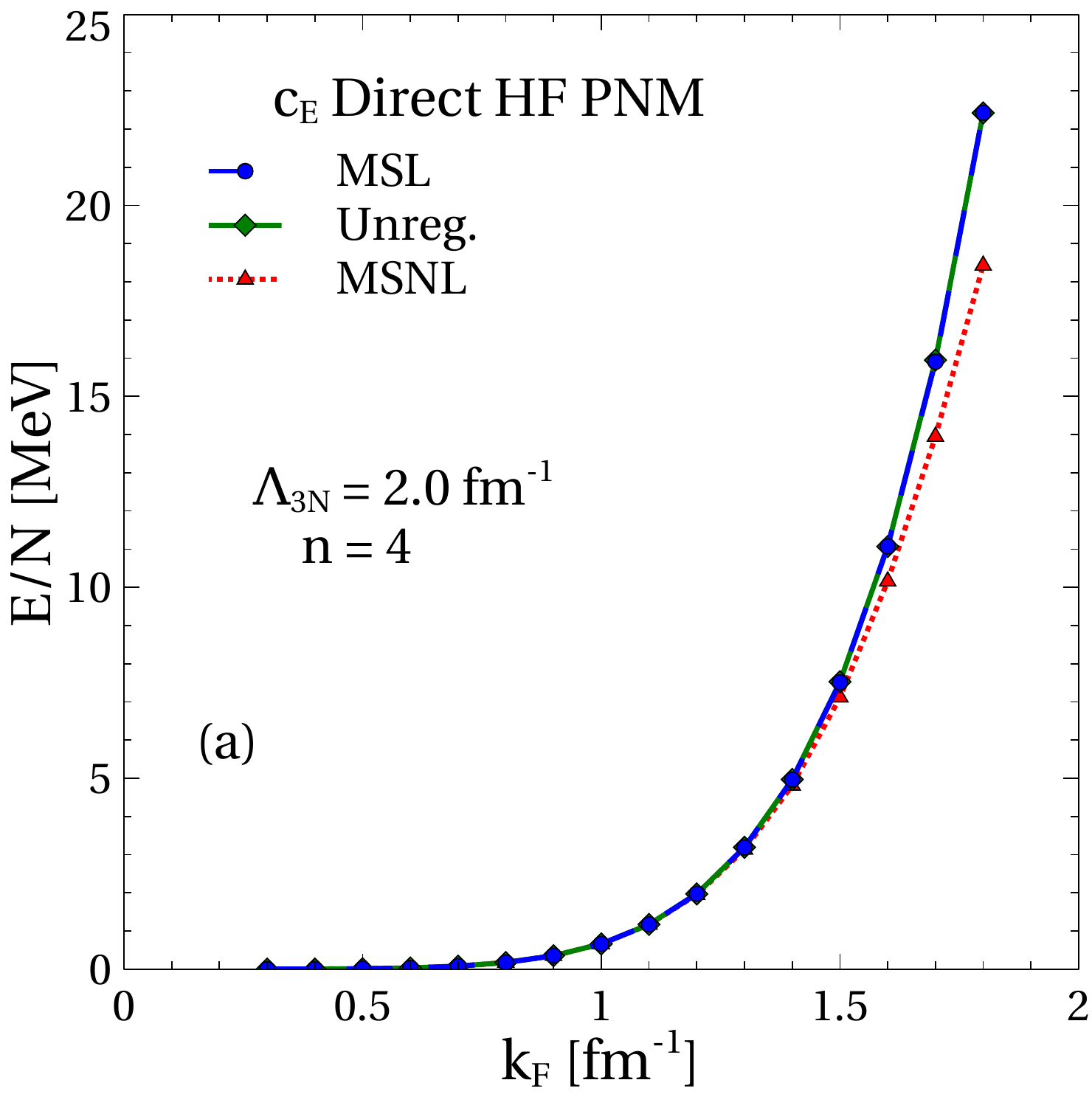}~
\includegraphics[width=0.32\textwidth]{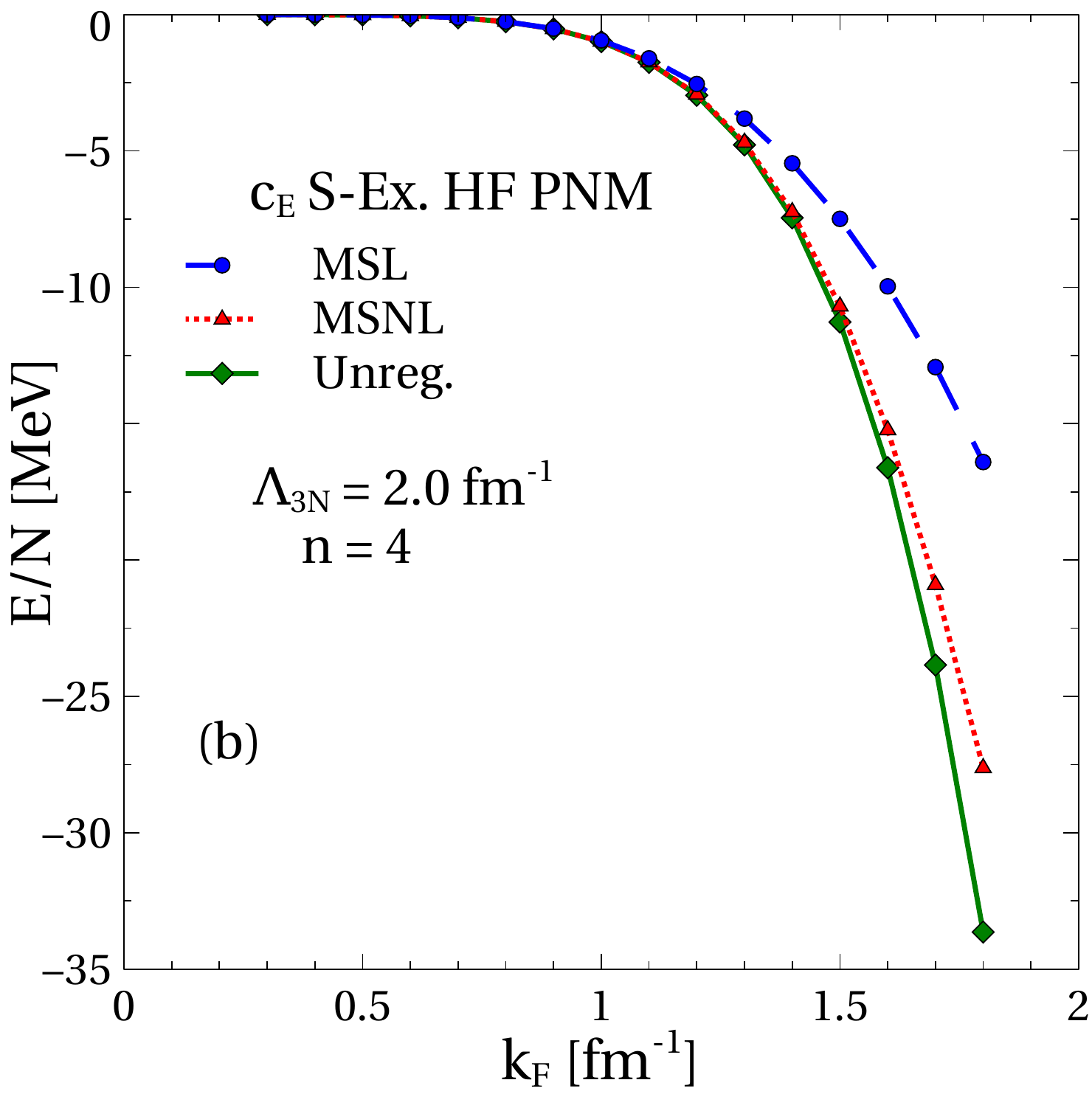}~
\includegraphics[width=0.32\textwidth]{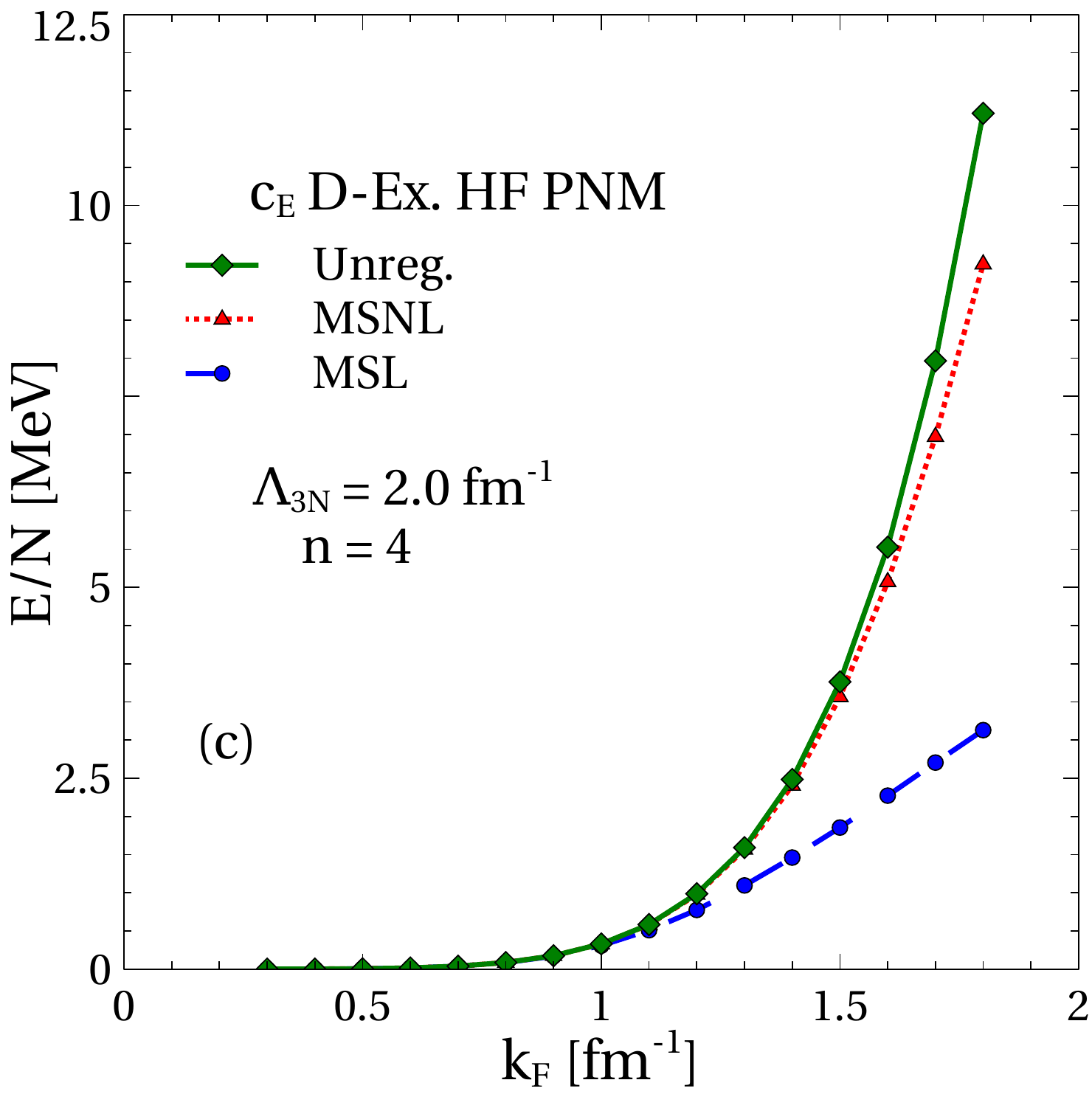}
 
\caption{Plots (a), (b), and (c) show the energy per particle of the $c_E$ term for the direct, single-exchange, and double-exchange topologies respectively in neutron matter with $\TNcut = 2.0~\fmi$ and $n = 4$.} 

\label{fig:HF_3N_ce_energies} 
\end{figure*}

The second-order energy values for the $C_S$--$C_S$ topology and the OPE--OPE 
topology are given in Fig.~\ref{fig:NN_SO_energy}.
(Diagrams with mixed vertices such as $C_S$--OPE will mix regulator effects; we
do not consider them here.) 
The $C_T$--$C_T$ term has similar behavior to the $C_S$--$C_S$ term. 
	In contrast to NN HF energy values in Fig.~\ref{fig:NN_HF_energy}, here the contact CSL regulator in Fig.~\ref{fig:NN_SO_energy}(a) and (b) deviates from the MSL scheme at large $\kf$. 
	We attribute this to the oscillating functional form of the CSL regulator in Fig.~\ref{fig:CSL_LOCAL_COMPARE}(a); the particle states at large $\kf$ probe the `ringing' of the CSL contact regulator function at large $|\qvec|$.
	We also note the large scheme dependence seen in the second-order OPE--OPE energy values, especially with respect to the coordinate space regulators. 

	Having established the utility of the phase space histograms for Hartree-Fock, we use them as diagnostics at second-order to study where the action of the regulator becomes important for the MSL and MSNL schemes. 
	Representative examples are given in Fig.~\ref{fig:NN_SO_histograms_REGS}.
	Because the choice of scheme dominates the phase space distribution, for simplicity, we consider the integrand magnitude weighted only by the regulator functions (cf.\ with \eqref{eq:NN_HF_histogram_equation} at NN HF),
\begin{align*}
I_{2} = |f_{\text{reg}}|
 \;
&n({\bf P}/2 + {\bf k}) \;
n({\bf P}/2 - {\bf k}) \;
\\
&\bar{n}({\bf P}/2 + {\bf k})  \; 
\bar{n}({\bf P}/2 - {\bf k})  \; .
\label{eq:NN_2nd_order_regs_only}
\numberthis
\end{align*}
Weighting by the more complicated full energy integrand does not alter the qualitative features of the histograms (see supplemental material and discussion below).	

	As in NN HF, single-particle momenta ${\bf p}_1, {\bf p}_2, {\bf p}_3, {\bf p}_4$ 
are randomly generated (subject to the momentum conservation constraint) and the 
corresponding energy integrand $I_{2}$ is calculated. 
	The resulting magnitude is then binned in the histogram.
	The plots are now two-dimensional, with color serving as a third degree of freedom to indicate the integrand magnitude $I_2$. 
	Hole and particle relative momentum are plotted on the x- and y-axis respectively, both normalized with respect to $\kf$. 
	After all momenta are generated, the plots are then normalized by the total number of $k',k$ pairs generated. 
		Additionally, a black horizontal line indicates the position of the cutoff and the sloping black line near the bottom of the plot separates out the inaccessible region due to Pauli blocking.

	A key distinction from NN HF for these second-order phase space plots is that the unregulated $k'$ can range up to arbitrarily high momenta. 
	Thus, while the HF plots in the previous section display a universal profile, the unregulated second-order phase space is infinite in extent and all regulated representations are \emph{inherently} scheme and scale dependent. 
	However, we do expect regulator dependencies to be less important in the lower density limit (see supplemental material).

	As the density is raised and $\kf$ starts to approach $\NNcut$, scheme artifacts will become more apparent. 
	For $\kf = 1.8~\fmi$ and $\NNcut = 2.0~\fmi$, we plot the second-order histograms in Fig.~\ref{fig:NN_SO_histograms_REGS} for the direct/exchange terms in the MSL and MSNL schemes.

	Looking first at the MSNL histograms in Fig.~\ref{fig:NN_SO_histograms_REGS}(c) and (d), we see that the distributions of the direct and exchange terms are equivalent. 
	This reflects the permutation symmetry of the nonlocal regulator in \eqref{eq:NN_reg_nonlocal}; direct and exchange terms are cut off in equivalent ways. 
We also note that the center of the MSNL distribution is at $k \approx 0.55 \kf$.
	This is similar to the center of the distribution at NN HF (cf.\ Fig.~\ref{fig:NN_HF_histograms}) and at lower densities (see supplemental material).
	This implies that as the density is raised, the phase space for the MSNL terms are primarily cut off at large $k'$.
	
	Different behavior is seen for the MSL scheme in Fig.~\ref{fig:NN_SO_histograms_REGS}(a) and (b).
	The phase space for the exchange term is suppressed compared to the phase space for the direct term as anticipated above. 
	The exchange term's phase space comes primarily from regions below the cutoff and is much more constrained in magnitude. 
	In contrast, a substantial portion of the direct term's phase space comes from relative momenta $k'$ which are above the cutoff $\NNcut$. 
	Furthermore, it is seen in each case that the central profile of the phase space is shifted away from $k \approx 0.55 \kf$.
	In the direct term, the center is shifted towards large $k$,$k'$ reflecting the potential cancellation between $\kvec$ and $\kvecp$. 
	In the exchange term, the center is shifted towards small $k$,$k'$.
	
		We make equivalent plots of the full integrand magnitude for the $C_S$--$C_S$ and OPE--OPE histograms in the supplemental at $\kf = 1.8~\fmi$.
	These do not display any qualitative differences compared to Fig.~\ref{fig:NN_SO_histograms_REGS}.
	This again emphasizes that the regulators are primarily determining the phase space distribution.

	We do not address the large scheme dependence seen for the second-order OPE--OPE energy values between the coordinate space regulators.
	Our histogram approach is not easily adapted to the use of the long-range coordinate space regulator functions at second-order and cannot offer intuition about which parts of the phase space are most relevant.

\subsection{3N Forces at HF}
\label{sec:3N_HF}

\begin{figure*}[tbh]
	\includegraphics[width=0.42\textwidth]
	{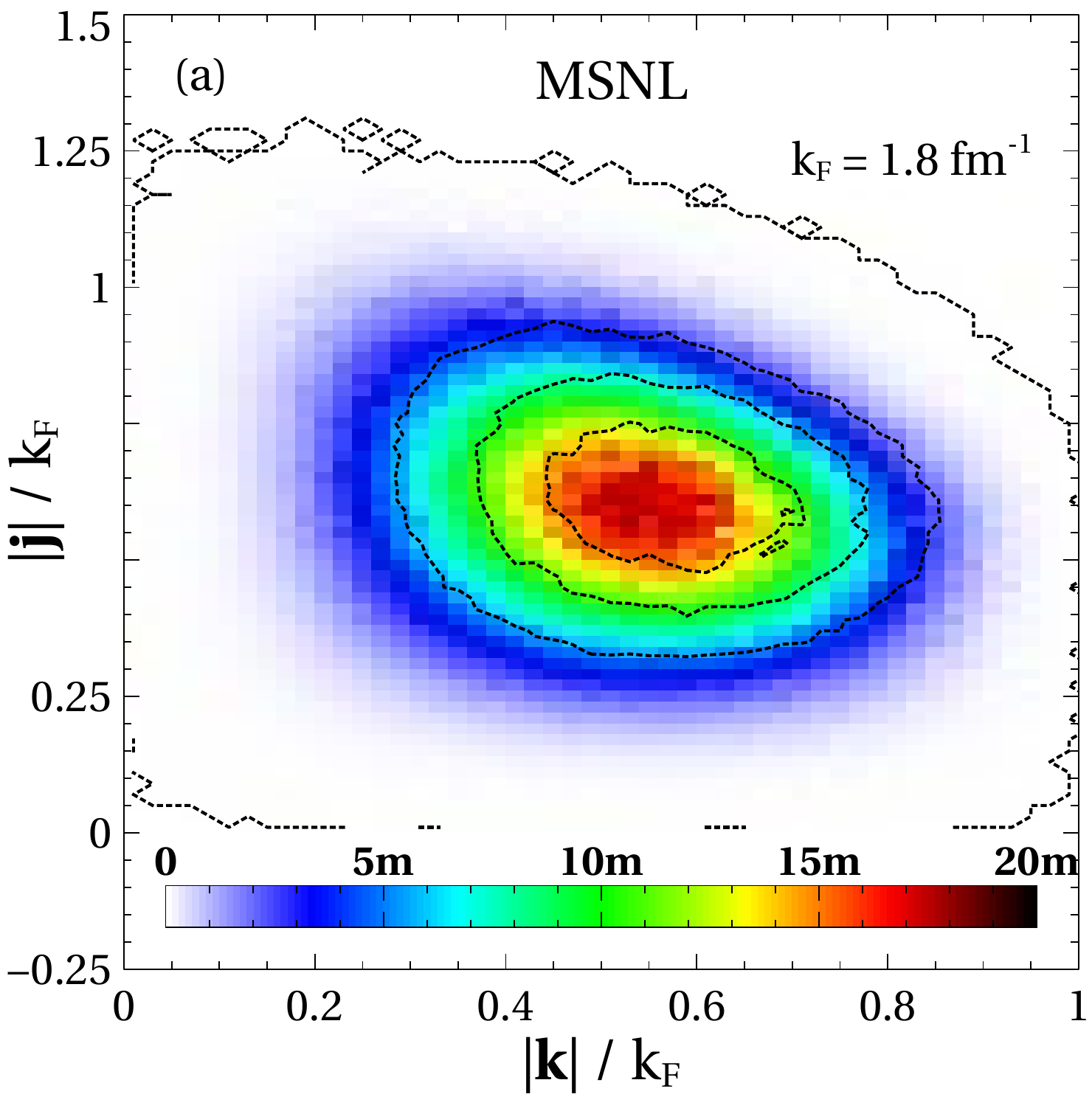}
	\includegraphics[width=0.42\textwidth]
	{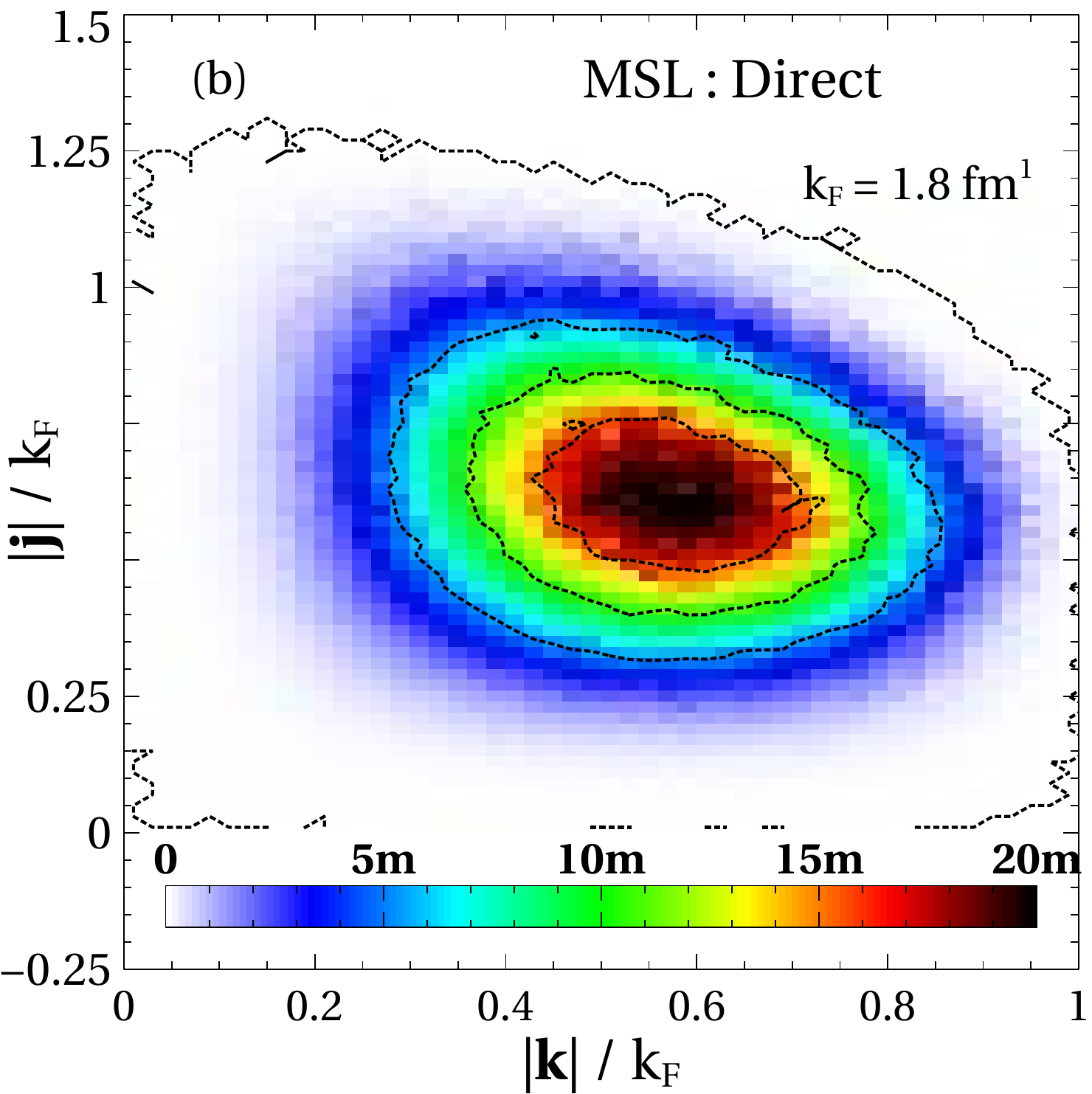}
	
	\medskip	
	
	\includegraphics[width=0.42\textwidth]
	{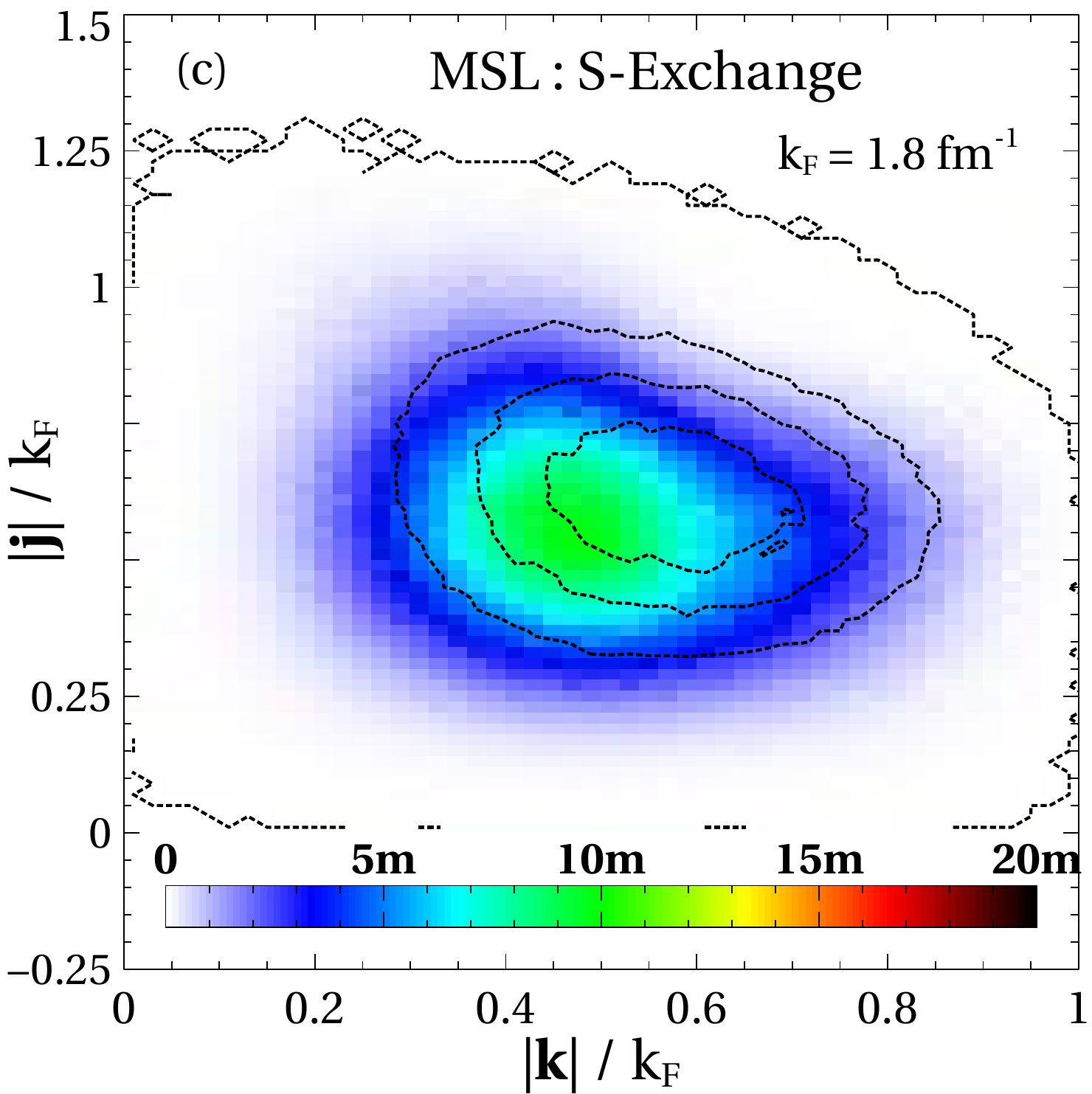}
	\includegraphics[width=0.42\textwidth]
	{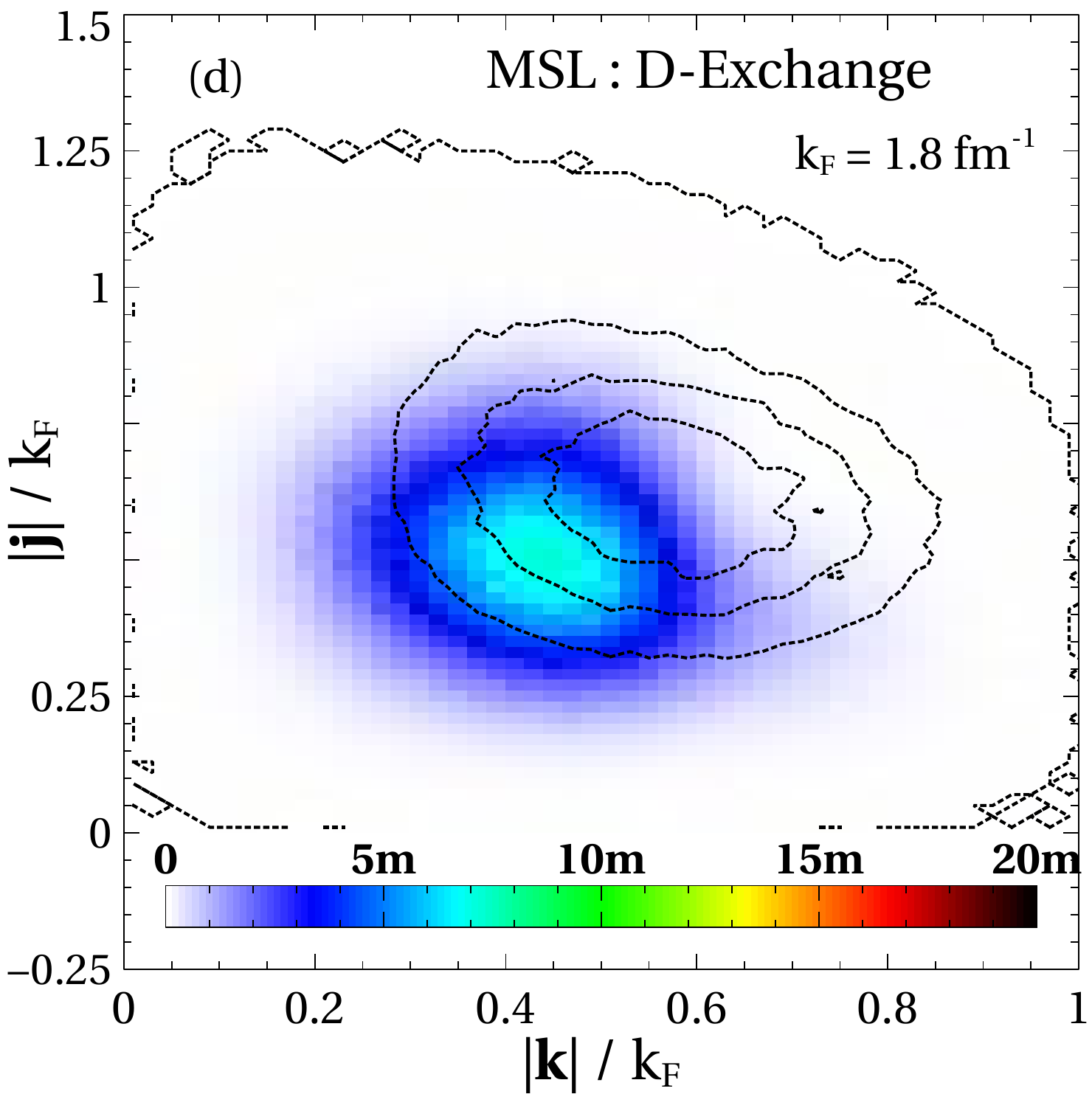}

	\caption	{(color online) Momentum histogram representing the 3N HF phase space for $\kf =
	1.8~\fmi$, $\TNcut = 2.0~\fmi$, and $n = 4$. The integrand magnitude $I_3$ in \eqref{eq:NNN_HF_histogram_equation} is plotted in color for a given $j,k$ pair normalized by $\kf$. 
	 Plot (a) shows the MSNL phase space for the direct term while (b), (c), and (d) show the MSL direct, single-exchange, and double-exchange spaces respectively.
	 The other MSNL terms have an equivalent distribution as the MSNL direct term absent magnitude rescaling.
	 The contour lines indicate the same histogram calculation but with no regulator ($f_{\text{reg}} \equiv 1$). 
	 The $m$ in the color bar stands for $10^{-3}$.}
	\label{fig:NNN_HF_histogram_ce}
\end{figure*}

For a 3-body interaction, the HF energy per
particle in terms of the single-particle momenta is
given by 
	\begin{multline}
	\frac{E^{\text{3N}}_{\text{HF}}}{N} = 
	\frac{1}{6 \rho}
	\sum_{\sigma_1 \sigma_2 \sigma_3}
	\sum_{\tau_1 \tau_2 \tau_3}
	\int \frac{d^3p_1}{(2\pi)^3}
	\int \frac{d^3p_2}{(2\pi)^3}
	\int \frac{d^3p_3}{(2\pi)^3}
	\\
	n({\bf p}_1)
	n({\bf p}_2)
	n({\bf p}_3)
	\la 1 2 3 | A_{123} 
	V_{\text{\NNLO}}^{\text{3N}}
	| 1 2 3 \ra \; ,
	\end{multline}
where $V_{\text{\NNLO}}^{\text{3N}}$ includes a regularization scheme.
The antisymmetrizer in \eqref{eq:3N_antisym} leads to three different classes 
of terms depending on the 
number of exchange operators $P_{ij}$: one term with no exchange operators, 
three terms with a single exchange operator, and two terms with two exchange operators. 
These components are respectively dubbed the direct, single-exchange, and double-exchange terms. 
Note that in this decomposition, single-exchange and double-exchange by our 
convention refer to \emph{all} the terms with the associated exchange operators 
(e.g., single-exchange energies include contributions from $P_{12}$, $P_{13}$, 
and $P_{23}$). 
Evaluating these different components with the $\NLTN$ regulator in \eqref{eq:3N_nonlocal} and the $\LTN$ regulator in \eqref{eq:3N_local} give the energies in Fig.~\ref{fig:HF_3N_ce_energies} for the contact $c_E$ term.
	The $\LTN$ scheme is equivalent to no regulator for the direct term while single-exchange and double-exchange terms are increasingly suppressed.
	The MSNL scheme has a similar relative effect on all contributions with respect to unregulated HF.
	Trends for the finite range $c_i, c_D$ pieces are similar (see supplemental material).
For both the $\NLTN$ and $\LTN$ schemes, the $c_E$ term dominates the energy per particle for natural choice of LECs.

As before, we analyze momentum histograms to describe the 3N HF phase space. 
Single-particle momenta ${\bf p}_1, {\bf p}_2, {\bf p}_3$ are randomly generated 
using Monte Carlo sampling and the 3N HF integrand magnitude, 
\begin{align*}
I_{3} =& |
f_{\text{reg}}| \;
k^2 \;
j^2 \;
P^2 \;
n({\bf P}/3 - {\bf j}/2 - {\bf k}) \;
\\
&\null\times
n({\bf P}/3 - {\bf j}/2 + {\bf k}) \;
n({\bf P}/3 + {\bf j})  \; ,
\numberthis
\label{eq:NNN_HF_histogram_equation}
%\bigg\{ 
%\frac{\left(\spinvec_i \cdot \qvec_i \right) 
%      \left(\spinvec_j \cdot \qvec_j \right)}
%  	  {\left(q_i^2 + \mpi^2 \right) 
%  	  \left(q_j^2 + \mpi^2 \right)}, 1
%\bigg\}
\end{align*}
is then calculated for the Jacobi momenta $\kvec$,$\jvec$ defined in \eqref{eq:3N_Jacobi}. 
	The integrand magnitude is then binned in a histogram with the moduli of the associated Jacobi momenta, normalized by $\kf$, plotted on the y- and x-axes. 
	The sampling process is then repeated and the final distribution is normalized by the total number of Monte Carlo iterations.
	As in Sec.~\ref{sec:NN_SO}, the resulting histograms are two-dimensional with color
intensity denoting integrand magnitudes.
	Note that in \eqref{eq:NNN_HF_histogram_equation} we do not weight the distribution by the different interactions $c_i, c_D, c_E$. 
	Such weighting is superfluous for our purposes as all the weightings generate similar plots (see supplemental material).
	As all the momenta in HF are on-shell, the phase space here is unambiguously 
well-defined, regardless of the cutoff or regulator. 
	As in NN HF, unregulated 3N HF serves as a touchstone to assess scale/scheme dependence via deviations from the unregulated result.

	In Fig.~\ref{fig:NNN_HF_histogram_ce} we plot representative examples of the full 3N HF phase space for the $\LTN$ and $\NLTN$\footnote{We only plot the direct term for the $\NLTN$ scheme in Fig.~\ref{fig:NNN_HF_histogram_ce}(a) as the single-exchange and double-exchange terms have the same distribution of points with rescaled magnitudes.} scheme.
	The color shows the integrand magnitude $I_3$ for the given regularization scheme while the contour lines indicate the same distribution with no regulator attached to the potential ($f_{\text{reg}} \equiv 1$).
%	Note that the color patterns are circular, demonstrating that $\jvec$ and $\kvec$ are uncorrelated independent variables. 
	As at NN second-order, the distribution of points in the weighted phase space is primarily determined by the choice of regulator function.

We make a few general comments:

\bi
	\I The hierarchy in energy values matches
	the volumes of the different phase spaces i.e., MSL direct $>$ MSNL $>$ MSL single-exchange $>$ MSL double-exchange.
	\I The MSL direct term is unaltered as the direct
	diagram has ${\bf q}_i = 0$ for all momentum transfers.
	\I The central profile of the MSNL term is slightly shifted towards smaller $k$. This reflects the regulator cutting into the hole phase space with exponential suppression of large $k,j$. Note that the factor of $\frac{3}{4}$ in \eqref{eq:3N_nonlocal_alt} means that large $k$ will cause more suppression compared to large $j$.
	\I The center of the MSL single-exchange histogram is shifted towards small $j$ and $k$. It also has an asymmetric shape extending out to large $|\kvec| / \kf$. 
	This results from the different parts of the 3N interaction not being regulated identically for the	different single-exchange	 components. 
	\I The MSL double-exchange histogram is also shifted to small $j$ and $k$ but to a larger extent than the MSL single-exchange term. 
	As in the single-exchange case, asymmetric features originate from the different momentum transfer possibilities for the two different double-exchange components.
	\ei
	%

%%%%%%%%%%%%%%%%%%%%%%%%%%%%%%%%%%%%%%%%%%%%%%%%%%
%
%%%%%%%%%%%%%%%%%%%%%%%%%%%%%%%%%%%%%%%%%%%%%%%%%%

\subsection{3N Forces at Second-Order} \label{sec:3N_SO}

\begin{figure}[tbh]
	\includegraphics[scale=0.45]{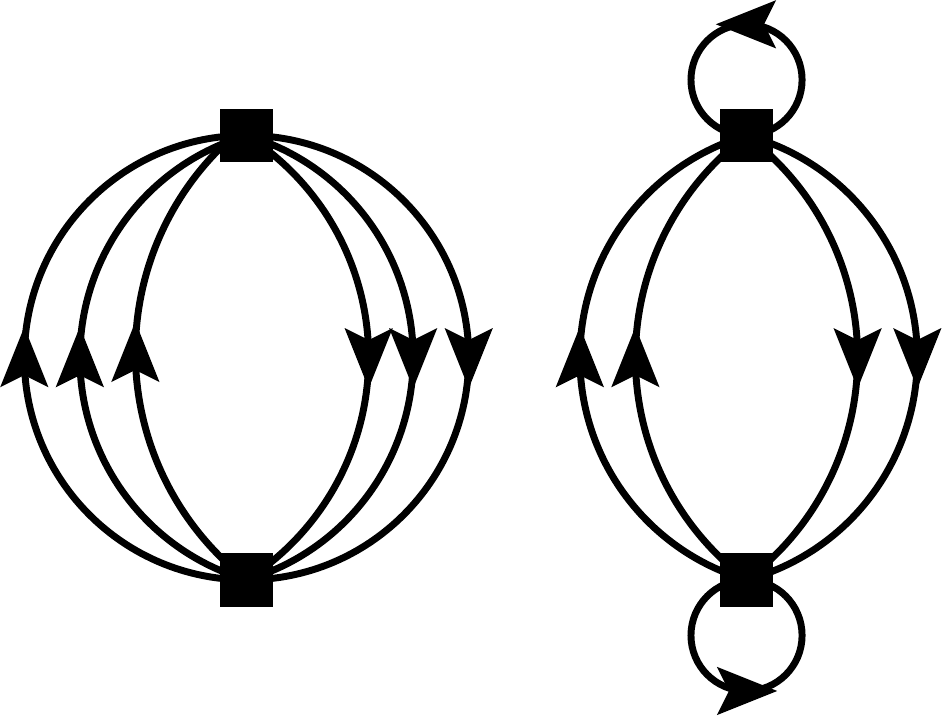}
	\caption{The residual (left) and normal-ordered (right) second-order diagrams
	arising from 3-body forces. The NO2B approximation discards the residual term while keeping the normal-ordered digram.}
	\label{fig:SO_3N_diagrams}
\end{figure}

	For MBPT at finite density, there exist two types 
	of diagrams resulting from 3-body forces at 
	second-order \cite{PhysRevC.81.024002,Hebeler:2009iv,
	Carbone:2014mja}. These can be found by normal-ordering the
	free-space second-quantized 3-body operators with respect to
	a finite density reference 
	state.%
	\footnote{We do not consider the second-order diagram with normal-ordered one-body interactions
	from the 3-body force because the diagram vanishes at zero temperature. The 0-body term is also not considered.}
	The first diagram is called normal-ordered or density-dependent (DD), and is found by closing a single-particle line at each 3-body vertex resulting in an effective 2-body interaction. 
	The other diagram, called the residual (RE) diagram, has three particles above and three holes below the Fermi surface and is a true 3-body term. 
	Both diagrams are shown in Fig.~\ref{fig:SO_3N_diagrams}.

	\begin{figure*}[tbh]
		\includegraphics[width=0.9\columnwidth]{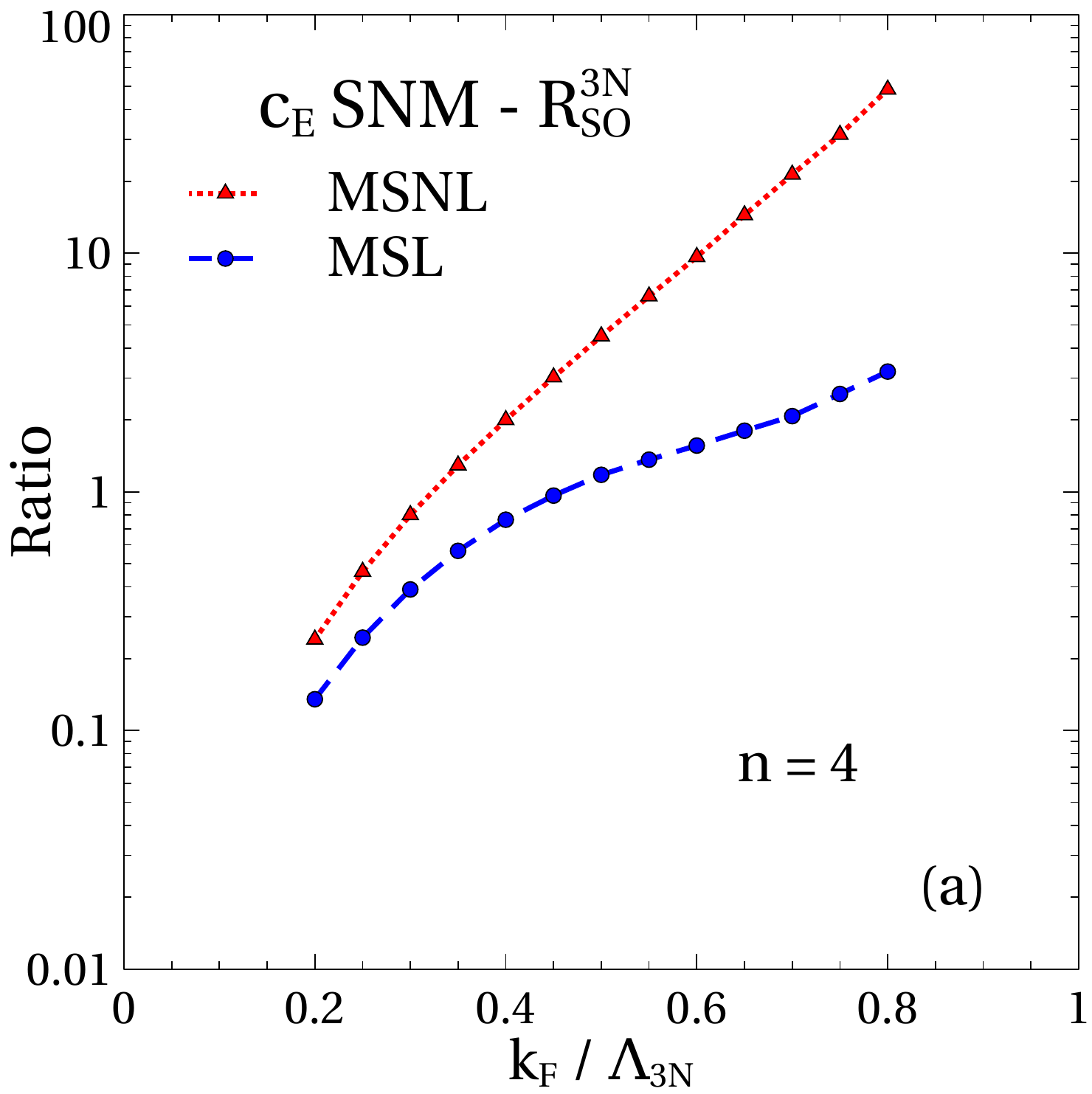}
		\includegraphics[width=0.9\columnwidth]{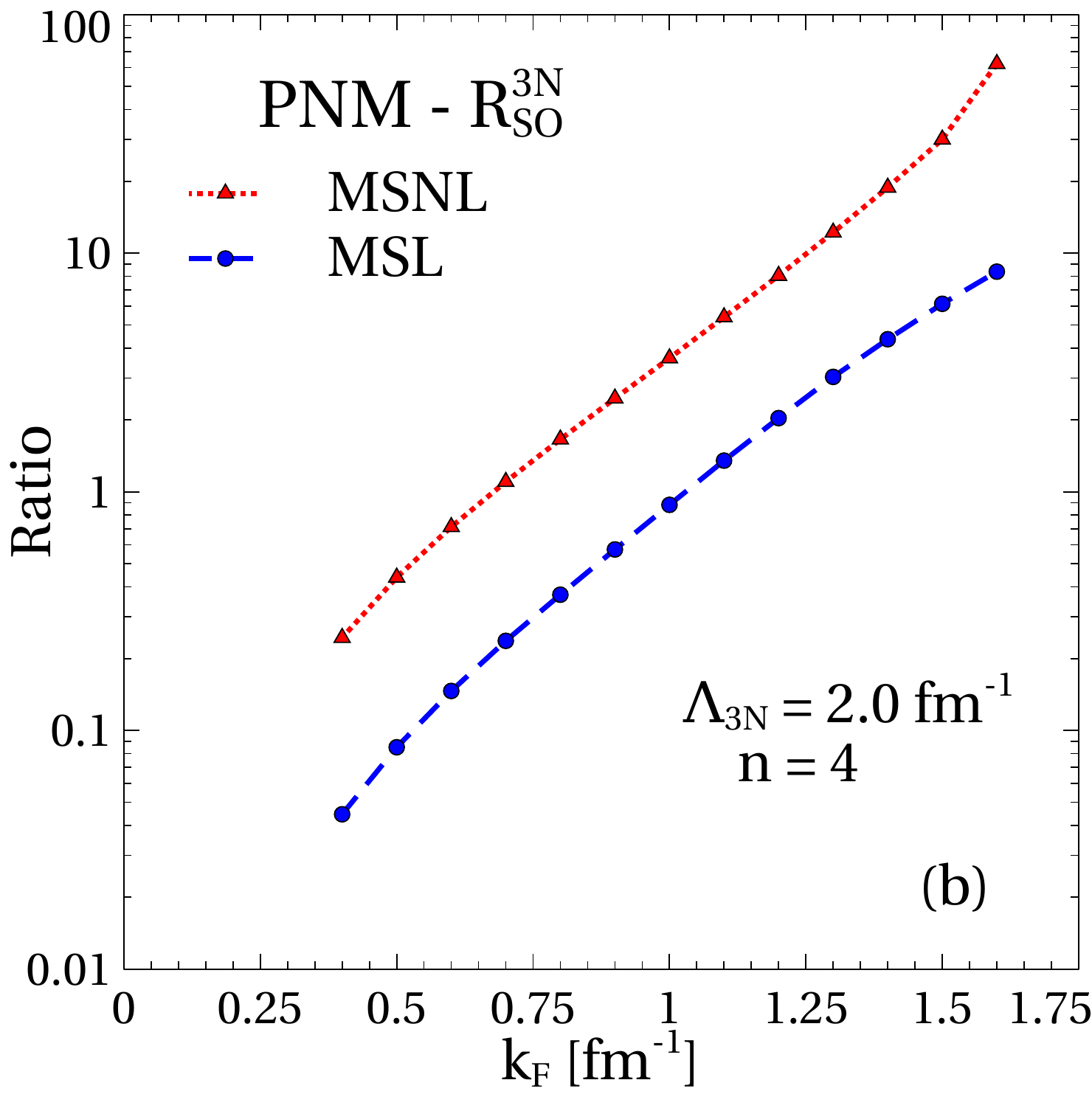}

	\caption{Plot (a) shows the
	ratio $\ratio$ in \eqref{eq:3N_ratio} for the 3N contact term evaluated in SNM. Plot (b) shows the same ratio $\ratio$ only now including all 3-body interactions in PNM. The trend in the ratio is very similar to plot (a). Both calculations are done with $\TNcut = 2.0~\fmi$ and $n=4$.}

		\label{fig:3N_2nd_ratio}
	\end{figure*}

	For the DD diagram,
	we treat the interaction coming
	from the 3N sector as an effective
	2-body force, so our 
	previously defined formula for
	the second-order NN energy in \eqref{eq:NN_SO_energy} applies,
	\begin{align*}
		\frac{E^{\text{3N}}_{\text{DD}}}{N} =
		&\frac{1}{4 \rho} 
		\bigg[ 
		\prod_{i=1}^{4} 
		\sum_{\sigma_i}
		\sum_{\tau_i}
		\int \frac{d^3p_i}
		{(2\pi)^3}
		\bigg] 
		n({\bf p}_1)
		n({\bf p}_2)
		\bar{n}({\bf p}_3)
		\bar{n}({\bf p}_4)
		\\
		& \null\times \frac{\la 12 | A_{12} 
		\overline{V}^{\text{3N}}_{\NNLO}
		|34 \ra \la 34 | A_{12} 
		\overline{V}^{\text{3N}}_{\NNLO}
		|12 \ra}
		{\varepsilon_{{\bf p}_1}
		+ \varepsilon_{{\bf p}_2}
		- \varepsilon_{{\bf p}_3}
		- \varepsilon_{{\bf p}_4}} 
		\\
		& \null\times (2\pi)^3
		\delta^3 ({\bf p}_1 
		+ {\bf p}_2
		- {\bf p}_3
		- {\bf p}_4)		
		\; ,
	\numberthis
	\end{align*}
	where here we have added an 
	overline to	the potential to indicate this 
	normal-ordering 	prescription with
	respect to the third particle,
	\begin{multline}
	\la 12 | \overline{V}^{\text{3N}}_{\NNLO}
	|45 \ra = \sum_{\sigma_3, \tau_3}
	\int \frac{d^3p_3}{(2\pi)^3}
	\; n({\bf p}_3)
	\\
	 \null\times \la 1 2 3| 
	(1 - P_{13} - P_{23}) 
	V^{\text{3N}}_{\NNLO}
	| 4 5 3 \ra \; .
	\end{multline}

\begin{figure*}[tbh]

\includegraphics[width=0.32\textwidth]{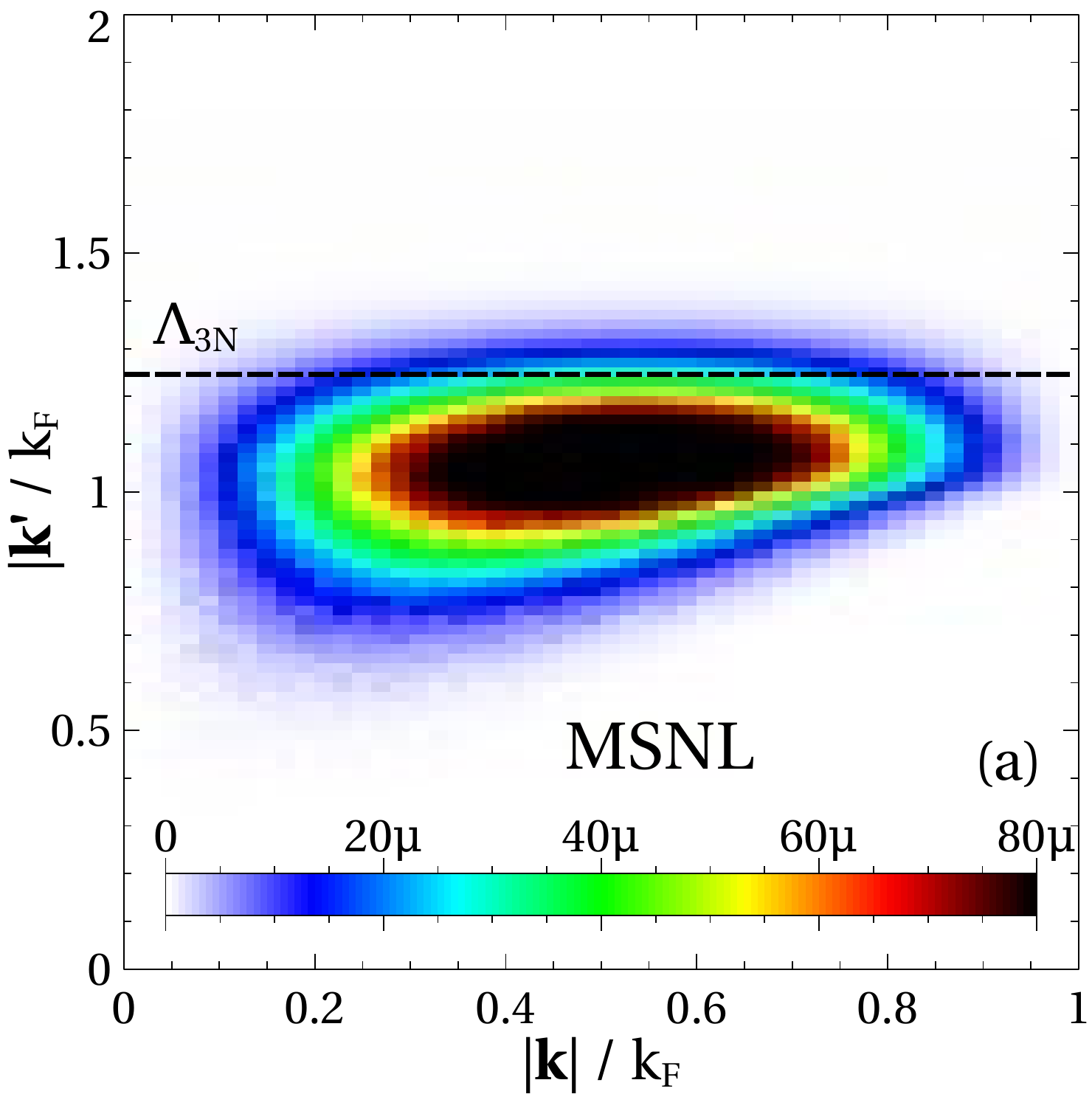}
\includegraphics[width=0.32\textwidth]{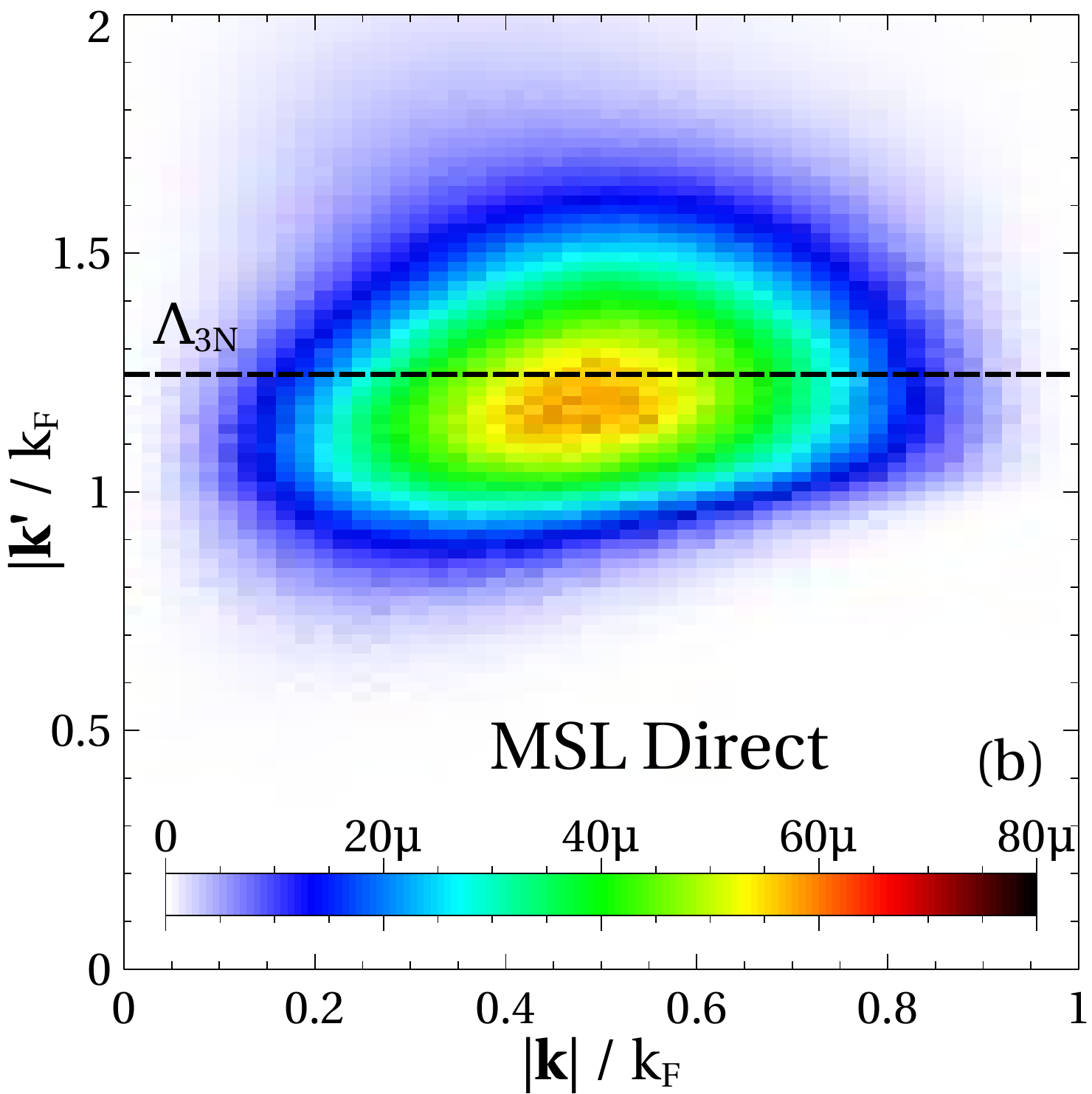}
\includegraphics[width=0.32\textwidth]
{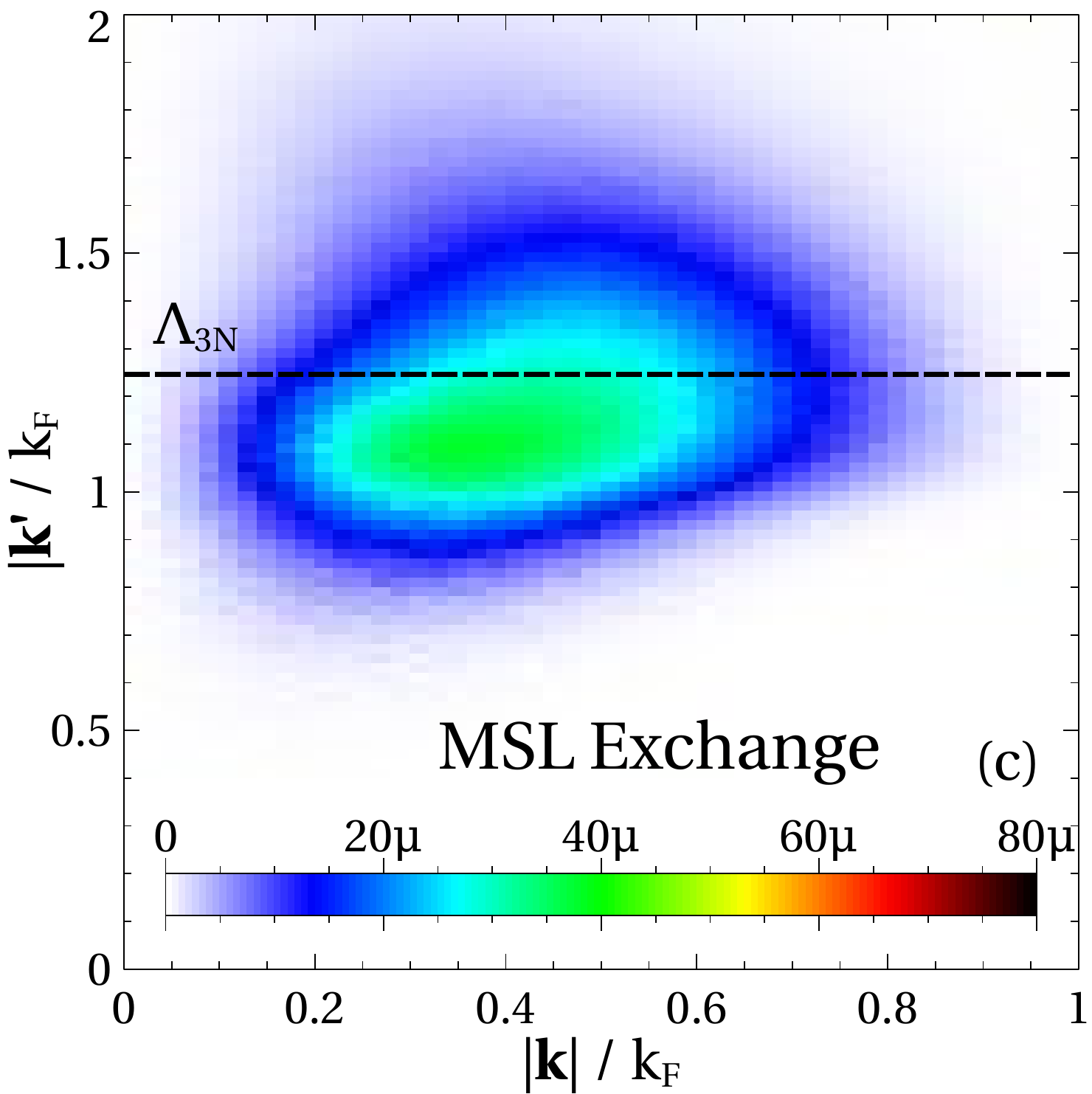}

\caption{(color online) Momentum histograms for the 3N second-order normal-ordered term where colors indicate the integrand magnitude $I_{4,\text{DD}}$ in \eqref{eq:NNN_SO_DD_histo_eq}. Plots done for the MSNL (a) term and the MSL direct (b) and MSL exchange (c) terms. The direct/exchange MSNL histograms are equivalent due to regulator permutation symmetry. Plotted for $\kf = 1.6~\fmi$, $\TNcut = 2.0~\fmi$, $n = 4$. The dashed line indicates the location of the cutoff $\TNcut$. The $\mu$ in the color bar stands for $10^{-6}$.}

\label{fig:3N_2nd_DD_histos}
\end{figure*}

	For the second-order 3N RE
	diagram, the energy per particle 
	is given by
	\begin{align*}
		\frac{E^{\text{3N}}_{\text{RE}}}{N} =
		&\frac{1}{36 \rho} 
		\bigg[ 
		\prod_{i=1}^{6} 
		\sum_{\sigma_i}
		\sum_{\tau_i}
		\int \frac{d^3p_i}
		{(2\pi)^3}
		\bigg] 
		\\		
		& \null\times n({\bf p}_1) 
		n({\bf p}_2)
		n({\bf p}_3)
		\bar{n}({\bf p}_4) 
		\bar{n}({\bf p}_5)
		\bar{n}({\bf p}_6)
		\\
		& \null\times \frac{\la 123 | A_{123} 
		V^{\text{3N}}_{\NNLO}
		|456 \ra \la 456 | A_{123} 
		V^{\text{3N}}_{\NNLO}
		|123 \ra}
		{\varepsilon_{{\bf p}_1}
		+ \varepsilon_{{\bf p}_2}
		+ \varepsilon_{{\bf p}_3}
		- \varepsilon_{{\bf p}_4}
		- \varepsilon_{{\bf p}_5}
		- \varepsilon_{{\bf p}_6}}
		\\ 
		& \null\times (2\pi)^3
		\delta^3 ({\bf p}_1 
		+ {\bf p}_2
		+ {\bf p}_3
		- {\bf p}_4
		- {\bf p}_5
		- {\bf p}_6) \; .
		\numberthis
		\end{align*}
	Calculations from different
	many-body methods (e.g. coupled-cluster) have
	indicated that the DD diagram is larger than the RE diagram~\cite{Hagen:2007ew}. As such, the
	RE diagram is usually excluded in
	the normal-ordered 2-body (NO2B) 
	approximation for reasons of computational
	efficiency. 
	If this approximation is to be	well-founded, the contribution of the DD term to the energy density should be much larger than the RE term. 
	That is, the ratio of the contribution of the DD diagram to the RE diagram, 
	\beq
	\ratio \equiv 
	\frac{E^{\text{3N}}_{\text{DD}}}
	{E^{\text{3N}}_{\text{RE}}}
	\; ,
	\label{eq:3N_ratio}
	\eeq
	must be much greater than one.
	The assessment of the NO2B approximation has practical consequences for calculations of finite nuclei and for calculating theoretical error bars. There are also implications for power counting at finite density and the general organization of the many-body problem.
	
	Here we take a simplest first look at the ratio $\ratio$ using only the $c_E$ 3N contact term. 
	 As a benchmark, the ratio $\ratio$ can be evaluated using dimensional regularization. 
	 Assuming the subtraction point is of the same order as $\kf$, the ratio is found to be $\ratio \approx 2$ \cite{Kaiser:2012mm}.
	
	For cutoff regularization, we find a significant scale and scheme dependence for $\ratio$.
	Evaluating $\ratio$ for the $c_E$ term in SNM%
	\footnote{The $c_E$ term vanishes in PNM for the $\NLTN$ scheme so here we switch to using SNM.} using the $\LTN$ and $\NLTN$ regulator 	results in the points in Fig.~\ref{fig:3N_2nd_ratio}(a). 
	Here the ratio is plotted	against the Fermi momentum $\kf$ scaled by the cutoff $\TNcut$. 
	Including \emph{all} the $\NNLO$ 3N interactions in PNM results in the plot in Fig.~\ref{fig:3N_2nd_ratio}(b).
	The qualitative and semi-quantitative features of Fig.~\ref{fig:3N_2nd_ratio}(a) and (b) are similar, establishing that the inclusion of the finite-range forces	and isospin does not appreciably alter this picture. 
	
	First, $\ratio$ in Fig.~\ref{fig:3N_2nd_ratio}(a) exhibits an obvious scale dependence for both schemes. 
	Staying in a particular scheme at a fixed density, changing the cutoff causes one to move left or right on this plot. 
	At a large cutoff $\TNcut$ compared to $\kf$, the particle phase space is not sufficiently cut off and dominates over the hole phase space. 
	The RE diagram has one fewer hole and one extra particle compared to the DD diagram and so consequently, a small $\kf / \TNcut$ amplifies the importance of the RE term. 
	 Looking at Fig.~\ref{fig:3N_2nd_ratio}(a) at $\kf / \TNcut \approx 0.3$, the diagram ratio $\ratio$ is $\mathcal{O}(1)$ for the $\NLTN$ scheme and already less than 1 for the $\LTN$ scheme.

\begin{figure*}[tbh]

\includegraphics[width=0.42\textwidth]{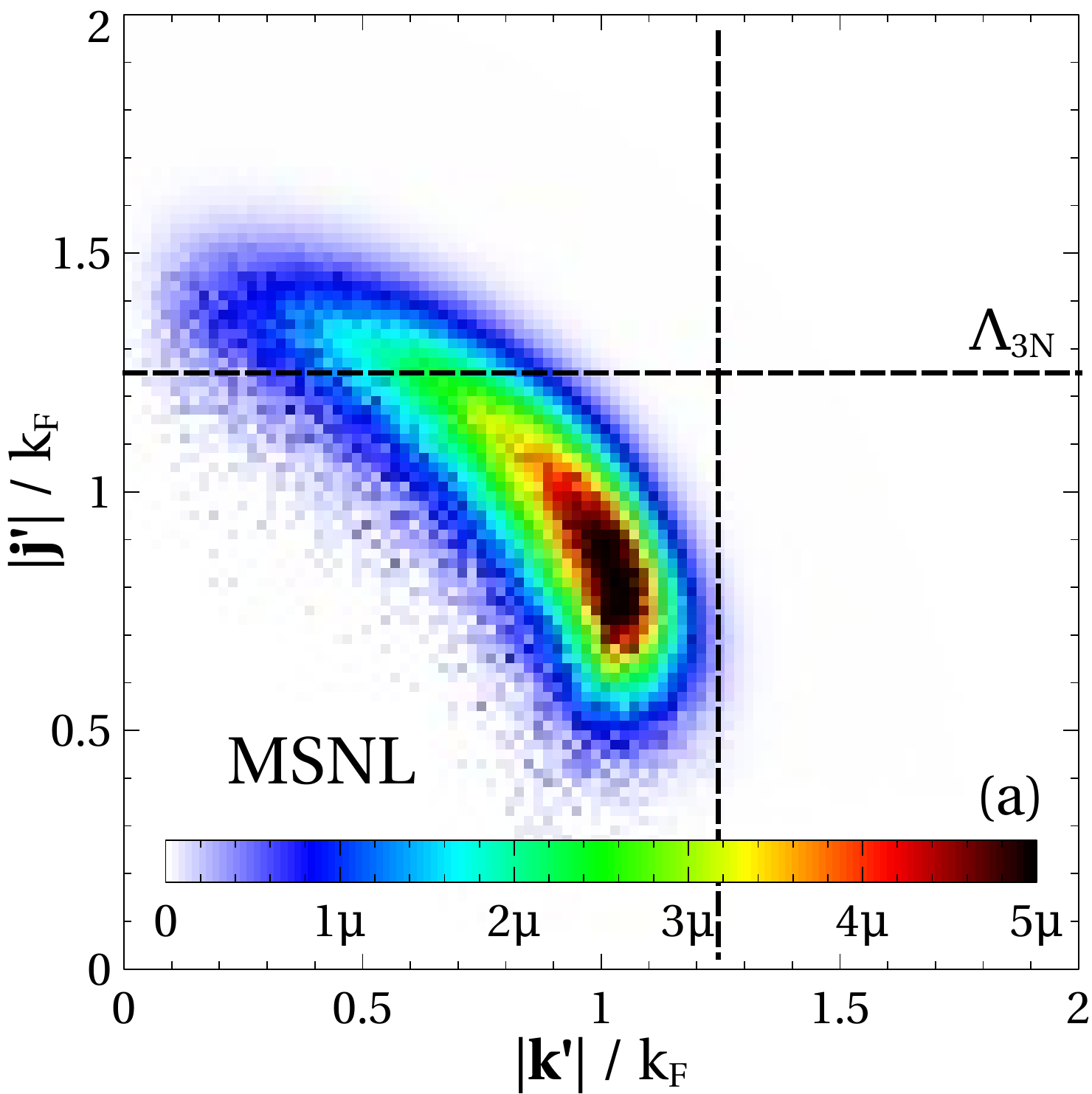}
\includegraphics[width=0.42\textwidth]{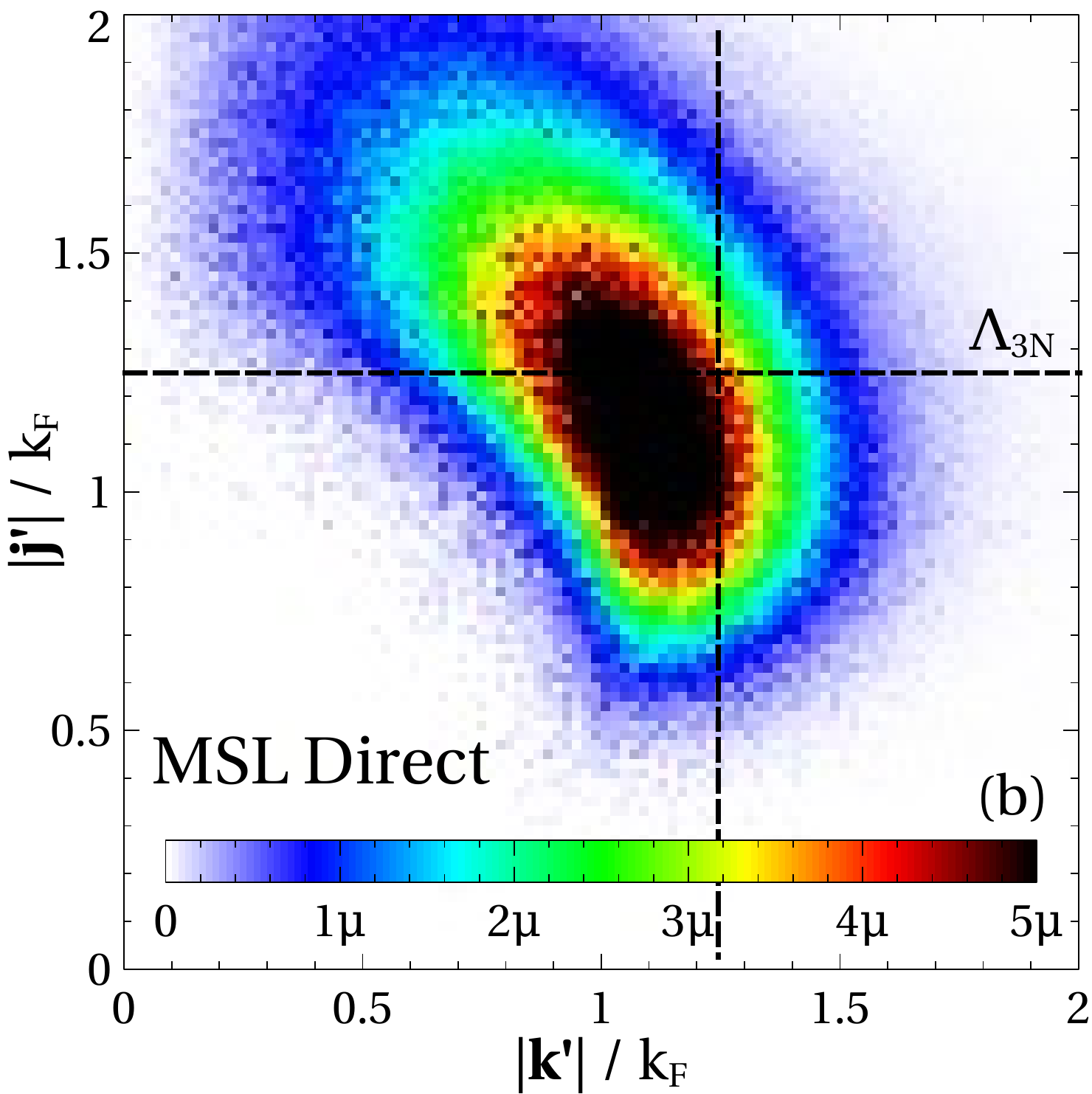}

\medskip

\includegraphics[width=0.42\textwidth]
{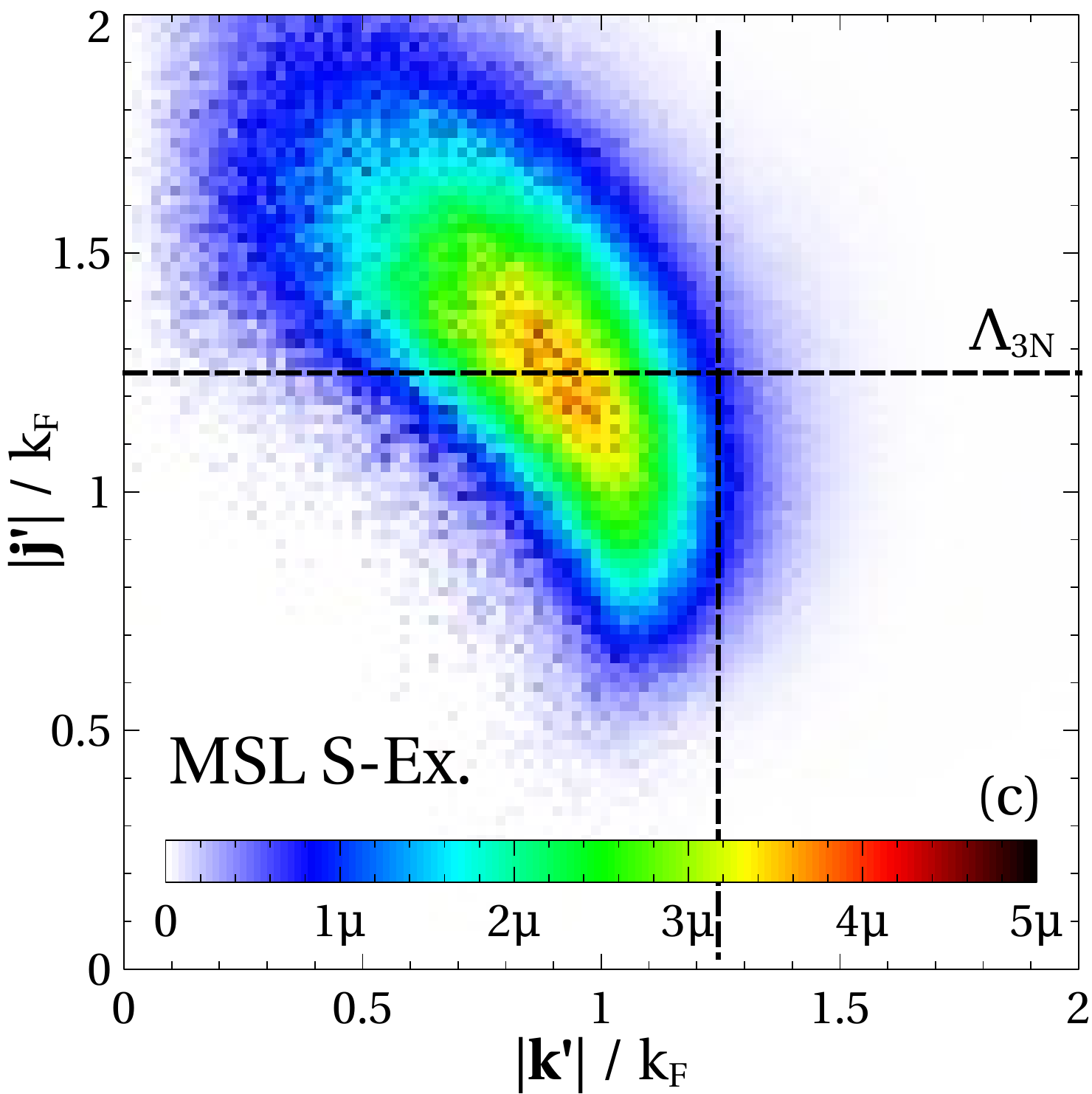}
\includegraphics[width=0.42\textwidth]
{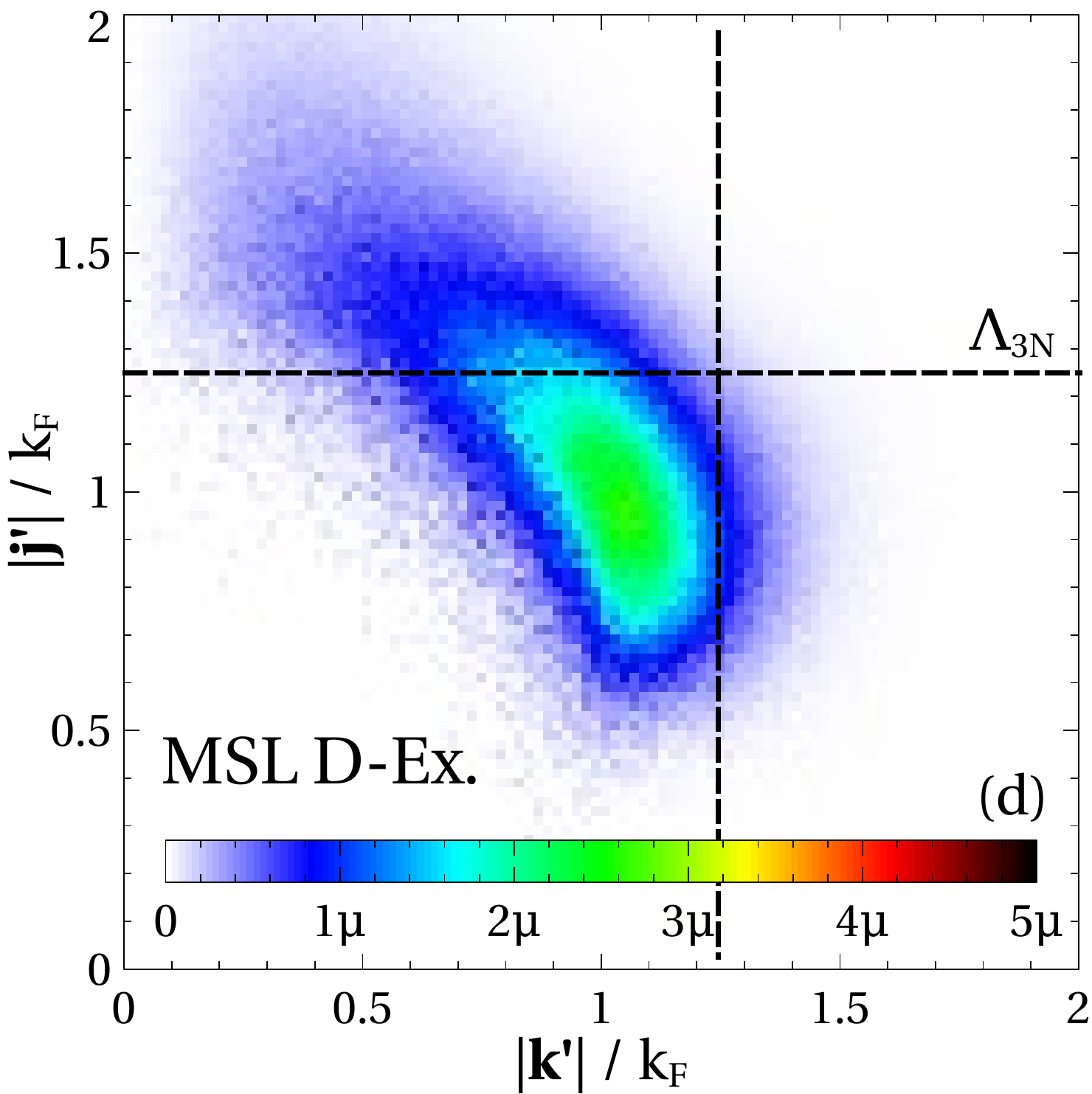}

\caption{(color online) 
Momentum histograms for the 3N second-order residual term where colors indicate the integrand magnitude $I_{4,\text{RE}}$ in \eqref{eq:NNN_SO_RE_histo_eq}. Plots done for the MSNL (a) term and the MSL direct (b), single-exchange (c), and double-exchange (d) terms. The different MSNL antisymmetric histograms are equivalent due to regulator permutation symmetry. Plotted for $\kf = 1.6~\fmi$, $\TNcut = 2.0~\fmi$, $n = 4$. The dashed lines indicate the location of the cutoff $\TNcut$. The $\mu$ in the color bar stands for $10^{-6}$.}

\label{fig:3N_2nd_RE_histos}
\end{figure*}

	Second, there is a scheme dependence for $\ratio$; for $\kf$ near the cutoff $\TNcut$, the relative importance of the different 3N diagrams in the two schemes differs by almost an order of magnitude.
	This difference between the two schemes can be understood by examining the effect of the regulator on the different 3N antisymmetric components. 
	As before, we use momentum histograms to highlight the action of the regulator on the phase space.

  The relevant integrand magnitudes for the second-order 3N energy, including only Pauli blocking and the regulators, is,
	\begin{align*}
	&I_{4,\text{DD}} = 
    |f_{\text{reg}}| \;
    n({\bf p}_5) \;
    n({\bf p}_6) \;
	n({\bf P}/2 + {\bf k}) \; 
	\\
    &\null\times
	n({\bf P}/2 - {\bf k}) \;
	\bar{n}({\bf P}/2 + {\bf k'}) \; 
	\bar{n}({\bf P}/2 - {\bf k'}) \; ,
	\label{eq:NNN_SO_DD_histo_eq}
	\numberthis
	\end{align*}
  for the DD term and, 
  \begin{align*}
  &I_{4,\text{RE}} = 
   |f_{\text{reg}}| \;
     n({\bf P}/3 + {\bf j}) \;
     \bar{n}({\bf P}/3 + {\bf j'}) \;
  \\
  &\null\times
  n({\bf P}/3 - {\bf j}/2 - {\bf k}) \;
  n({\bf P}/3 - {\bf j}/2 + {\bf k}) \;
  \\
  &\null\times\bar{n}({\bf P}/3 - {\bf j'}/2 - {\bf k'}) \;
  \bar{n}({\bf P}/3 - {\bf j'}/2 + {\bf k'}) 
\; ,
  \label{eq:NNN_SO_RE_histo_eq}
  \numberthis
  \end{align*}
  for the RE term. 
	However, now the relevant space is 4-dimensional due to the different momenta moduli which can vary when plotting $I_4$.
	We arbitrarily choose to plot $I_4$ as a function of the two relative momenta moduli $k$, $k'$ for the DD histogram and the two particle Jacobi momenta moduli $k'$, $j'$ for the RE histogram to better illustrate the effect of the regulator. 
	The histograms for the different antisymmetric components of the DD term are plotted in Fig.~\ref{fig:3N_2nd_DD_histos} for the $\NLTN$ and $\LTN$ schemes.
	As can be seen, the distribution is similar to the 2nd order NN histograms (cf.\ Fig.~\ref{fig:NN_SO_histograms_REGS}). 
	The $\NLTN$ integrand is cut off at large $k'$ (squeezed from above) while the $\LTN$ integrand to some extent includes $k’$ above the cutoff $\TNcut$.
	This similarity in structure is expected in that the DD term is an effective 2-body interaction. 
	The key difference between the NN second-order and the DD case is the magnitude of the DD MSNL term compared with the DD MSL term.
	That is, the magnitude of the $\NLTN$ term in the DD case is enhanced compared with the $\LTN$ term.
	 
	Now we examine the residual histograms in Fig.~\ref{fig:3N_2nd_RE_histos}. 
	The $\NLTN$ scheme in Fig.~\ref{fig:3N_2nd_RE_histos}(a) has no $k'$ points above the cutoff and few $j'$ points above $\TNcut$, a difference coming from the factor of $3/4$ in the regulator in \eqref{eq:3N_nonlocal_alt}.
	Due to regulator permutation symmetry, the different $\NLTN$ antisymmetric terms (direct, single-exchange, double-exchange) have equivalent distributions.
	In contrast, the direct $\LTN$ term in (b) shows a distinct enhancement coming from small momentum transfers ${\bf q}_i$, ${\bf q}_j$ in \eqref{eq:3N_reg_potential_local}. 
	Note also that the range of the direct $\LTN$ distribution extends far above the cutoff $\TNcut$. 
	Going to the other $\LTN$ antisymmetric pieces in Fig.~\ref{fig:3N_2nd_RE_histos}(c) and (d), we see increasing suppression. 

	Therefore, we can explain the difference in the ratio $\ratio$ between the two schemes in Fig.~\ref{fig:3N_2nd_ratio}(a).
	Relative to the $\NLTN$ scheme, the suppression of the DD $\LTN$ terms and the enhancement of the direct RE $\LTN$ term doubly act to keep $\ratio$ small for the $\LTN$ scheme.

%%%%%%%%%%%%%%%%%%%%%%%%%%%%%%%%%%%%%%%%%%%%%%%%%%
%
%%%%%%%%%%%%%%%%%%%%%%%%%%%%%%%%%%%%%%%%%%%%%%%%%%

\subsection{Fierz Rearrangements}
\label{sec:fierz}

	\begin{figure*}[tbh]
		\includegraphics[width=0.42\textwidth]{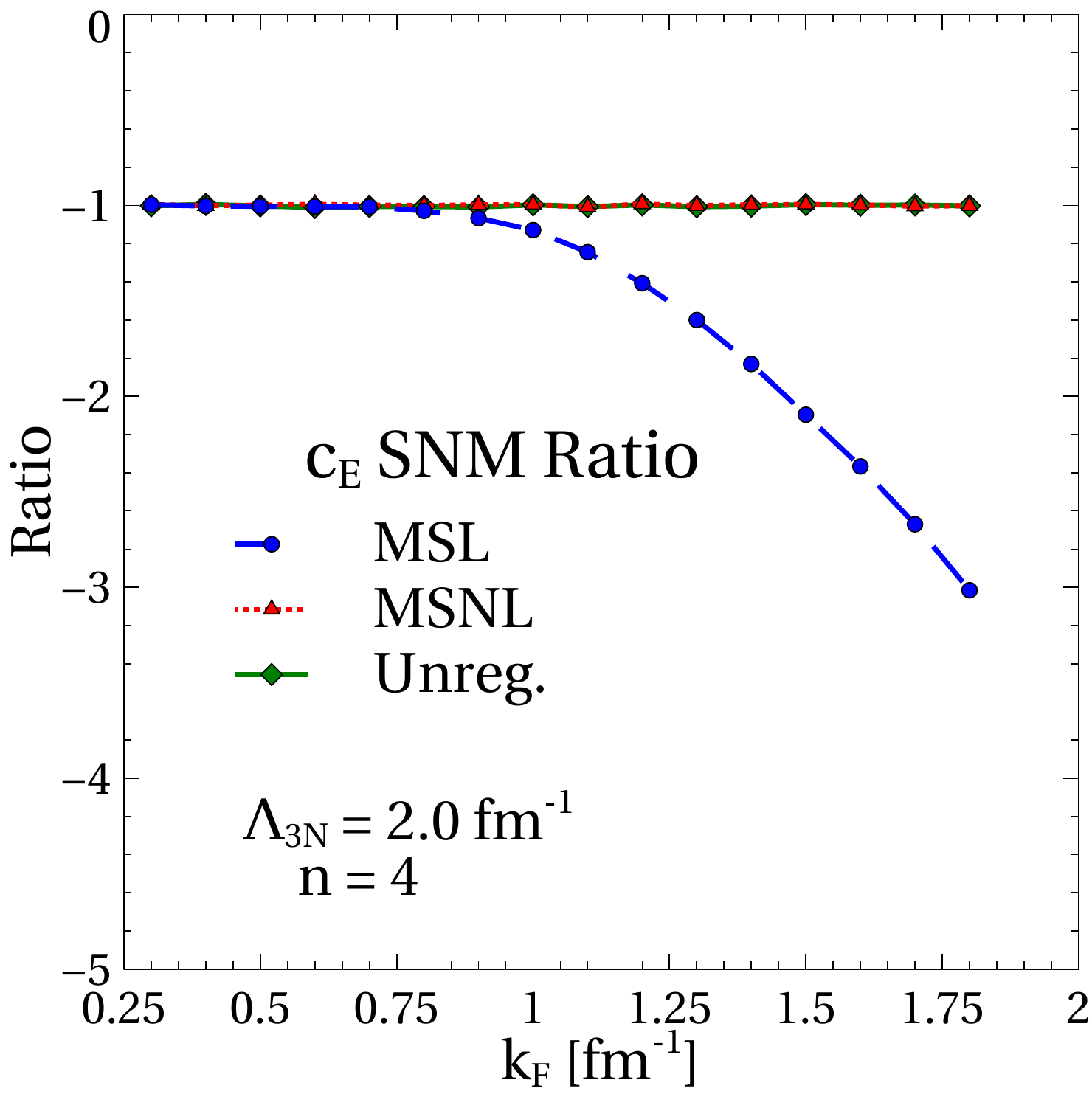}~~%
		\caption{The ratio for the 3N HF energy in \eqref{eq:fierz_HF_ratio} of the $c_E$ contact term in SNM 
		calculated with two different operator structures. The numerator (denominator) was calculated with the operator corresponding to $\beta_1(\beta_3)$ in \eqref{eq:ce_6_choices}.}
		\label{fig:fierz_ce}
	\end{figure*}

When constructing the  pure 3-body contact coming in at \NNLO, 
there are six different possible spin-isospin structures which satisfy 
all the relevant symmetries of the low-energy theory~\cite{Epelbaum:2002vt},
\begin{align*}
	V_{\text{con}}^{\text{3NF}} =& \sum_{i \neq j \neq k}
	\big[ 
	\beta_1 
	+ 
	\beta_2 \spinvec_i \cdot \spinvec_j 
	+ 
	\beta_3 \isovec_i \cdot \isovec_j 
	\\		
	& \null + \beta_4 \left( \spinvec_i 
	\cdot \spinvec_j \right) \left(\isovec_i \cdot \isovec_j \right)
	+ 
	\beta_5 \left( \spinvec_i \cdot \spinvec_j \right) 
	\left(\isovec_j \cdot \isovec_k \right) 
	\\
	& \null + 
	\beta_6 \left( 
	[ \spinvec_i \times \spinvec_j ] \cdot \spinvec_k \right) \; 
	\left( [ \isovec_i \times \isovec_j ] \cdot \isovec_k \right) \big] \; .
	\numberthis
\label{eq:ce_6_choices}
\end{align*}
	Using Fierz rearrangements it can be shown that, up to numerical prefactors, only one of the above operator structures is linearly independent.
	As such, it is only necessary to include one of the six operator structures in \eft~when fitting LECs and doing calculations. 
	The typical choice made in current applications is to use $\isovec_i \cdot \isovec_j$ corresponding to $\beta_3$ in \eqref{eq:ce_6_choices}.
	However, a complication enters when the regulator is no longer symmetric under individual nucleon permutation e.g., the $\LTN$ regulator in \eqref{eq:3N_local} \cite{PhysRevC.93.024305,PhysRevC.85.024003}.
	The Fierz relations establishing equivalence between the different operator structures are spoiled when the antisymmetric pieces of the 3N interaction are regulated differently. 
	This ambiguity of the 3N contact operator for local regulators has recently been explored in Ref.~\cite{PhysRevLett.116.062501}.
	
	This point can be seen in our perturbative approach to the uniform system. 
	In Fig.~\ref{fig:fierz_ce}, we plot a calculation of the ratio of the 3N HF energy for the operator choices corresponding to $\beta_1$ and $\beta_3$ in \eqref{eq:ce_6_choices}:
\beq
E^{\text{HF}}_{1} / E^{\text{HF}}_{\isovec_i \cdot \isovec_j} \; .
\label{eq:fierz_HF_ratio}
\eeq
	Using the $\NLTN$ regulator, or no regulator at all, the ratio of the two different HF energy calculations in \eqref{eq:fierz_HF_ratio} is constant with respect to density.
	This reflects the pure numerical prefactor between the different operators one 
gets upon Fierz rearrangement. 
	However, when using the $\LTN$ regulator, the ratio between the two calculations is now density-dependent. 
	Whether one then needs to keep \emph{all} the operator structures in \eqref{eq:ce_6_choices} when working with regulators that don't respect permutation symmetry is an open question.

%%%%%%%%%%%%%%%%%%%%%%%%%%%%%%%%%%%%%%%%%%%%%%%%%%
%
%%%%%%%%%%%%%%%%%%%%%%%%%%%%%%%%%%%%%%%%%%%%%%%%%%

%\clearpage
%\newpage

\section{Summary and Outlook}
\label{sec:outlook}

Recent progress in nuclear many-body methods has led to increasingly precise ab initio calculations
of observables over a growing range of nuclei.  This in turn has shifted focus to the input
\eft~Hamiltonian in the quest for more accurate calculations and a systematic
understanding of theoretical uncertainties.
A major source of variation among Hamiltonians currently considered 
stems from the regularization scheme chosen, because
	\eft~implemented using Weinberg power counting is not renormalizable order by order.
	As such there remains residual
cutoff dependences in the theory to all orders and regulator artifacts, which are
scheme dependencies that remain after implicit renormalization, are inevitable. 
	
	In this work, we characterized the impact of various NN and 3N regulator choices by analyzing perturbative energy calculations in the uniform system at Hartree-Fock and second-order using the leading NN/3N chiral interactions.
	This allows us to test both long-range and contact potentials,
and both the on-shell and off-shell parts.

	We find significant scale and scheme dependence for perturbative energy calculations at finite density using chiral forces and the scheme choices outlined in Table~\ref{tab:reg_table}. 
	In particular, we have identified characteristic regulator artifacts resulting from the differing regulator functional forms. 
	To uncover the origins of the differing behavior of energy calculations, we adopted an approach based on analyzing the phase space available at each order in MBPT using a Monte Carlo sampling of momenta.
	In all cases, it is this phase space that serves as a guidepost to the effect of different schemes. 
	The momentum histograms in section~\ref{sec:results} are used to show:
	\bi
		\I the extent and shape of the phase space;
		\I the connection between the size of the phase space and the total computed energy;
		\I which parts of the phase space are suppressed by the regulator;
		\I how the regulator cuts off the phase space.
	\ei
	We anticipate that this histogram diagnostic will have wider applications, such as in assessing finite-density power counting or in guiding the implementation of
	long-range chiral forces in nuclear density functionals via the density matrix expansion \cite{Negele:1972zp,
	Negele:1975zz,
	Gebremariam201117,
	Stoitsov:2010ha,
	Bogner:2008kj,
	Bogner:2011kp}.

	Here we summarize some of our observations from Sec.~\ref{sec:results} about scale and scheme dependencies:
\bi	
	 
	\I In special cases where the regulators can be directly related to one another, scheme dependence translates simply to a different effective cutoff. 
	For example, the MSNL \eqref{eq:NN_reg_nonlocal} and MSL \eqref{eq:NN_reg_local} schemes at NN HF can be put into equivalence \eqref{eq:q_k_relation_HF} due to the relation between momentum transfer and relative momentum.
	Likewise, the MSL and CSL contact \eqref{eq:smeared_delta} regulator for $n = 1$ allows $\NNcut$ and $R_0$ to be directly related \eqref{eq:fourier_trans_local} to each other. 
	But in general regulators cannot be put into a direct correspondence.

	\I Coordinate space regulators (usually) lead to oscillatory behavior when Fourier
	 transformed (see Fig.~\ref{fig:CSL_LOCAL_COMPARE}). 
	 In contrast to the smooth cutoff behavior of the momentum space regulators, this manifests as zero points in the interaction phase space, 
	 see CSL/EKM in Fig.~\ref{fig:NN_HF_histograms}(c).
	
	\I Our primary analysis tool are phase space histograms, which are used to understand the effect of the regulator at different orders of MBPT.
	The analytic form of the energy integrand in MBPT is only easily found at NN HF. 
	The expression in \eqref{eq:HF_phasespace} and the plot in Fig.~\ref{fig:NN_HF_phase_space_local} shows this analytic form plotted for the HF exchange term with the $n\rightarrow\infty$ limit for the MSNL and MSL schemes as given in \eqref{eq:sharp_nonlocal} and \eqref{eq:sharp_local}.
	Our histogram approach reproduces this picture in the sharp regulator limit as demonstrated in Fig.~\ref{fig:NN_HF_histograms_N_vary}. 
	Likewise, examining the NN HF energy per particle calculations in Fig.~\ref{fig:NN_HF_energy}, we find an exact matching in the energy hierarchy to the phase space volume for the histograms in Fig.~\ref{fig:NN_HF_histograms}. 
	The same observation can also be made for the MSL and MSNL NN second-order energies (Fig.~\ref{fig:NN_SO_energy}) and histograms (Fig.~\ref{fig:NN_SO_histograms_REGS}) along with the 3N HF energy (Fig.~\ref{fig:HF_3N_ce_energies}) and histograms (Fig.~\ref{fig:NNN_HF_histogram_ce}).

	\I The qualitative scale dependence of all the regulators is similar, with softer cutoffs (i.e., those with smaller $\NNcut$, $\TNcut$, and larger $R_0$) generating larger energy differences at a fixed density. 
	In regions where the Fermi momentum $\kf$ is small compared to the cutoff, scheme artifacts are generally small. 
	However, finite range coordinate space regulators (CSL/EKM) have modifications that persist even at small $q$ (Fig.~\ref{fig:CSL_LOCAL_COMPARE}) leading to differences at small $\kf$ (Fig.~\ref{fig:NN_HF_histograms}(c) and Fig.~\ref{fig:COOR_DISTORTION}). 	
	Note that to highlight scheme effects, in this paper we worked at soft cutoffs of $\NNcut = 2.0~\fmi$, $R_0= 1.2~\fm$, and $\TNcut = 2.0~\fmi$. 
	Calculations with harder cutoffs present quantitative smaller artifacts but are qualitatively similar (see supplemental material). 
	
	\I At higher densities, regulators cut into the hole phase space  and at second-order (and beyond) the regulators squeeze the particle phase space, making artifacts more apparent.
	This can be seen in the NN HF histograms of Fig.~\ref{fig:NN_HF_histograms} where scheme differences become larger as $\kf$ increases. 
	Likewise in the NN second-order histograms, large differences exist at large $\kf$ between different schemes (Fig.~\ref{fig:NN_SO_histograms_REGS} and Figs.~\ref{fig:NN_SO_histograms_CS},~\ref{fig:NN_SO_histograms_OPE}). 
	The corresponding effects are seen for 3N HF (Fig.~\ref{fig:NNN_HF_histogram_ce}) and at 3N second-order (Figs.~\ref{fig:3N_2nd_DD_histos},~\ref{fig:3N_2nd_RE_histos}).
	 
	\I The behavior of the regulator under permutation symmetry, the interchanging of nucleon labels $i,j$ due to the exchange operator $P_{ij}$, affects how the different parts of the potential are affected i.e., direct vs.\ exchange. 
	Stark differences in behavior can occur when the regulator does not respect permutation symmetry.
	Certain regulator schemes respect (MSNL) or do not respect (MSL) permutation symmetry.
	At the NN HF level, the phase space histograms can clearly demonstrate this fact (cf.\ Fig.~\ref{fig:NN_HF_histograms}(a) and (b) at $\kf = 1.8~\fmi$ for MSL and MSNL). 
	The direct/exchange components in the MSL scheme are very different but they are identical in the MSNL scheme.
	This manifests at second-order as well, as can be seen in comparing Fig.~\ref{fig:NN_SO_histograms_REGS}(a) and (b) for the MSL scheme along with (c) and (d) for the MSNL scheme. 
	3-body contributions at HF and second-order also display these differences between antisymmetric components in different schemes in Figs.~\ref{fig:NNN_HF_histogram_ce}, \ref{fig:3N_2nd_DD_histos}, and \ref{fig:3N_2nd_RE_histos}.
	
	\I Additionally, how regulators behave under permutation symmetry can affect Fierz rearrangements between operator structures. 
	In particular, when constructing the \NNLO~3N contact term, six different possible operator structures exist which respect the relevant symmetries of \eft, see \eqref{eq:ce_6_choices}. 
	However, upon Fierz rearrangement, only one operator is shown to be linearly independent. 
	As a result, different operator structures can be related to one another and differ only by a pure numerical prefactor. 
	However, these rearrangements depend on relations between different antisymmetric components of the operators. 
	For regulators which do not respect permutation symmetry (e.g., the $\LTN$ scheme), these Fierz rearrangements are no longer automatic. 
	In Sec.~\ref{sec:fierz}, we show that the Fierz relation is spoiled for two operator choices in a 3N HF energy calculation using the $\LTN$ scheme (see Fig.~\ref{fig:fierz_ce}). 
							
	\I Approximations of many-body perturbation theory (MBPT) also exhibit scheme and scale dependence. 
	For 3N forces, a common technique is to normal-order the free-space second-quantized operators with respect to a finite density ground state. 
	At second-order in MBPT, this results in an effective 2-body term (called the normal-ordered term) and a remaining 3-body piece (called the residual term). 
	The residual term is a true 3-body term and is computationally expensive to calculate. In the NO2B approximation, the residual term is discarded and the computationally simple normal-ordered term is retained.
	Such an approximation is only valid if the contribution of the normal-ordered term to the energy is greater than the residual i.e., if the ratio of the former to latter is greater than one.

	In Sec.~\ref{sec:3N_SO}, we demonstrated that the ratio of the normal-ordered term to the residual term has a distinct scheme and scale dependence. 
	The scale dependence comes from changing the extent and importance of the hole/particle phase space as the cutoff is changed. 
	As the cutoff is raised, the particle phase space increases and the residual term dominates.
	We use our momentum histograms to understand the scheme dependence for the $\NLTN$ and $\LTN$ schemes.
	In the $\LTN$ scheme, the direct residual term is enhanced due to small momentum transfers $\qvec_{i,j}$ (Fig.~\ref{fig:3N_2nd_RE_histos}(b)) while the normal-ordered terms are suppressed compared to the $\NLTN$ scheme (Fig.~\ref{fig:3N_2nd_DD_histos}). 
	This residual term enhancement and normal-ordered term suppression in the $\LTN$ scheme results in very different ratios for local and nonlocal schemes, as seen in Fig.~\ref{fig:3N_2nd_ratio}.
	
	\I While we have emphasized the dominant role of the phase space, there are also quantitative differences in calculations due to the role of the interaction and how it interplays with the chosen scheme.  
		For example, the 3N $c_1$ term weights states lower in the Fermi sea (see supplemental material and Fig.~\ref{fig:ci_COMPARE}).
		Consequently, $c_1$ scheme dependence and regulator artifacts are less pronounced than compared with $c_3$, $c_D$, and $c_E$. 
	
\ei
	A critical but open question is the ultimate impact of the regulator artifacts.
		For example, it has been seen that two-pion exchange regulator artifacts can affect the chiral power counting  in uniform matter \cite{PhysRevC.91.044001}.
	However, recent research has indicated that these artifacts are better controlled using certain position space local regulators \cite{Epelbaum:2014efa}. 
	Whether local regulators are the only way to control finite range artifacts, and avoid distorting analytic structures, remains an open question.
	
	If \eft\ is to be model independent and follow the chiral power counting, regulator artifacts at one order must be absorbed at higher order consistent with the power counting.
	But how the regulator dependence
is absorbed (if it is) by implicit renormalization is not manifest. 		
		Furthermore, a systematic comparison of uncertainties due to truncation of the chiral expansion and truncation in MBPT still needs to be explored.
	
	The significant regulator artifacts observed here and for two-pion exchange motivate exploration of a wider range of functional forms for regulators, such as those commonly used for the functional renormalization group (RG)~\cite{Pawlowski:2015mlf} and nuclear low-momentum RG evolution (e.g., see Ref.~\cite{Bogner:2006vp}). 
	For example, there are regulators with an independent dimensional scale parameter to set the smoothness of the cutoff, instead of relying on a super-Gaussian suppression.
	This may provide greater control over artifacts.
	The analysis tools introduced here are being applied to these alternatives in an ongoing investigation.

\section*{Acknowledgements}

	We would like to thank Christian Drischler for numerical comparisons and Achim Schwenk for useful discussions.
	We would also like to thank Joel Lynn, Stefano Gandolfi, Alessandro Lovato, and other colleagues in the NUCLEI collaboration.
	This work was supported in part by the National Science Foundation under Grant No. PHY{1306250 and Grant No. PHY-1430152 (JINA Center for the Evolution of the Elements), the NUCLEI SciDAC Collaboration under DOE Grant de-sc0008533 and DOE Grant No. DE-FG02-00ER41132,  and by the ERC Grant No. 307986 STRONGINT.

\bibliography{srg_refs}

\clearpage
\newpage

\appendix

\section{NN Energy Values at Harder Cutoffs}

	In this appendix we show plots for the energy per particle at NN HF using more common cutoffs of $\NNcut = 2.5~\fmi$ and $R_0 = 0.9~\fm$ for position space regulators.
	The antisymmetric terms for the energy per particle for NN HF are given in Fig.~\ref{fig:NN_HF_energy_ALT}.
	These can be compared to the energy calculations at the softer cutoffs $\NNcut = 2.0~\fmi$, $R_0 = 1.2\fm$ in Fig.~\ref{fig:NN_HF_energy}.
	Note that energy values between the different schemes are more similar here compared with the $\NNcut = 2.0~\fmi$ case; i.e., regulator artifacts for $\NNcut = 2.5~\fmi$ are less pronounced at a given $\kf$.

\begin{figure*}[t]
	\includegraphics[width=0.33\textwidth]{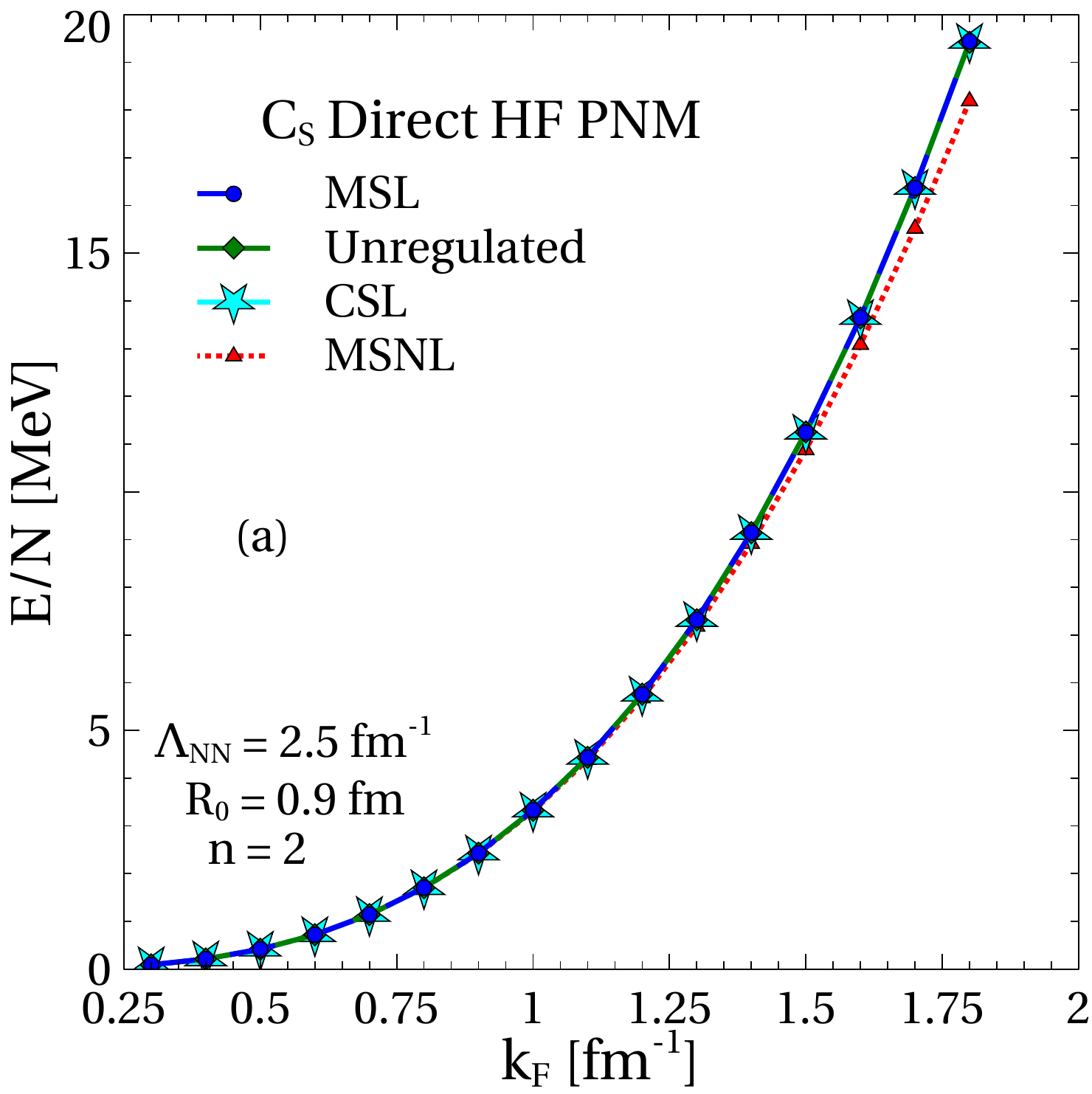}~%
	\includegraphics[width=0.33\textwidth]{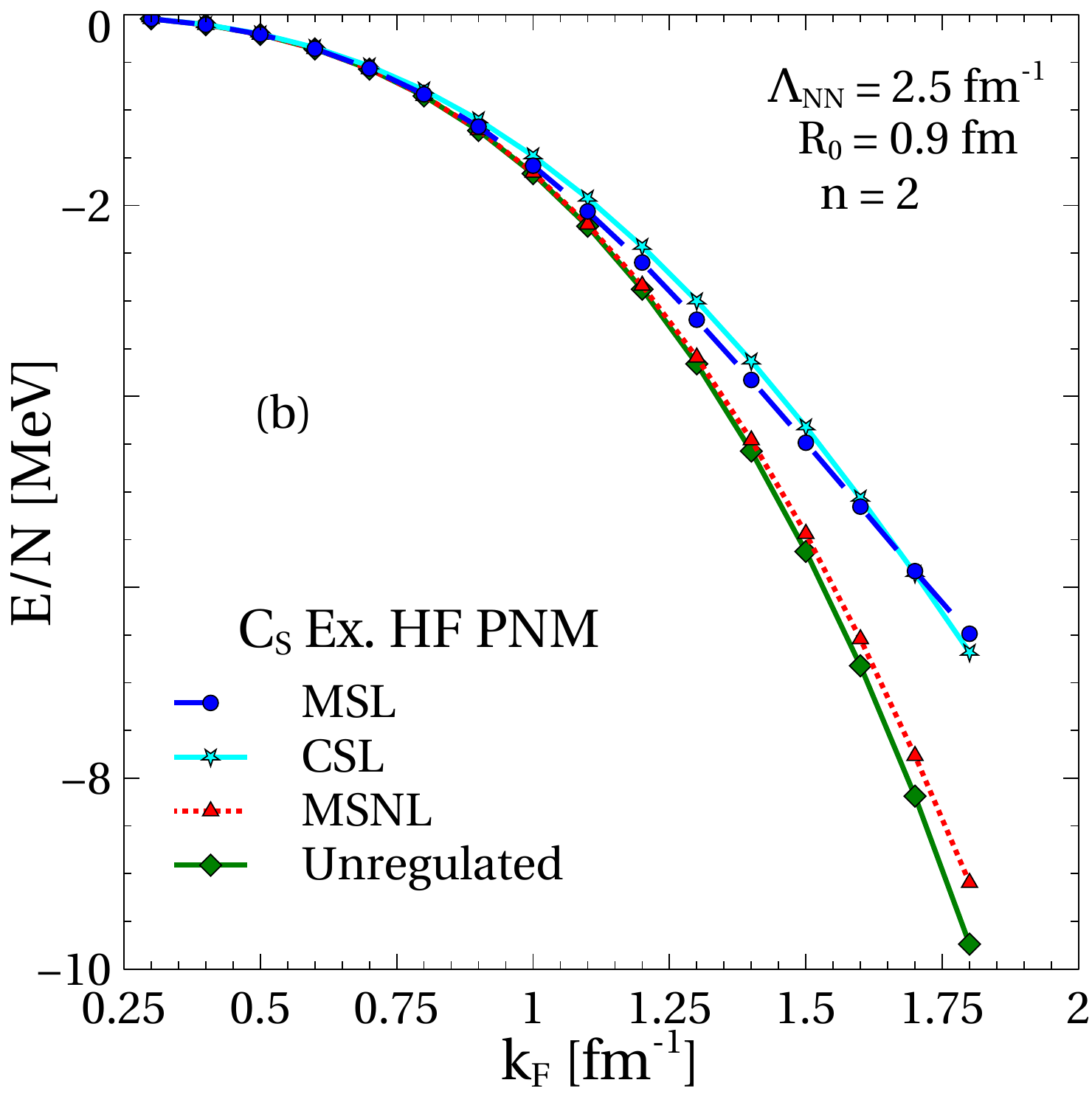}~%
	\includegraphics[width=0.33\textwidth]{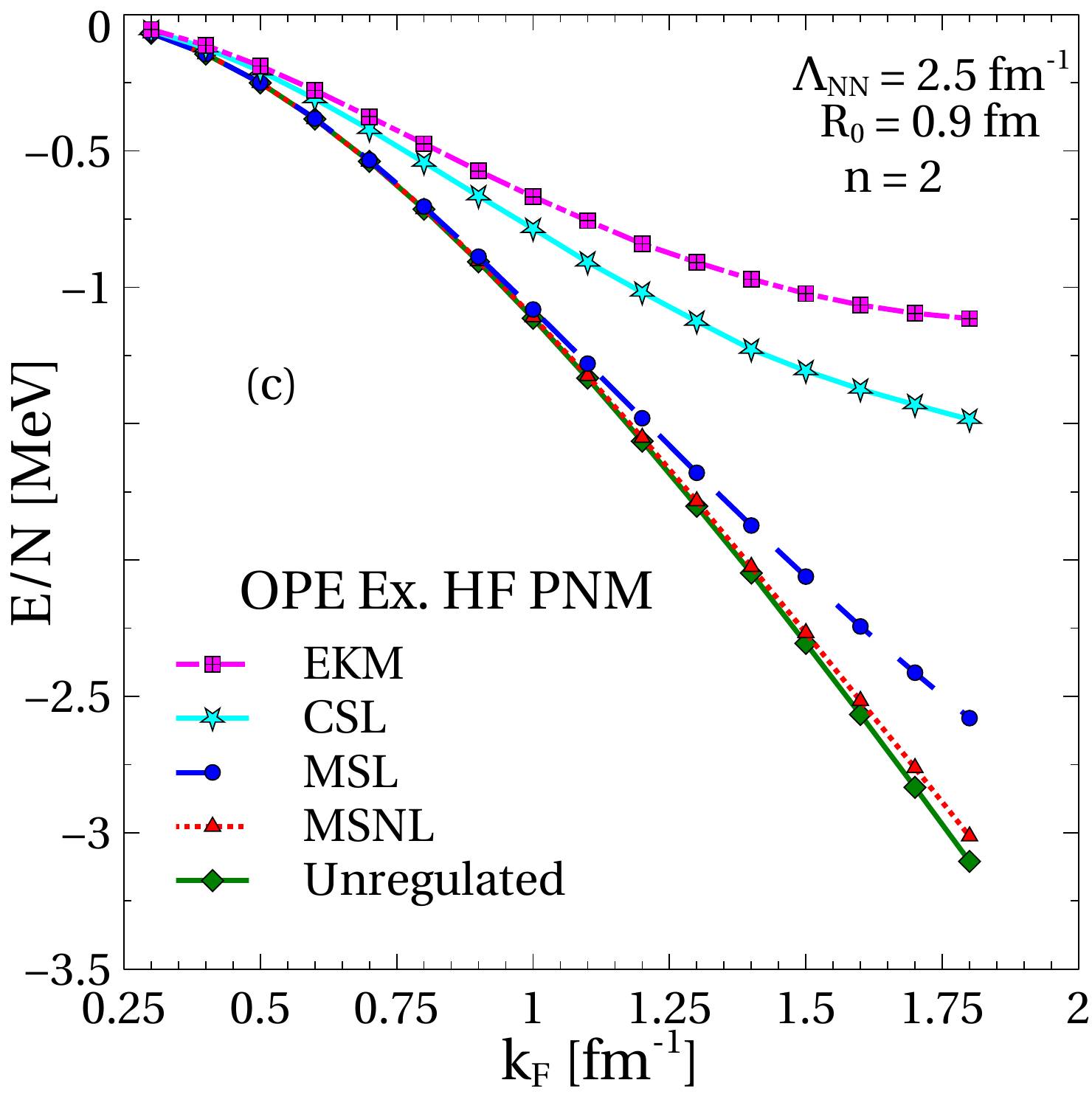}
	\caption{Neutron matter calculations of the HF energy per particle for the direct (a)
	and exchange (b) terms for
	$C_S$ and the OPE exchange term (c) using the regularization schemes in Table~\ref{tab:reg_table}.
	 The $C_T$ calculation has similar behavior to the $C_S$ exchange term. 
		The calculations use $C_S = 1.0~\csunits$, $n = 2$, $\NNcut = 2.5~\fmi$, and $R_{0} = 0.9~\fm$.}
	\label{fig:NN_HF_energy_ALT}
\end{figure*}

\begin{figure*}[tbh]
	\includegraphics[width=0.35\textwidth]{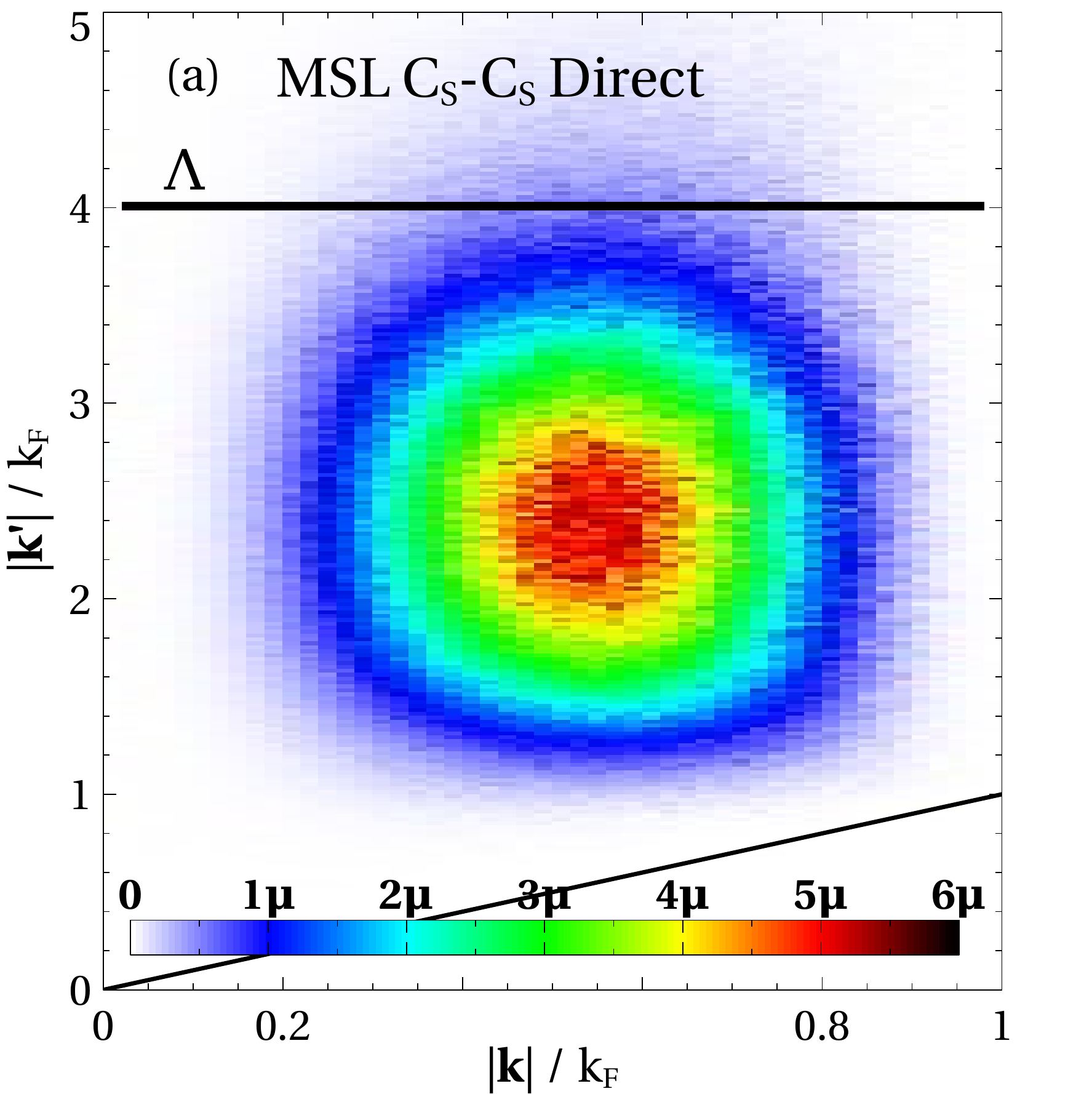}~~%
	\includegraphics[width=0.35\textwidth]{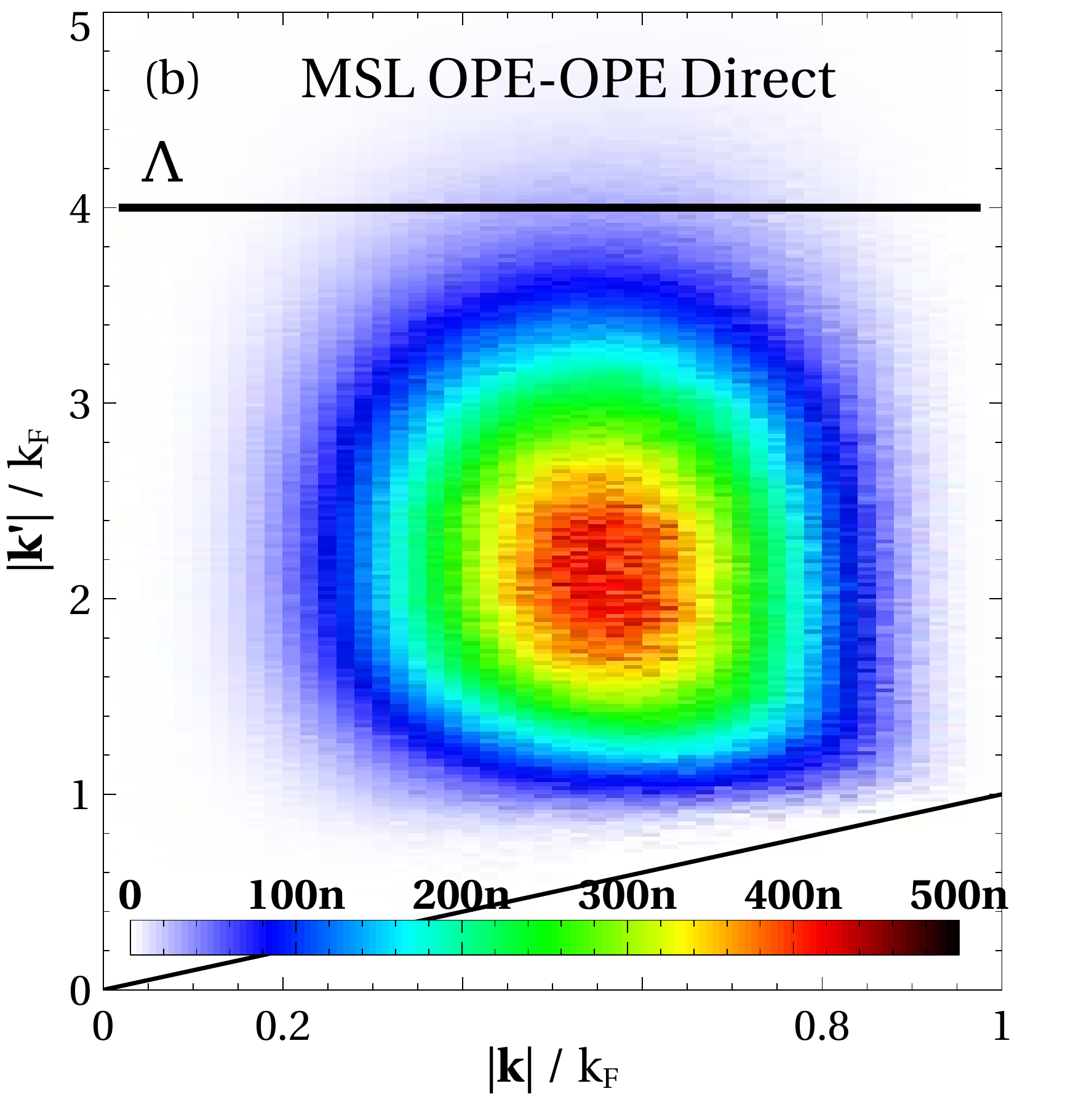}~~%
	
	\caption{(color online) Momentum histograms representing the second-order NN phase space for the  MSL direct $C_S$--$C_S$ term (a) and OPE--OPE term (b) at $\kf =%
	0.5~\fmi$, $n = 2$, and $\NNcut = 2.0~\fmi$.
	The exchange and MSNL plots have a similar distribution.
	The y-axis gives
	the particle relative momentum while the 
	x-axis gives the hole relative momentum both scaled by $\kf$.
	Colors indicate
	the $I_{2,\text{full}}$ magnitude for a particular
	$k,k'$ pair. The horizontal black line indicates the cutoff $\NNcut$ and the sloping black line separates out the inaccessible region due to Pauli blocking. ${\mu}$ and $n$ in the color bar stand for $10^{-6}$ and $10^{-9}$ respectively.}
	\label{fig:NN_SO_histograms_example}
\end{figure*}

\section{NN Second-Order $I_2$ Plots}

	In Sec.~\ref{sec:NN_SO}, the histogram plots are weighted by the phase space $I_{2}$ in \eqref{eq:NN_2nd_order_regs_only} rather than the full integrand magnitude.
The full second-order energy integrand $I_{2,\text{full}}$ is given by,
\begin{align*}
&I_{2,\text{full}} = | f_{\text{reg}} | \;
\frac{k^2 \;
k'^2 \;
P^2}
{\left(k'^2/m - k^2/m \right)}
\\
&\null \times
n({\bf P}/2 + {\bf k}) \; 
n({\bf P}/2 - {\bf k}) \;
\bar{n}({\bf P}/2 + {\bf k'}) \; 
\bar{n}({\bf P}/2 - {\bf k'})
	\\
	&\null \times
	\begin{dcases}
	1, &\text{Contact} \; ,
	\\
	\left| \frac{q^2 \; S_{12}(\qhat) - 
	\mpi^2 \spinone \cdot \spintwo}{q^2 + \mpi^2} \right| , &\text{OPE} \; ,
	\end{dcases}
	\numberthis
\label{eq:NN_SO_histogram_equation}
\end{align*}
where the first (second) term in brackets corresponds to weighting by the contact (OPE) interaction and all spin terms are summed over.
In Fig.~\ref{fig:NN_SO_histograms_example} we plot two examples 
of this full second-order phase space at a low density $\kf = 0.5~\fmi$ with a cutoff 
$\NNcut = 2.0~\fmi$.
	In Fig.~\ref{fig:NN_SO_histograms_example}(a) and (b), we show the MSL direct $C_S$--$C_S$ term and the MSL direct OPE--OPE term respectively. Here we have only plotted the MSL direct terms because the equivalent MSNL plots are nearly identical (scheme artifacts are small). 
	Likewise, the exchange plots have an equivalent distribution, but with smaller magnitudes.

	The circular shapes in the color in Fig.~\ref{fig:NN_SO_histograms_example} are interpreted as no correlation in the selection of $k$ and $k’$ at lower densities.
	This can be manifested by rewriting \eqref{eq:NN_SO_energy} using relative and center-of-mass coordinates.

	Looking at Figs.~\ref{fig:NN_SO_histograms_CS} and \ref{fig:NN_SO_histograms_OPE}, we see the histogram plots for the $C_S$--$C_S$ and OPE--OPE terms respectively. 
	The integrand magnitude $I_{2,\text{full}}$ of the two terms in \eqref{eq:NN_SO_histogram_equation} are given by the color intensity for a given $k'$,$k$ pair. 
		Comparing with Fig.~\ref{fig:NN_SO_histograms_REGS}, it is seen that there is little qualitative difference between plotting just the phase space ($I_2$) or the full energy integrand ($I_{2,\text{full}}$) absent magnitude rescaling. 
	Thus, it is the regularization scheme which primarily drives the distribution of points in the histograms.

\begin{figure*}[tbh]
	\includegraphics[width=0.42\textwidth]{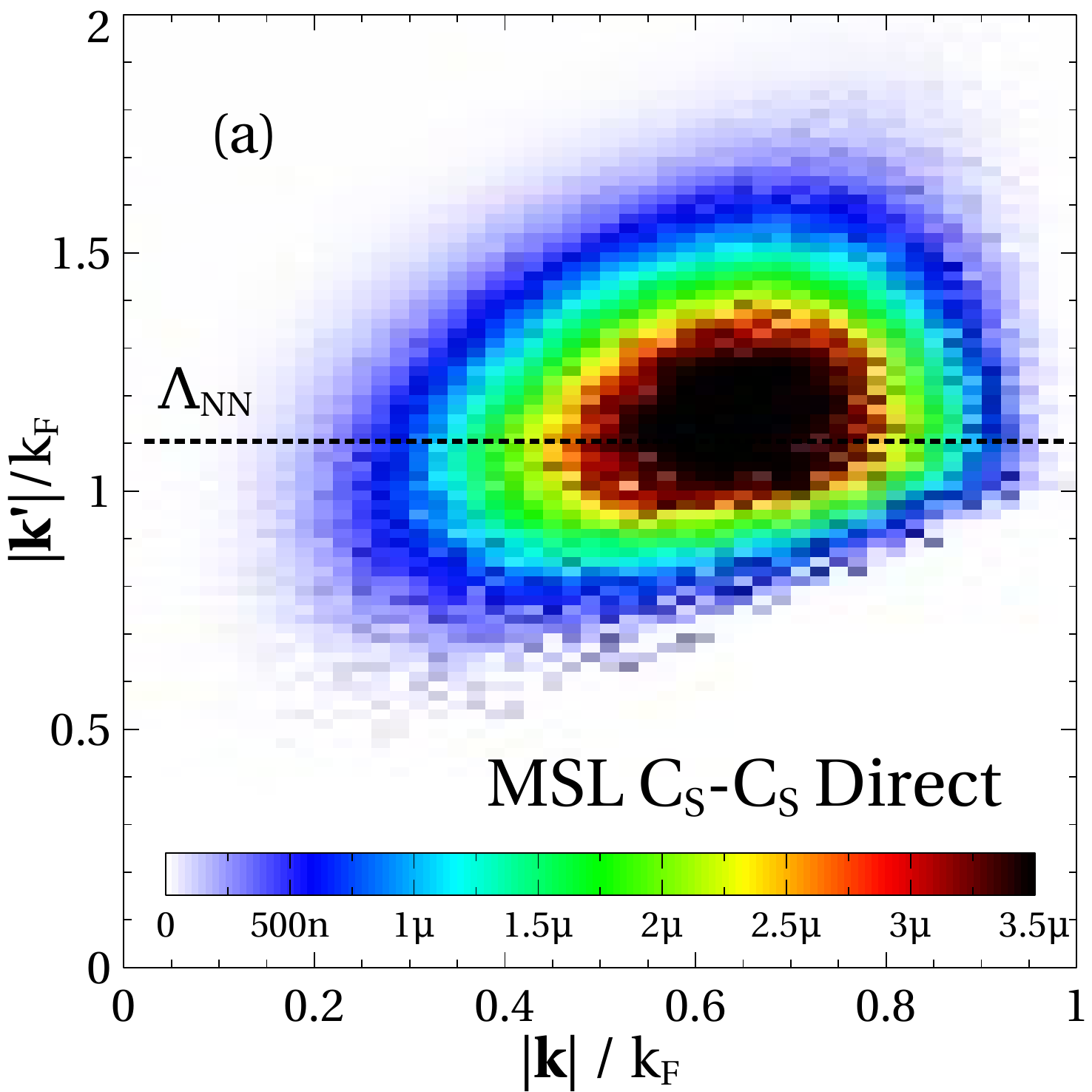}
	\includegraphics[width=0.42\textwidth]{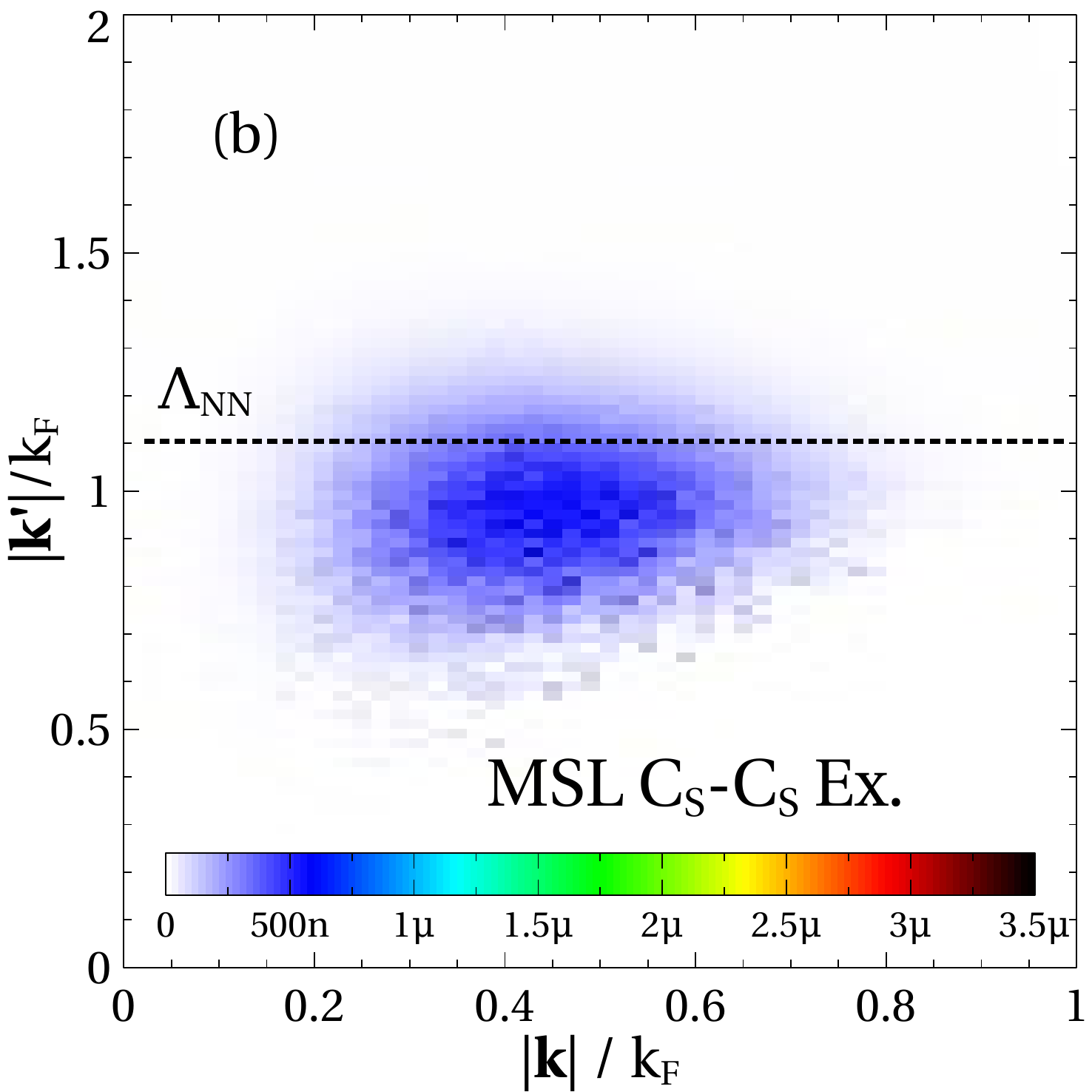}

	\includegraphics[width=0.42\textwidth]{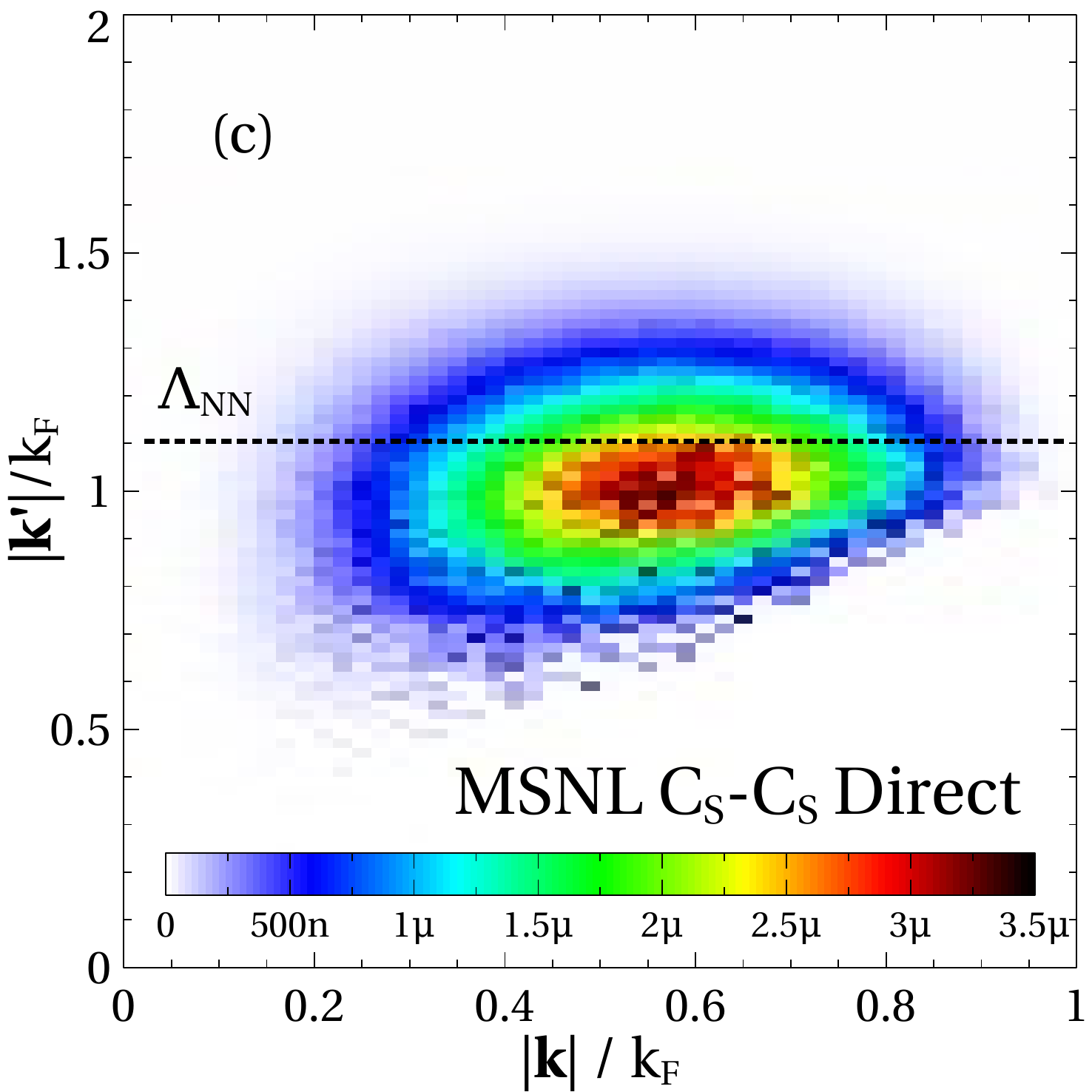}
	\includegraphics[width=0.42\textwidth]{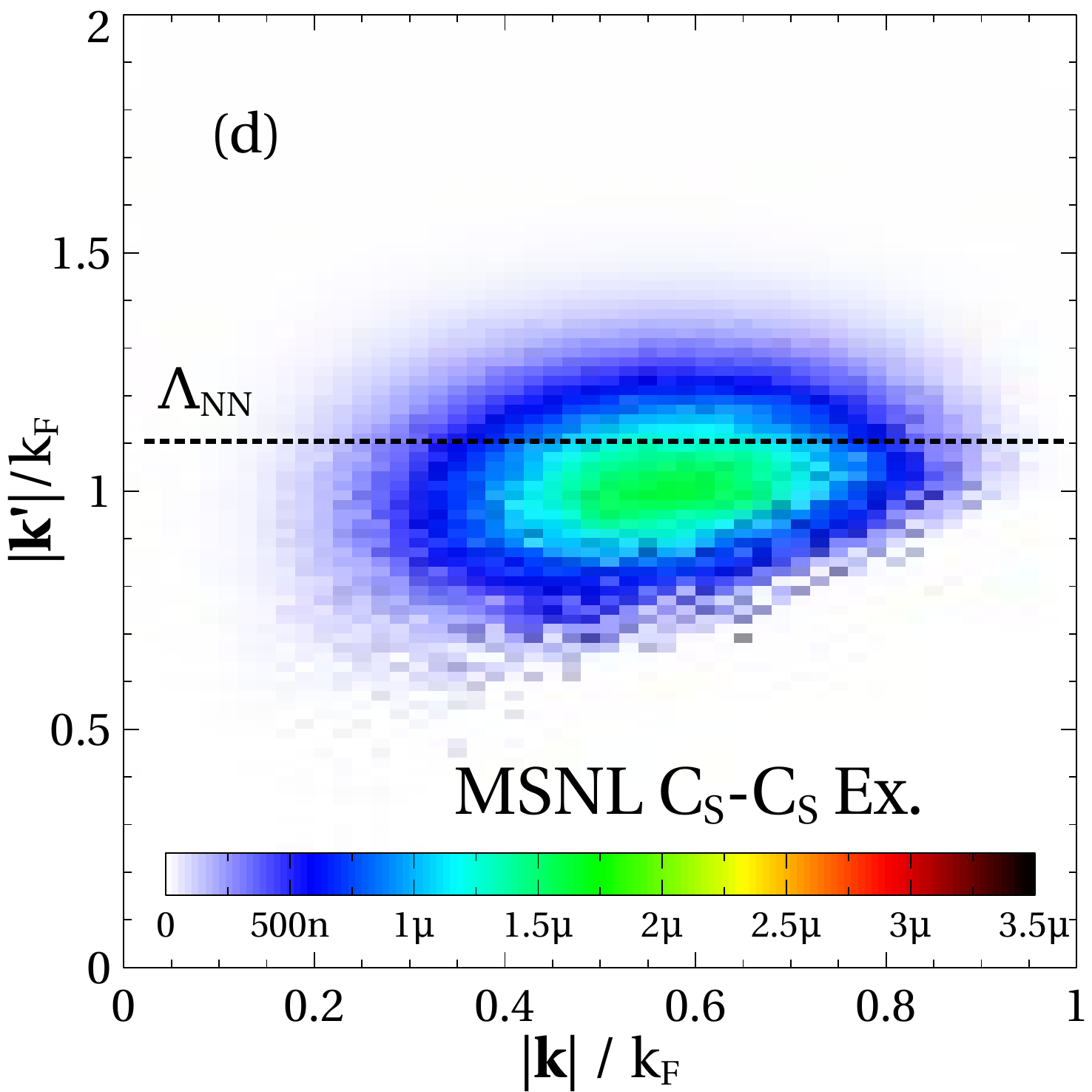}

	\caption{(color online) Momentum histograms representing the second-order NN phase space for the $C_S$--$C_S$ term. 
	The MSL direct (a) and exchange (b) terms are shown  along with the MSNL direct (c) and exchange (d) terms. 
	Plots done at $\kf = 1.8~\fmi$, $n = 2$, $\NNcut = 2.0~\fmi$.
	The y-axis gives the particle relative momentum $k'$ \eqref{eq:NN_particle_relative_mom} while the x-axis gives the hole relative momentum $k$ \eqref{eq:rel/tran_mom} both scaled by $\kf$.
	Colors indicate the $I_{2,\text{full}}$ magnitude in \eqref{eq:NN_SO_histogram_equation} for a particular $k,k'$ pair.}
	\label{fig:NN_SO_histograms_CS}
\end{figure*}	

\begin{figure*}[tbh]
	\includegraphics[width=0.42\textwidth]{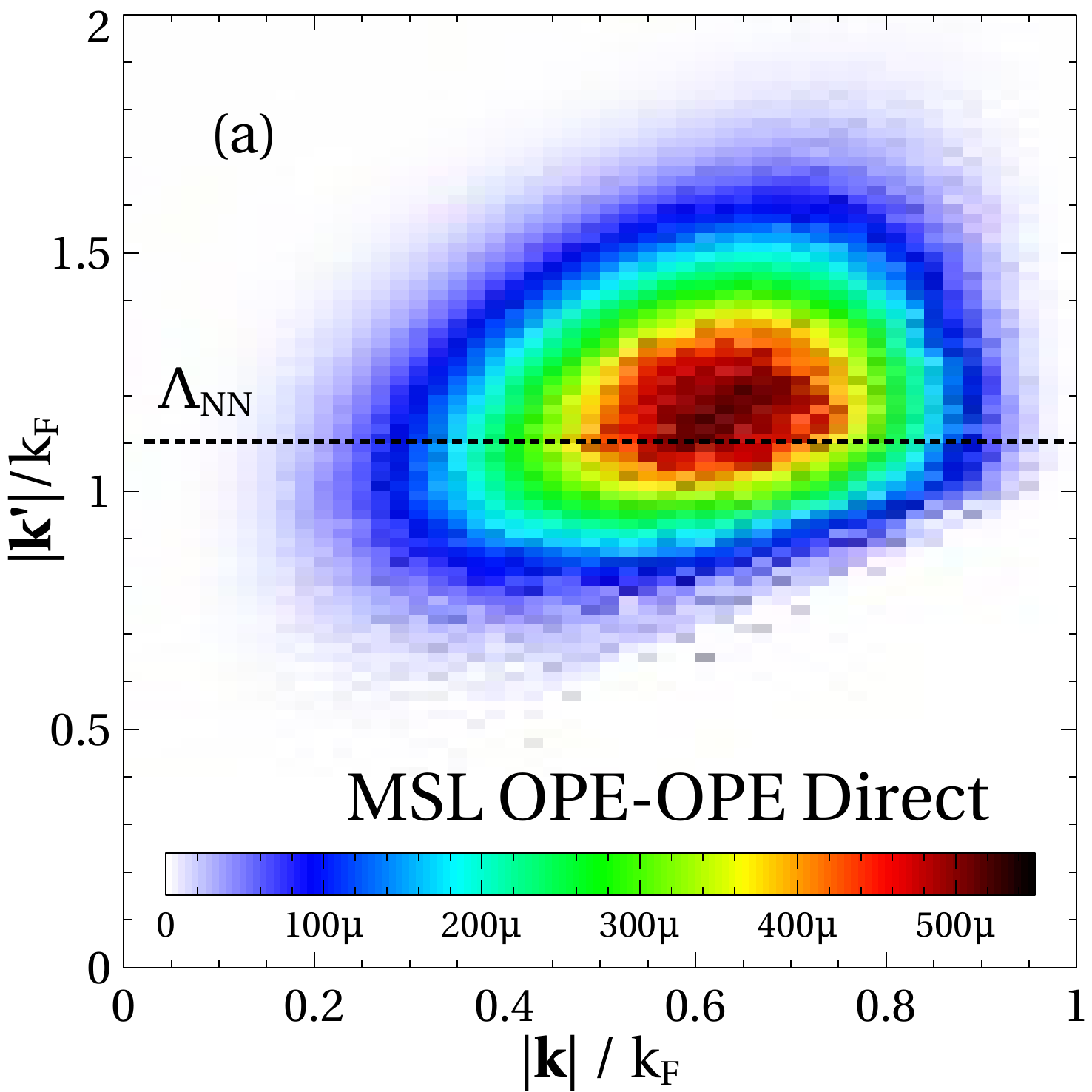}
	\includegraphics[width=0.42\textwidth]{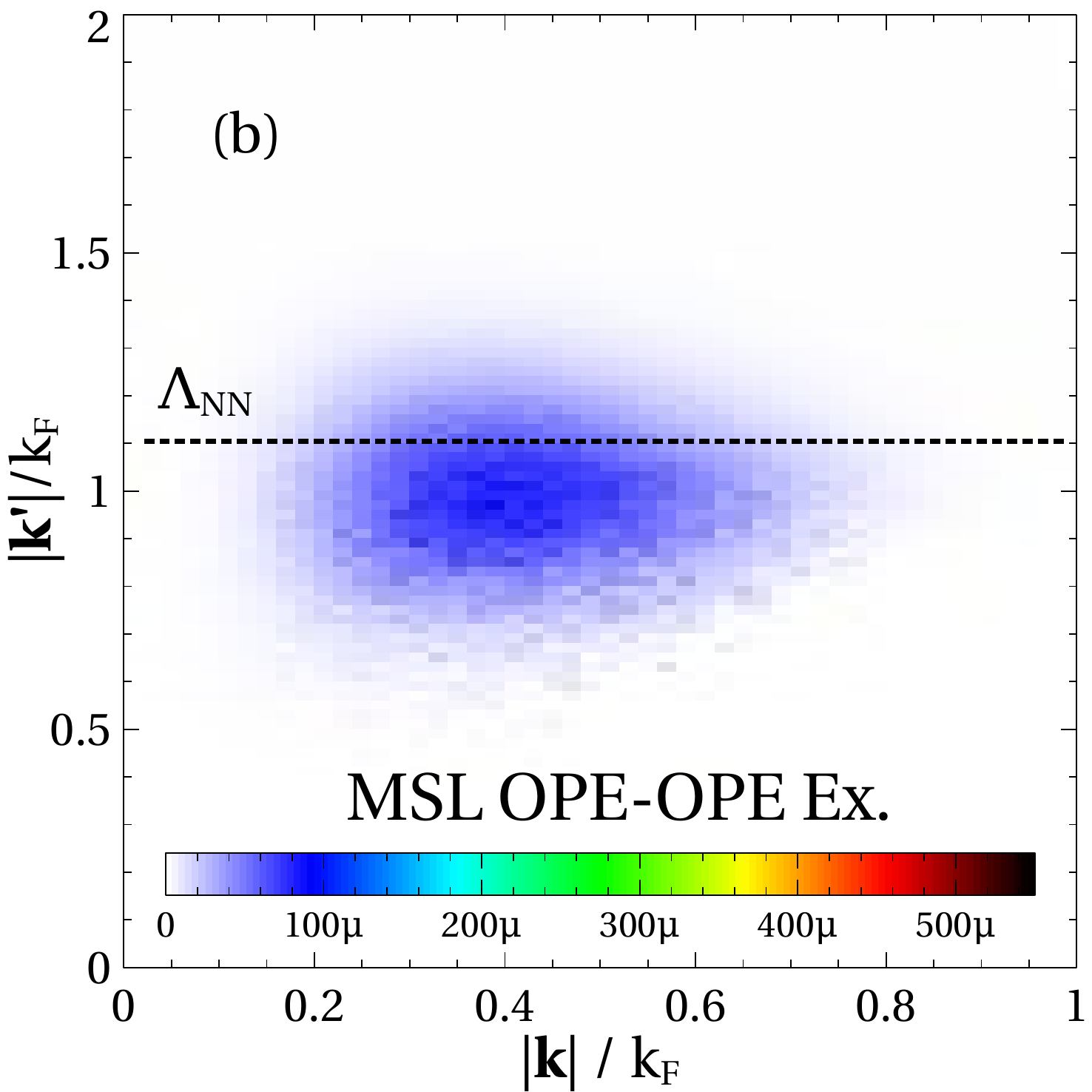}

	\includegraphics[width=0.42\textwidth]{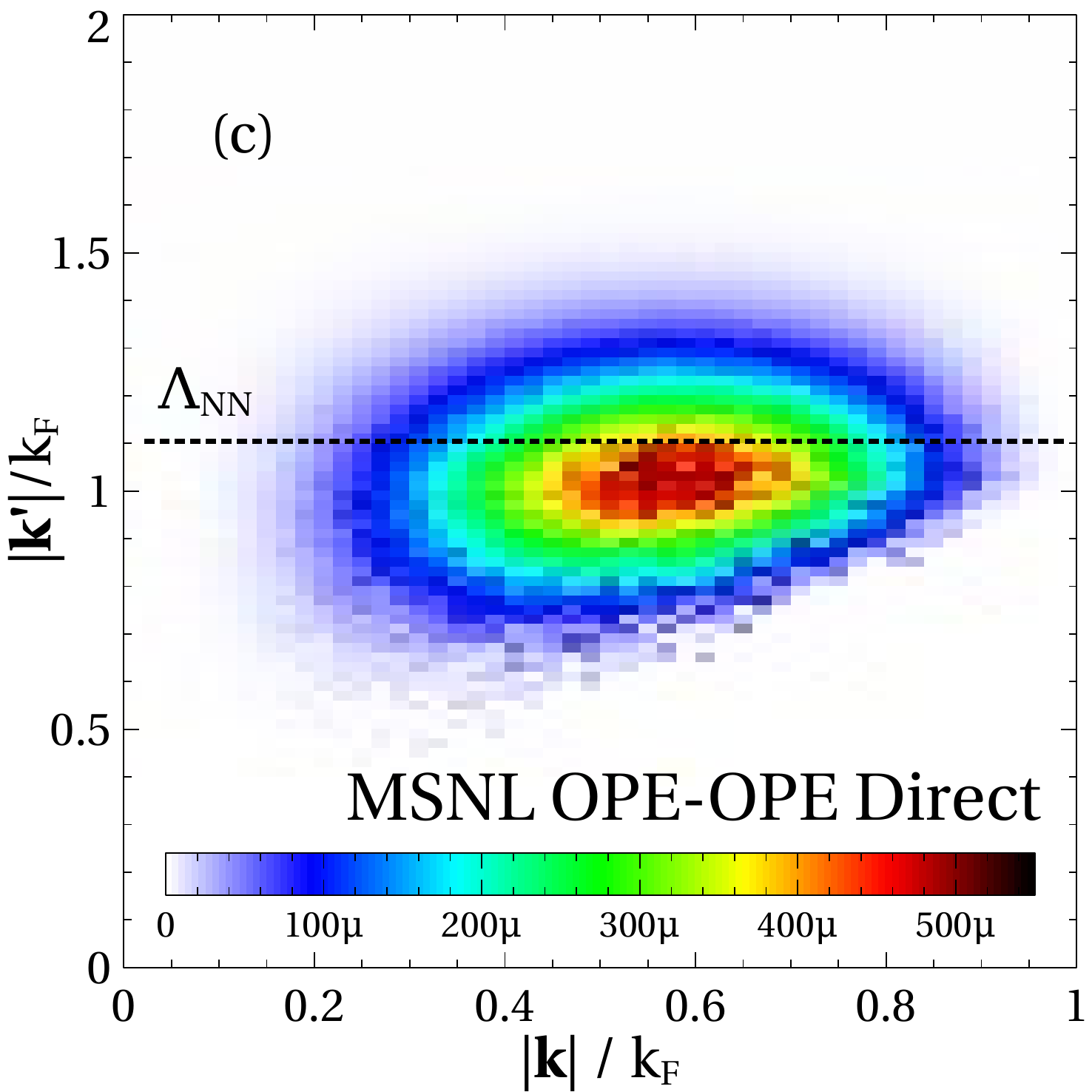}
	\includegraphics[width=0.42\textwidth]{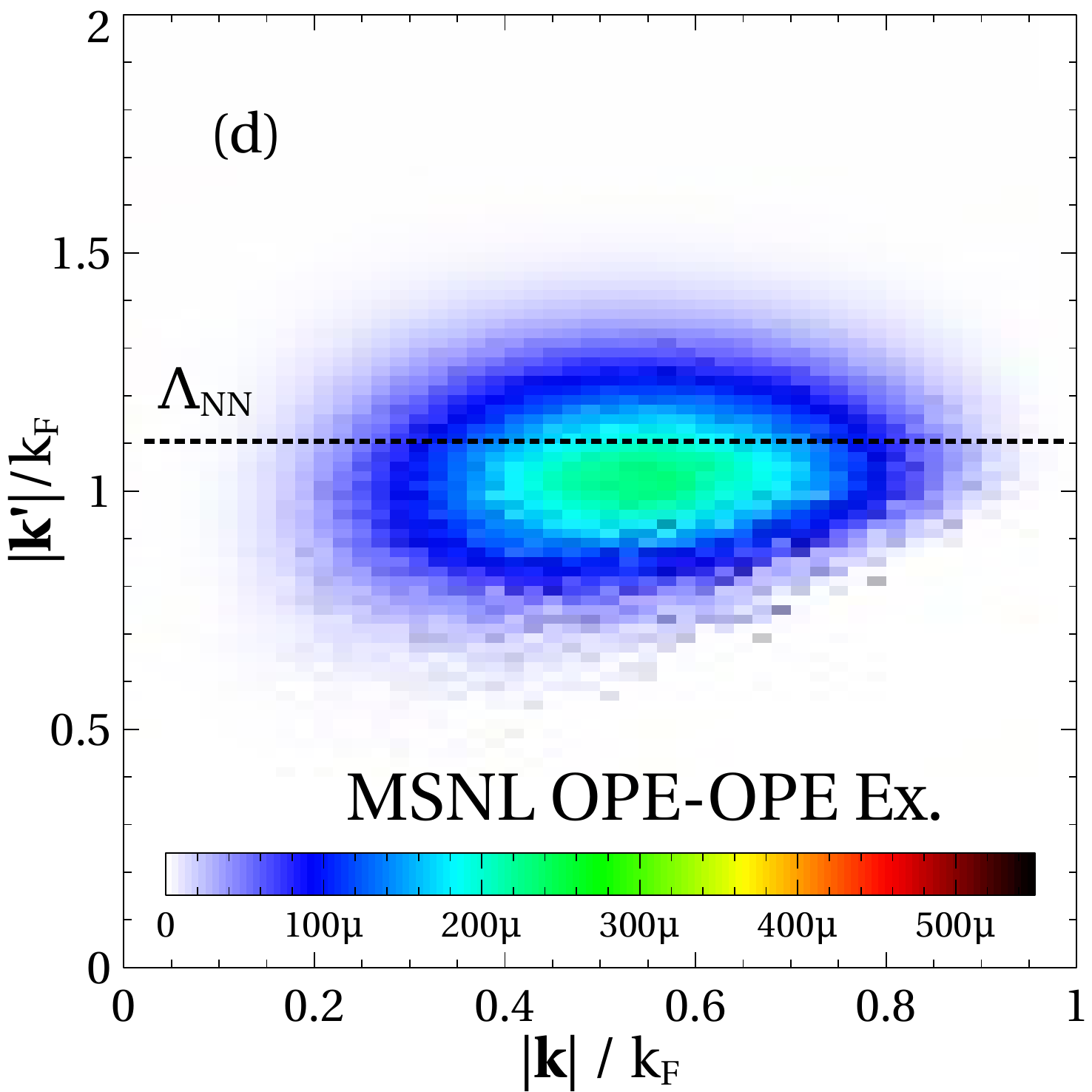}

	\caption{(color online) Momentum histograms representing the second-order NN phase space for the OPE--OPE term. 
	The MSL direct (a) and exchange (b) terms are shown  along with the MSNL direct (c) and exchange (d) terms. 
	Plots done at $\kf = 1.8~\fmi$, $n = 2$, $\NNcut = 2.0~\fmi$.
	The y-axis gives the particle relative momentum $k'$ \eqref{eq:NN_particle_relative_mom} while the x-axis gives the hole relative momentum $k$ \eqref{eq:rel/tran_mom} both scaled by $\kf$.
	Colors indicate the $I_{2,\text{full}}$ magnitude in \eqref{eq:NN_SO_histogram_equation} for a particular $k,k'$ pair.}
	\label{fig:NN_SO_histograms_OPE}
\end{figure*}

\section{Finite Range Interactions at 3N HF}

	In Sec.~\ref{sec:3N_HF}, only the HF energy per particle for the $c_E$ term was given. 
	Here, we show plots for the energy per particle for the finite range pieces as well. 
	Fig.~\ref{fig:HF_3N_s_ex_energies} shows the $c_1$, $c_3$, and $c_D$ single-exchange contributions while Fig.~\ref{fig:HF_3N_d_ex_energies} shows the double-exchange contributions to the energy per particle. 
	The direct terms for the finite range interactions vanish at HF from tracing over spin-isospin. 
	Comparing with Fig.~\ref{fig:HF_3N_ce_energies}, there is little qualitative difference in the scheme hierarchy for the different interactions (but see App.~\ref{sec:c1_term}).
	
\begin{figure*}
\includegraphics[width=0.35\textwidth]{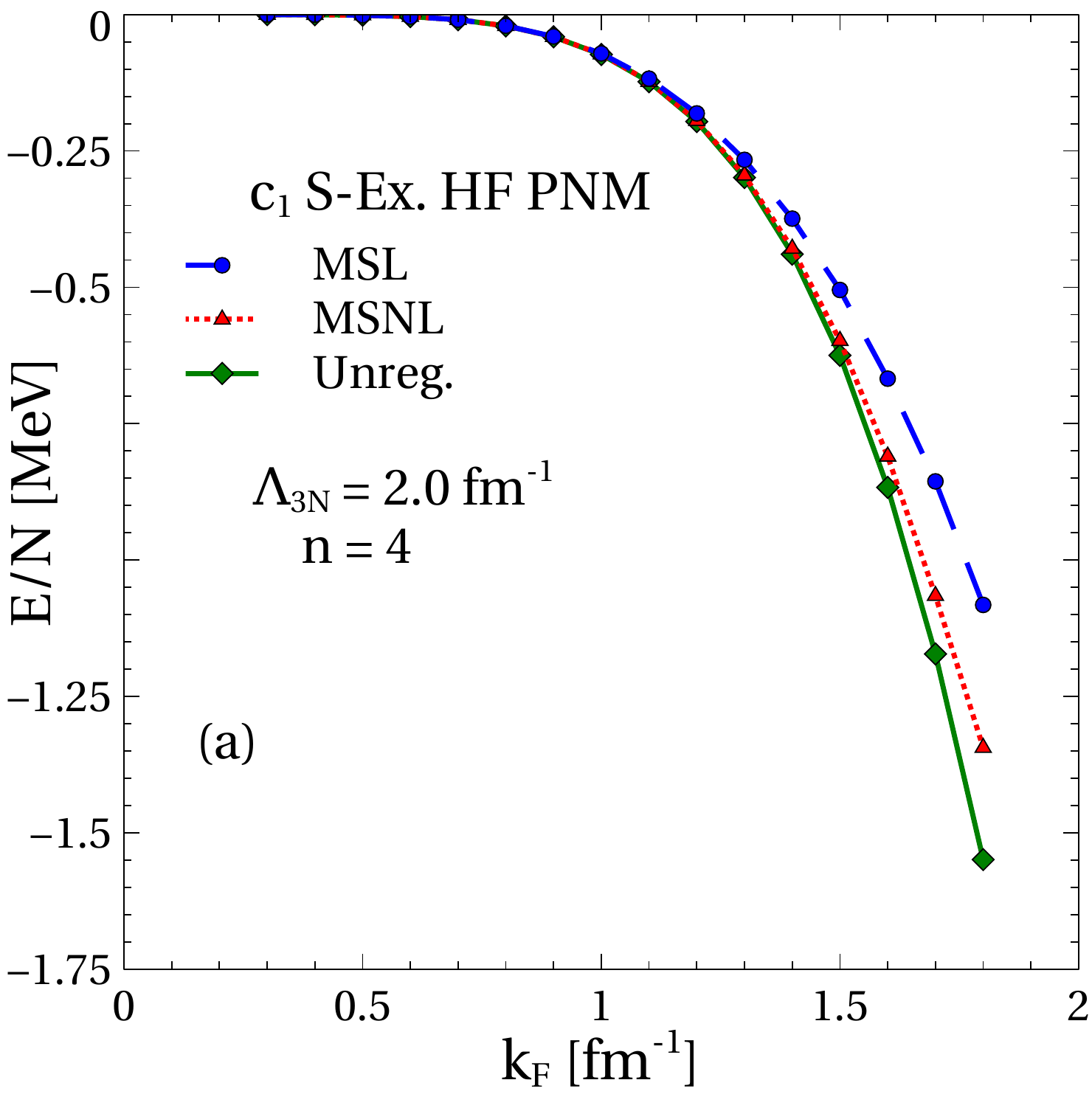}~
\includegraphics[width=0.35\textwidth]{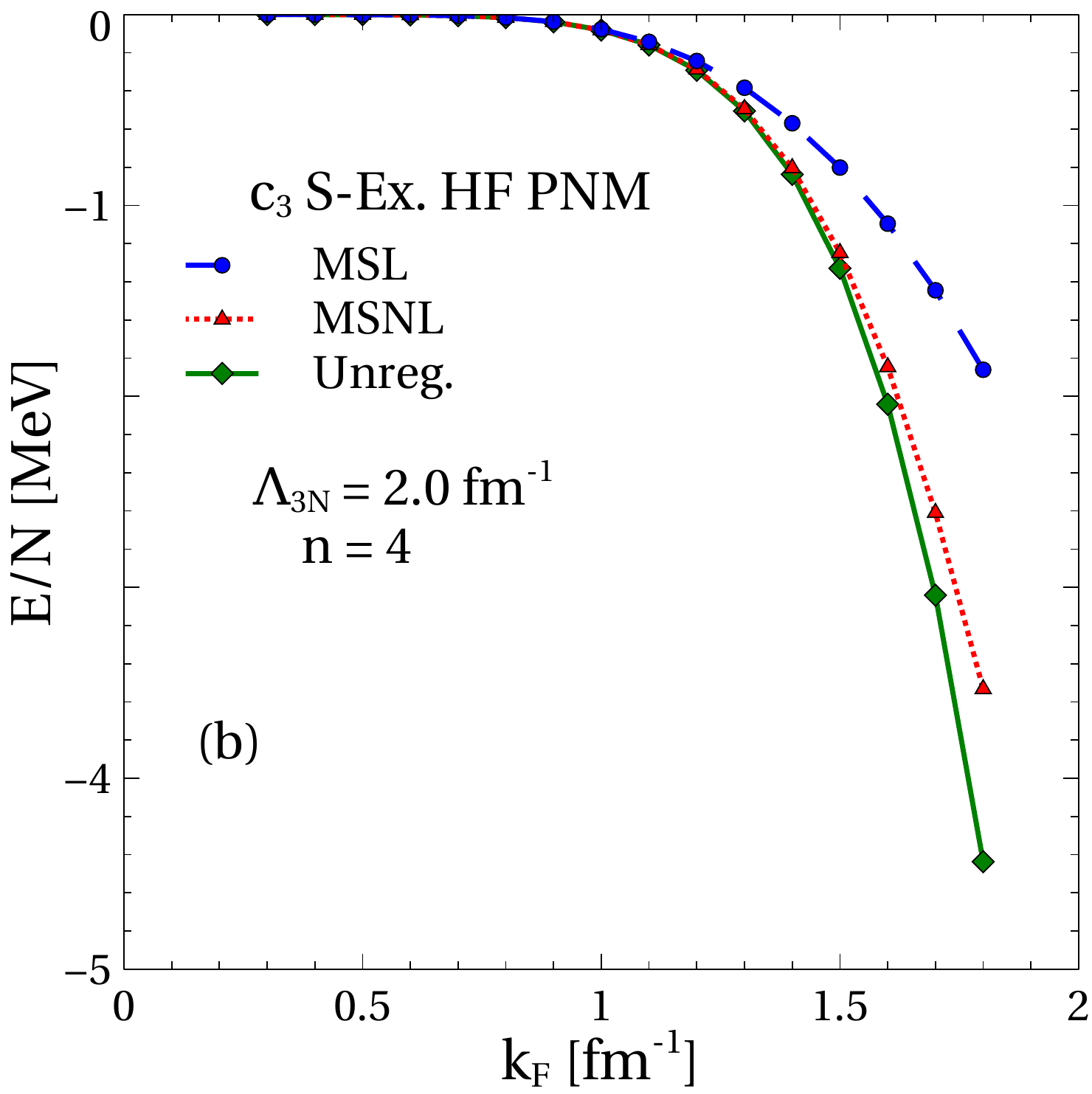}~
\includegraphics[width=0.35\textwidth]{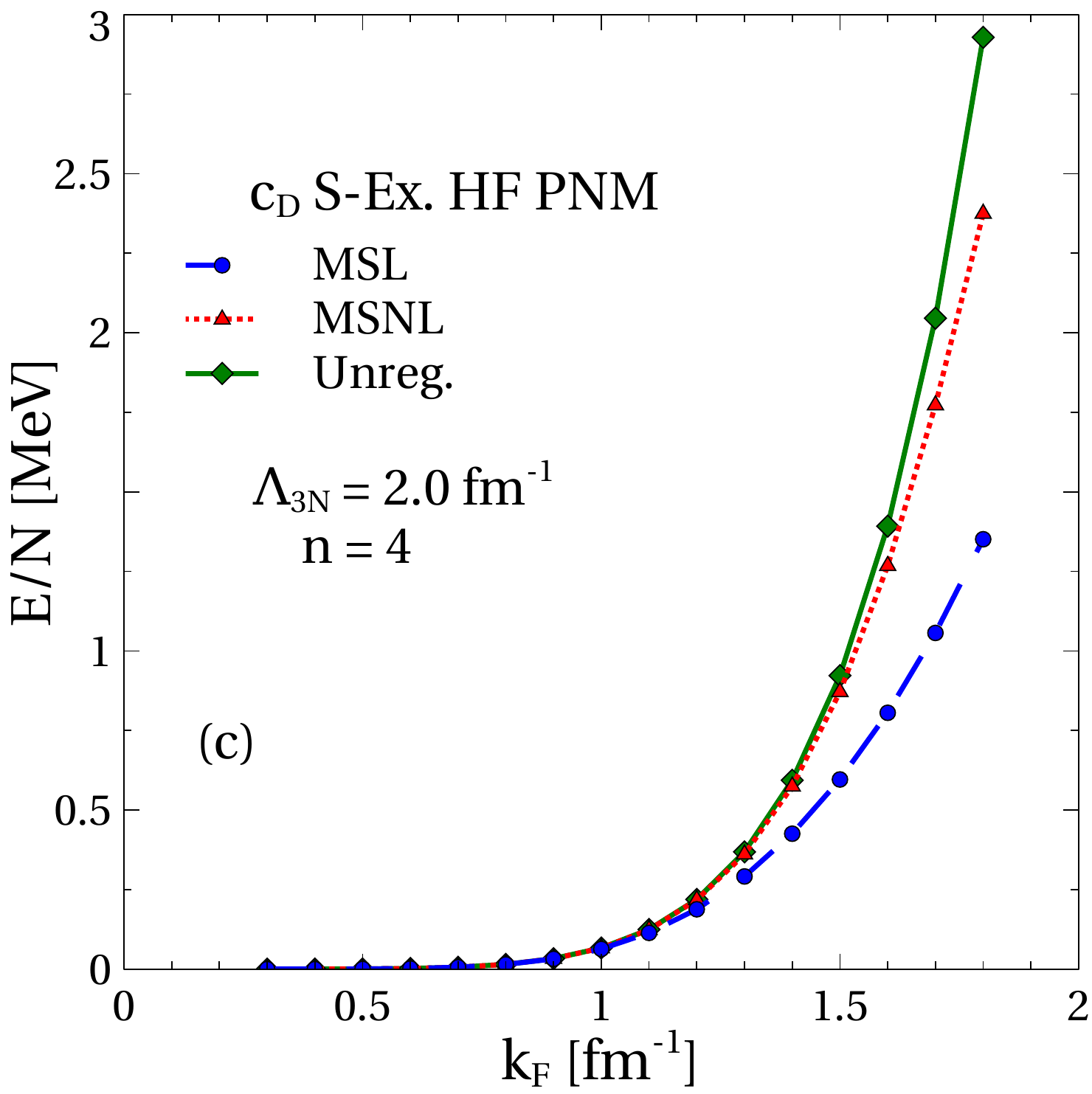}
 
\caption{Plots (a), (b), and (c) show the energy per particle for the single-exchange $c_1$, $c_3$, and $c_D$ terms respectively in neutron matter with $n=4$, $\TNcut = 2.0~\fmi$, $c_{i} = 1~\ciunits$, $c_{D} = 1$.} 

\label{fig:HF_3N_s_ex_energies} 
\end{figure*}

\begin{figure*}
\includegraphics[width=0.35\textwidth]{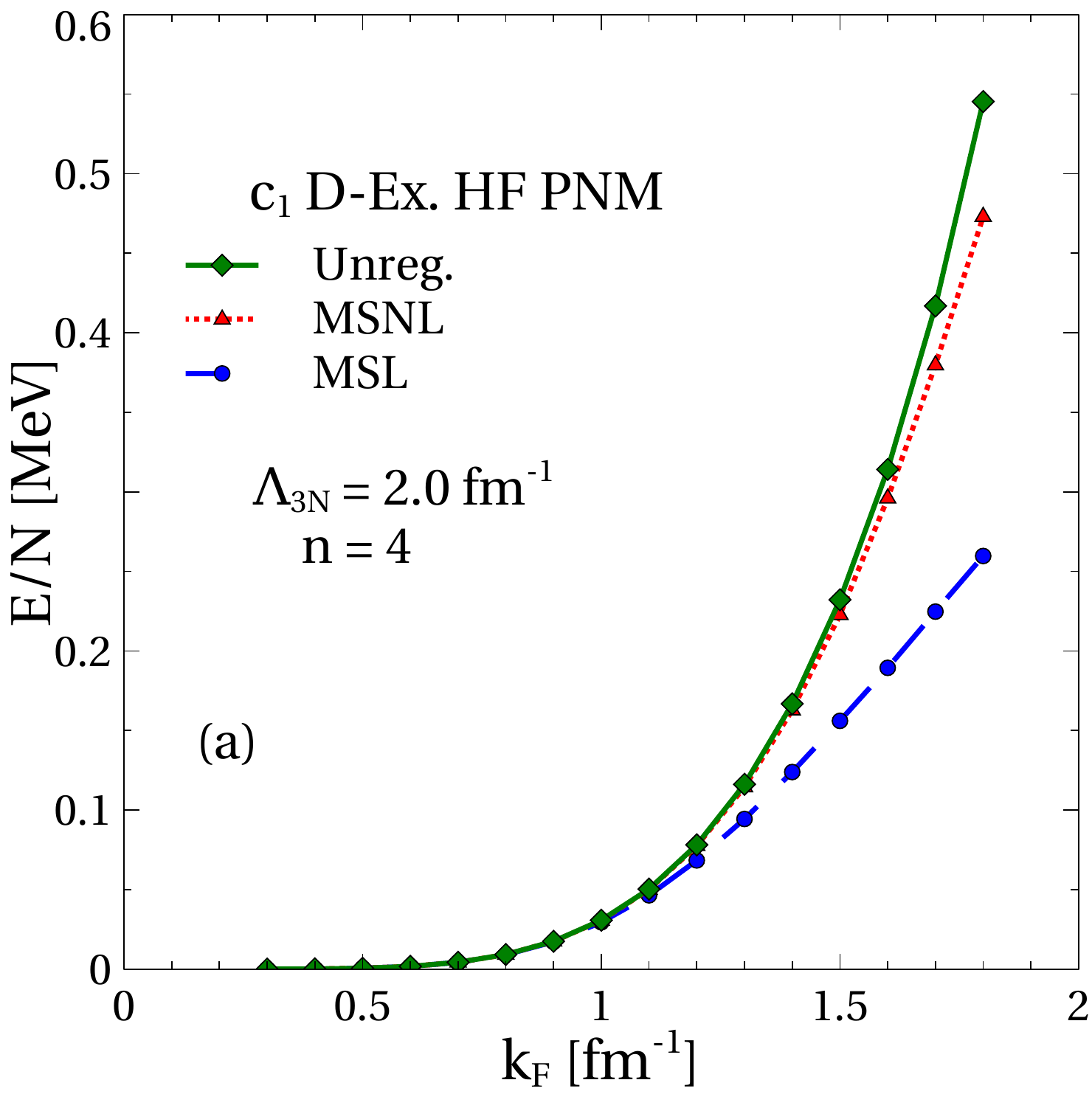}~
\includegraphics[width=0.35\textwidth]{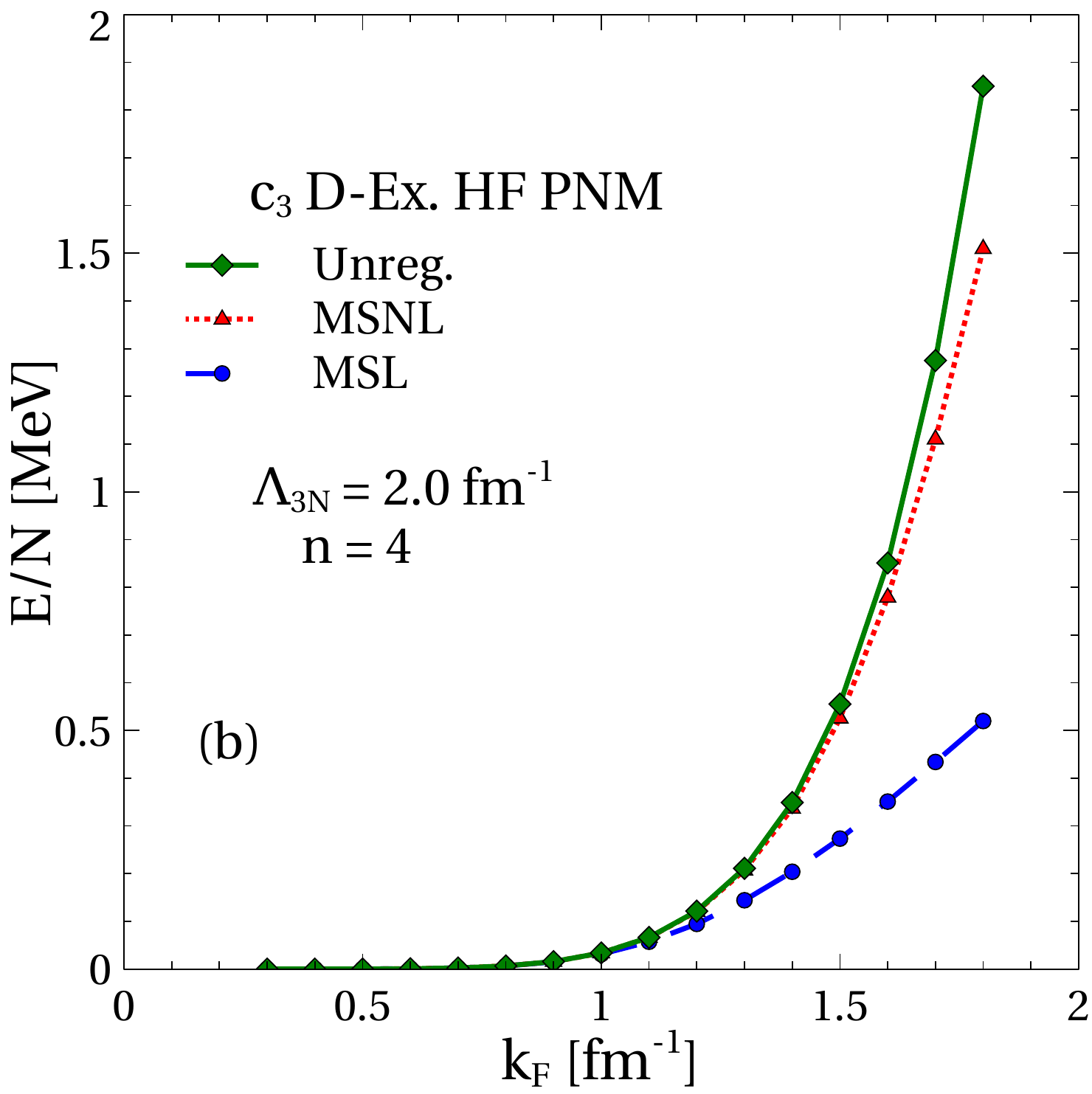}~
\includegraphics[width=0.35\textwidth]{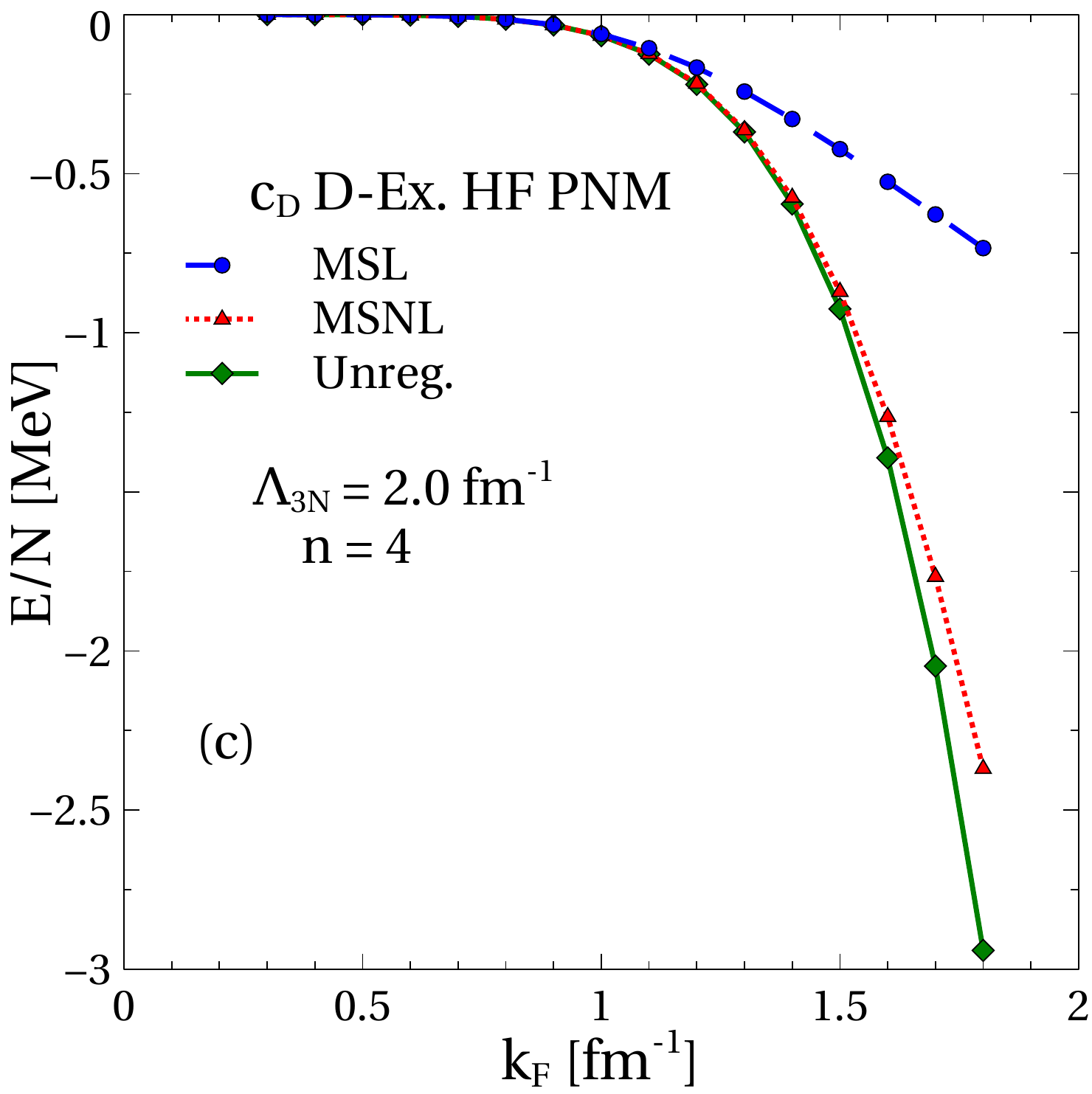}
 
\caption{Plots (a), (b), and (c) show the energy per particle for the double-exchange $c_1$, $c_3$, and $c_D$ terms respectively in neutron matter with $n=4$, $\TNcut = 2.0~\fmi$, $c_{i} = 1~\ciunits$, $c_{D} = 1$.} 

\label{fig:HF_3N_d_ex_energies} 
\end{figure*}

\section{3N HF $I_3$ Plots}

	Here, we demonstrate that weighting by the finite range interaction $c_1$ does not qualitatively change the phase space histograms. 
	In Fig.~\ref{fig:NNN_HF_histogram_c1}, we plot the integrand magnitude $I_3$ in \eqref{eq:NNN_HF_histogram_equation} including a $c_1$ weighting term from \eqref{eq:LO_3N_forces} analogously to what is done in \eqref{eq:NN_SO_histogram_equation} for NN second-order.
	The integrand magnitude is plotted as a function of the moduli of the Jacobi momenta defined in \eqref{eq:3N_Jacobi}. 
	Comparing to Fig.~\ref{fig:NNN_HF_histogram_ce}, the two plots are seen to be qualitatively the same. 

\section{3N HF Interaction Terms}
\label{sec:c1_term}
	
	In this appendix, we illustrate how different regularization schemes interplay with the form of the different $\NNLO$ 3N interactions. 
	As seen in the 3N HF energy per particle plot of Fig.~\ref{fig:HF_3N_ce_energies}, a clear hierarchy is established for the antisymmetric components of the $\NLTN$ and $\LTN$ schemes ($\LTN >$ $\NLTN$ for the direct term but $\NLTN > \LTN$ for the exchange terms).
	Although, Fig.~\ref{fig:HF_3N_ce_energies} only shows the $c_E$ term, this hierarchy is generic for the finite range interactions as well. 
	In Fig.~\ref{fig:3N_HF_NL_L_Ratio}, we plot the ratio of the energy per particle in PNM for the different schemes,
	\begin{equation}
	E^{\text{3N}	}_{\NLTN} \; / \; E^{\text{3N}	}_{\LTN} \; ,
	\label{eq:3N_HF_scheme_ratio}
	\end{equation}		
	that is the ratio of the HF energy per particle of the $\NLTN$ scheme to the $\LTN$ scheme.
	As can be seen, there are two different trends in the above ratio for the exchange terms, one for $c_1$ and one for $c_3$, $c_D$, and $c_E$.

	To see the origin of this difference, we count powers of the momentum transfer in \eqref{eq:TPE} and \eqref{eq:3N_OPE} and find one-dimensional variants of the $c_1$, $c_3$, $c_D$ interactions in Fig.~\ref{fig:ci_COMPARE} ignoring spin-isospin, 
	\begin{align*}
	f(q) = 
%	\bigg\{ \frac{q^2}{\left(q^2 + \mpi^2\right)^2}, \enspace 
%	\frac{q^4}{\left(q^2 + \mpi^2\right)^2}, \enspace 
%	\frac{q^2}{\left(q^2 + \mpi^2\right)} \bigg\}
	\begin{dcases}
	\frac{q^2}{\left(q^2 + \mpi^2\right)^2} 
	\quad &c_1
	\\
	\frac{q^4}{\left(q^2 + \mpi^2\right)^2} 
	\quad &c_3
	\\
	\frac{q^2}{\left(q^2 + \mpi^2\right)}
	\quad &c_D
	\end{dcases}
	\; .
	\label{eq:1d_ci}
	\numberthis
	\end{align*}
	These 1-D functions $f(q)$ are plotted in Fig.~\ref{fig:ci_COMPARE}. 
	It can be seen that the functional form of the $c_3$, $c_D$ terms is monotonically increasing in the momentum transfer $q$.
	Taking a large $q$ expansion of the $c_3$, $c_D$ terms in \eqref{eq:1d_ci}, where the contribution of the interaction is largest, reveals that $c_3$, $c_D$ should scale as $q^0 = 1$ or like the scalar term $c_E$. 
	This exactly matches the ratio behavior as seen in Fig.~\ref{fig:3N_HF_NL_L_Ratio}.

	In contrast, the $c_1$ interaction of Fig.~\ref{fig:ci_COMPARE} reaches a peak near $q \approx 0.7~\fmi$, in the vicinity of the pion mass. 
	This implies that the major contribution to the energy integrals with the $c_1$ term will come from this region as opposed to the large $|\qvec|$ area as one	would expect for $c_3$, $c_D$. 
	As the $\LTN$ regulator cuts off in the momentum transfer, we correspondingly	expect to see less suppression in the energy values involving the $c_1$ term.

\begin{figure*}[tbh]
	\includegraphics[width=0.42\textwidth]
	{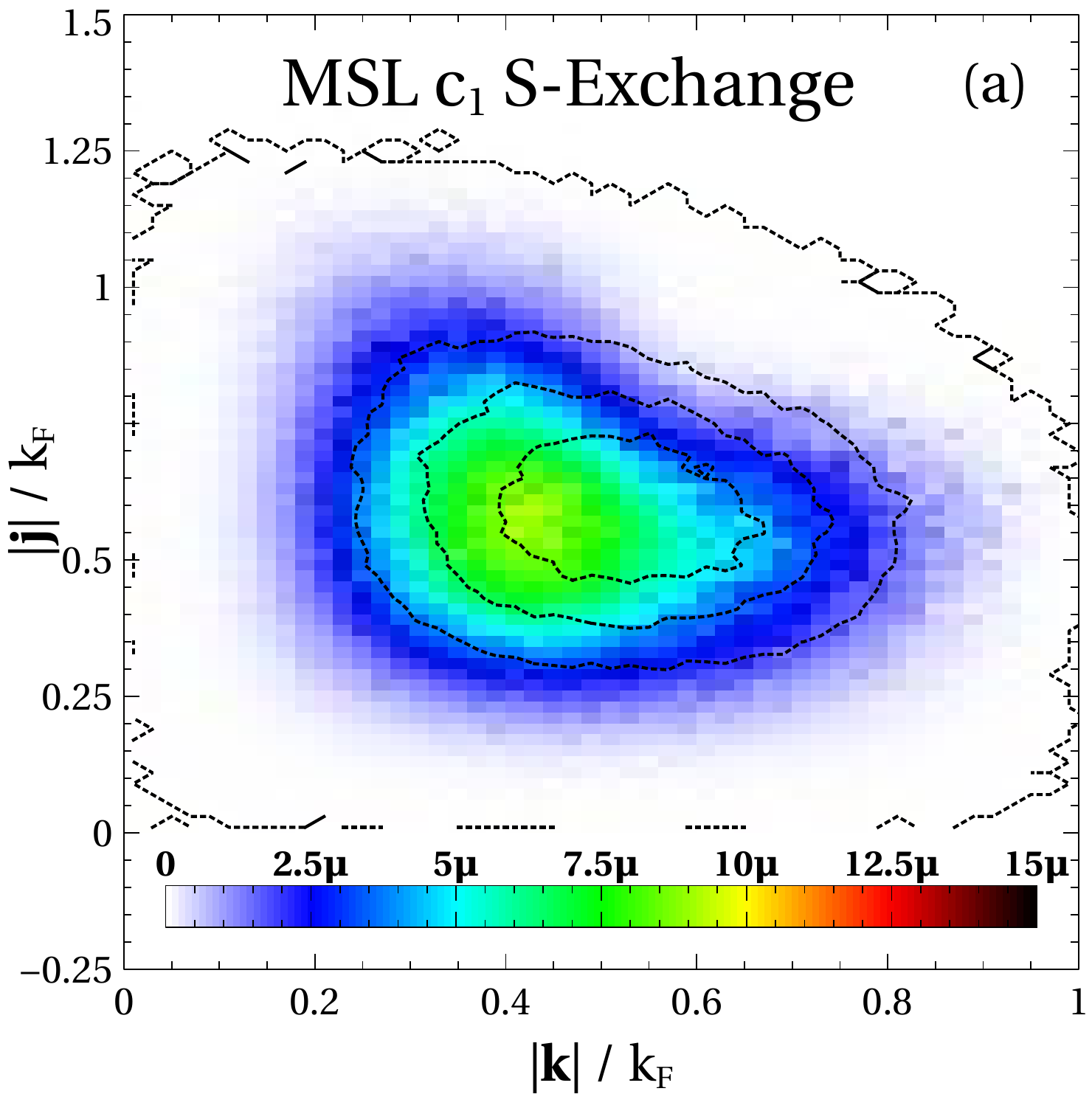}
	\includegraphics[width=0.42\textwidth]
	{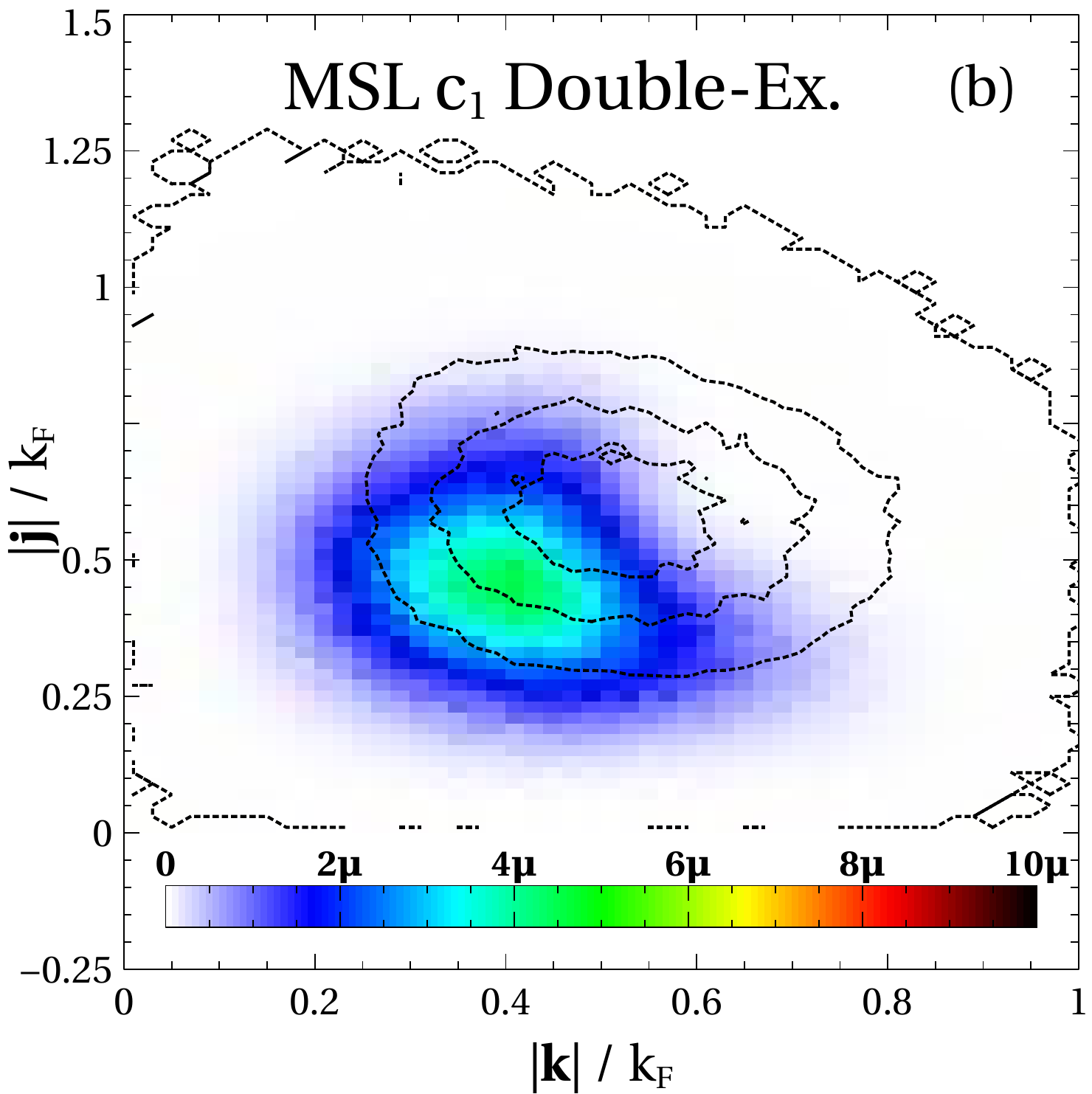}
	
	\medskip	
	
	\includegraphics[width=0.42\textwidth]
	{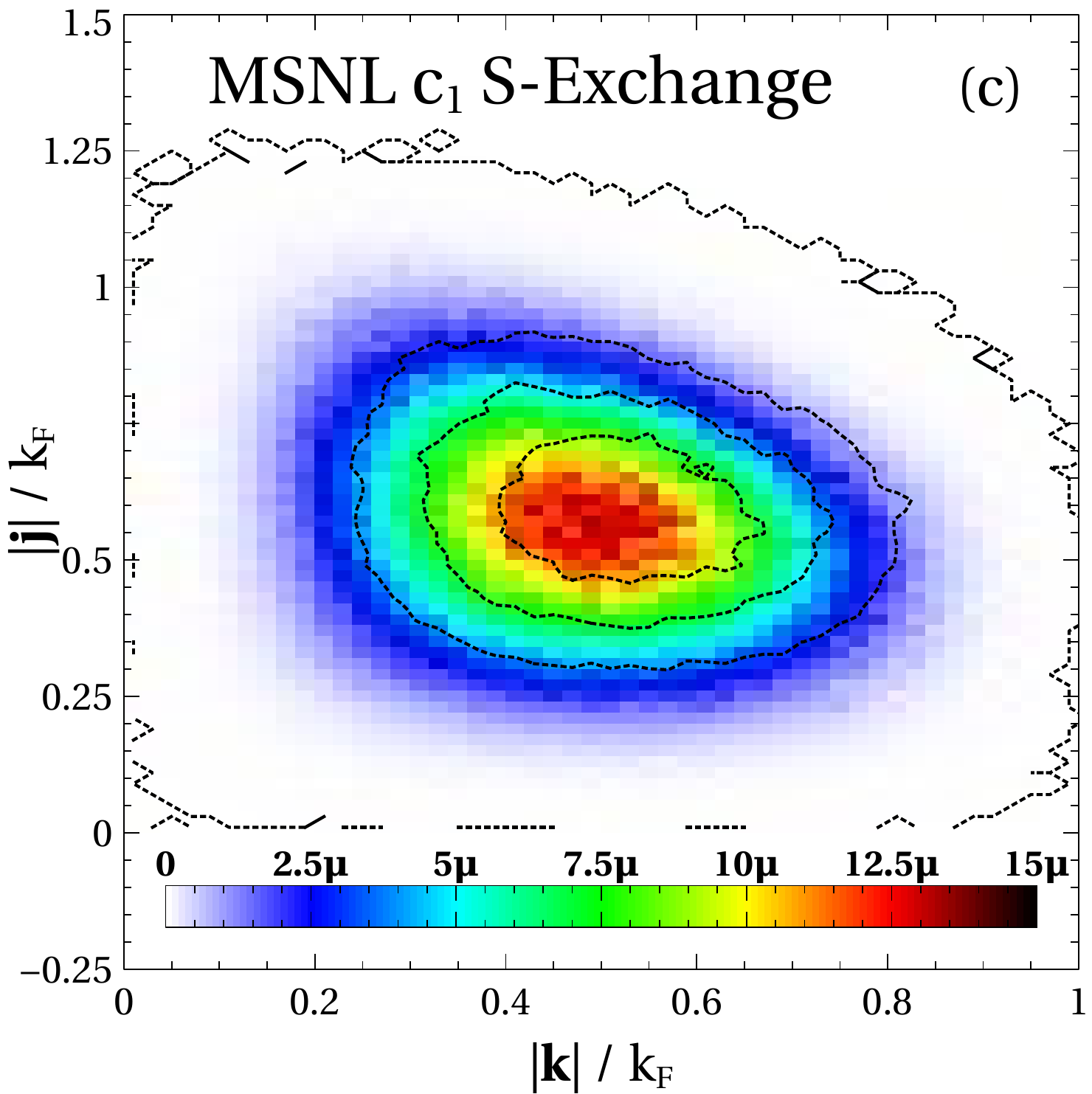}
	\includegraphics[width=0.42\textwidth]
	{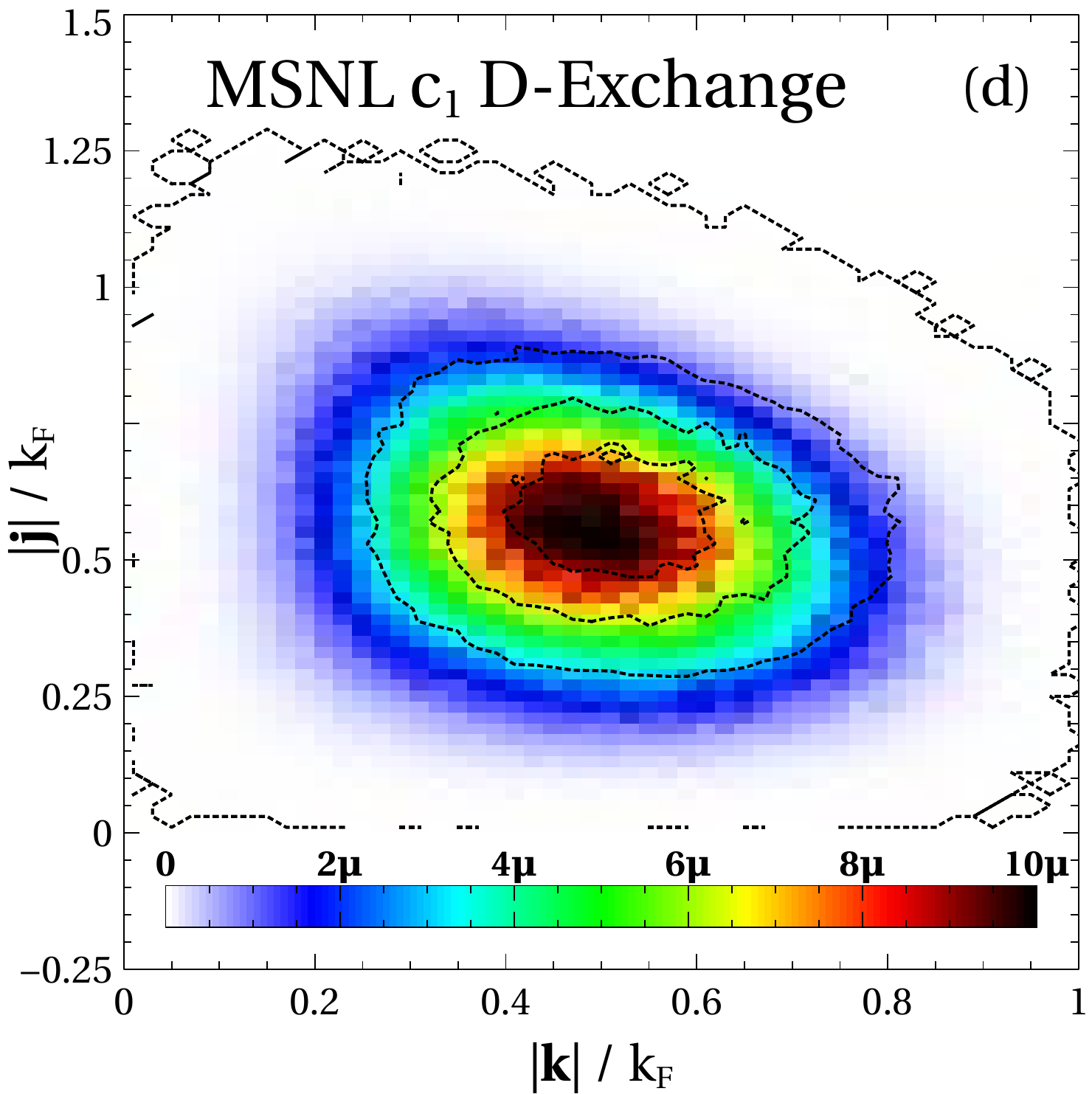}

	\caption	{(color online) Momentum histogram representing the 3N HF phase space for the $c_1$ term.
	Plots (a) and (b) show the MSL single-exchange and double-exchange terms respectively.
	Plots (c) and (d) show the MSNL single-exchange and double-exchange terms respectively.
	Plotted for $\kf =
	1.8, \TNcut = 2.0~\fmi,$ and $n = 4$.
	Colors indicate
	the integrand magnitude $I_3$ in \eqref{eq:NNN_HF_histogram_equation} with a $c_1$ weight term from \eqref{eq:LO_3N_forces}. Note the change in color scale between the plots.}
	\label{fig:NNN_HF_histogram_c1}
\end{figure*}

\begin{figure*}[tbh]
	\includegraphics[width=0.32\textwidth]{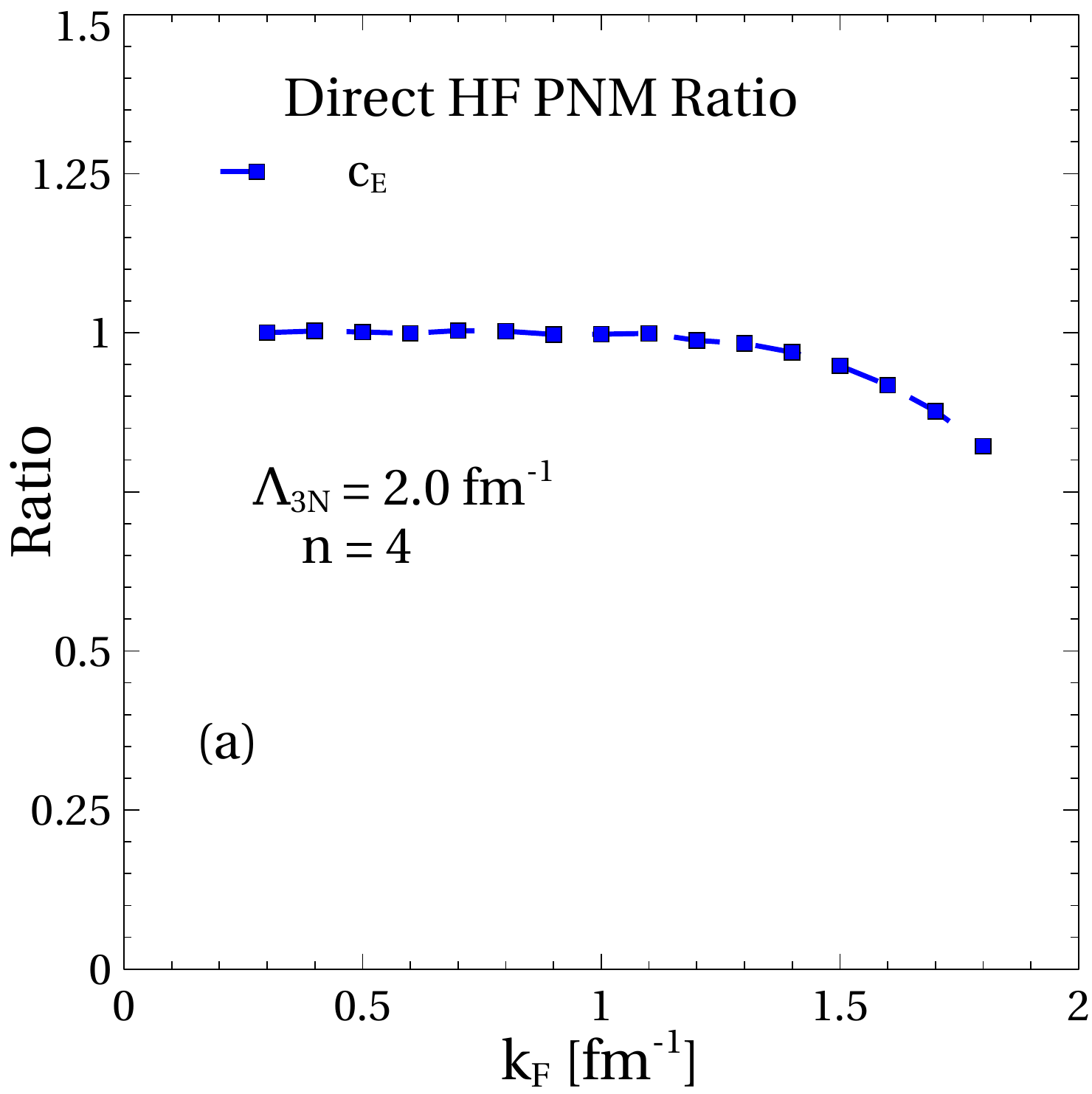}~~%
	\includegraphics[width=0.32\textwidth]{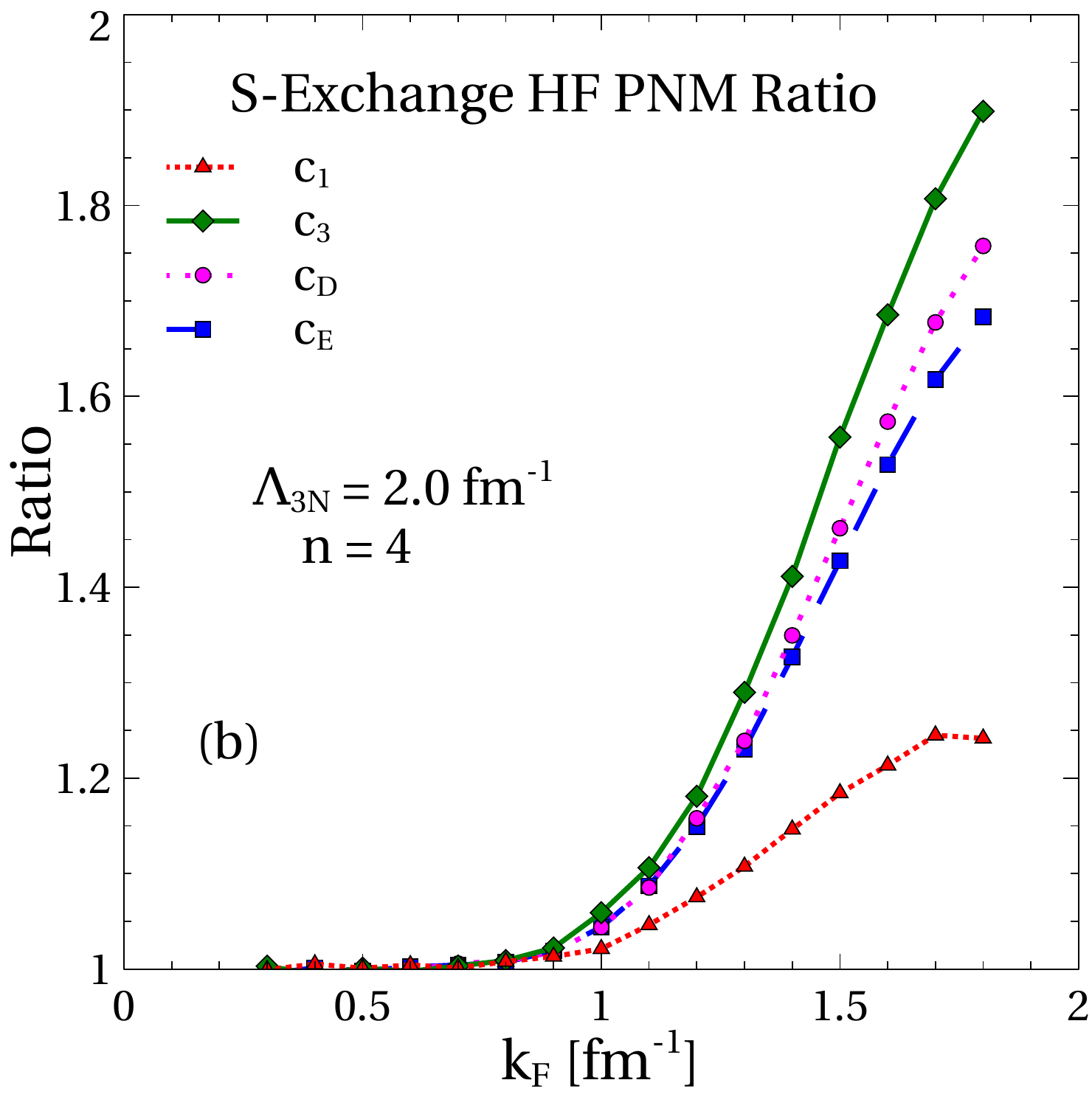}~~%
	\includegraphics[width=0.32\textwidth]{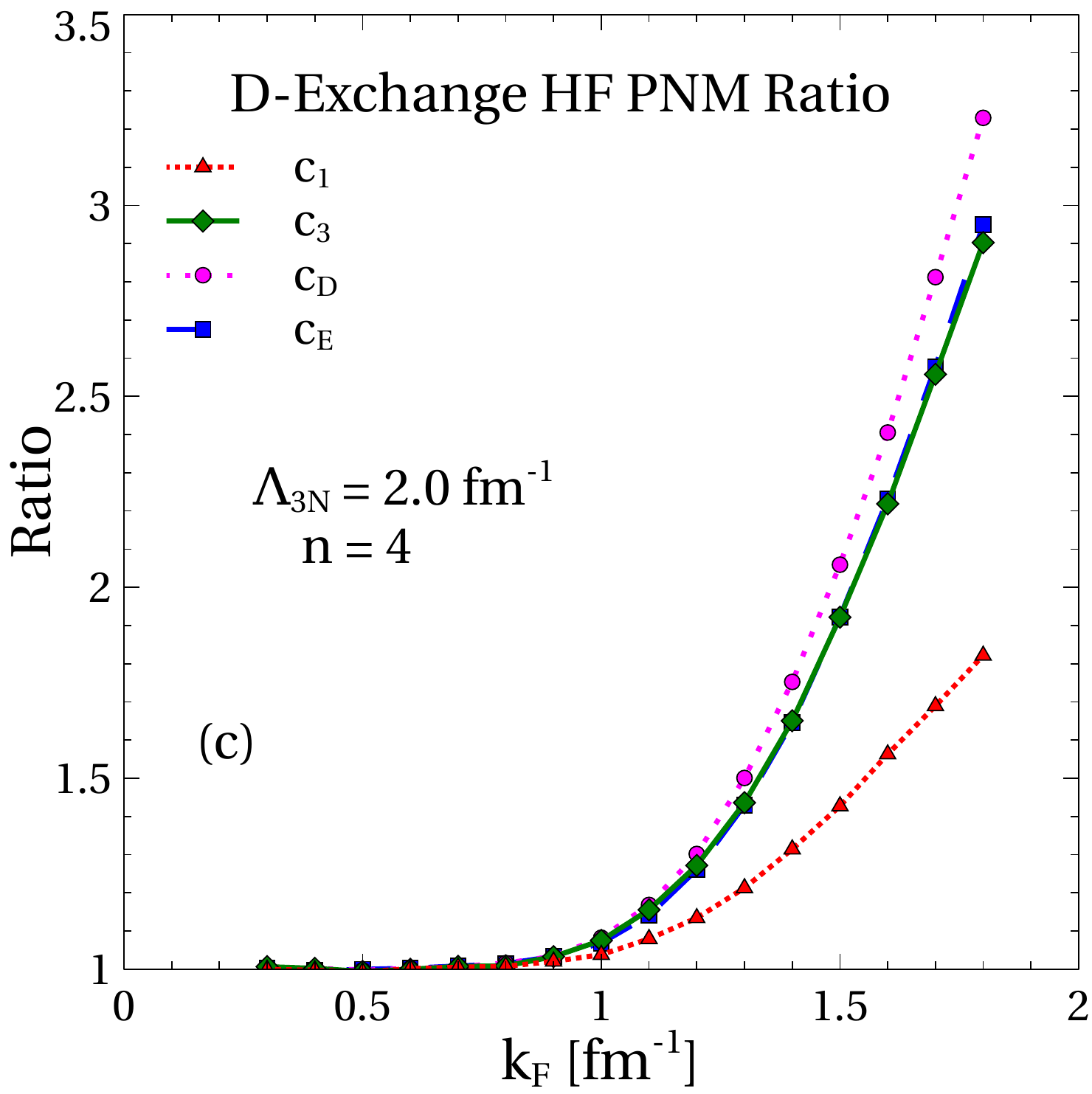}
	\caption{Ratio of the 3N HF energy, \eqref{eq:3N_HF_scheme_ratio}, calculated with the $\NLTN$ regulator in \eqref{eq:3N_nonlocal_alt} to the same energy with the $\LTN$ regulator in \eqref{eq:3N_local} for the individual 3N interaction terms in PNM. 
	Ratio is plotted for the direct (a), single-exchange (b), and double-exchange (c) terms of the antisymmetric 3N force. 
	The calculations use $c_i = 1.0~\ciunits$, $c_D = 1.0~\cdunits$, $c_E = 1.0~\ceunits$, $\TNcut = 2.0\fmi$, and $n = 4$.}
	\label{fig:3N_HF_NL_L_Ratio}
\end{figure*}

\begin{figure}[tbh]
	\includegraphics[width=0.42\textwidth]{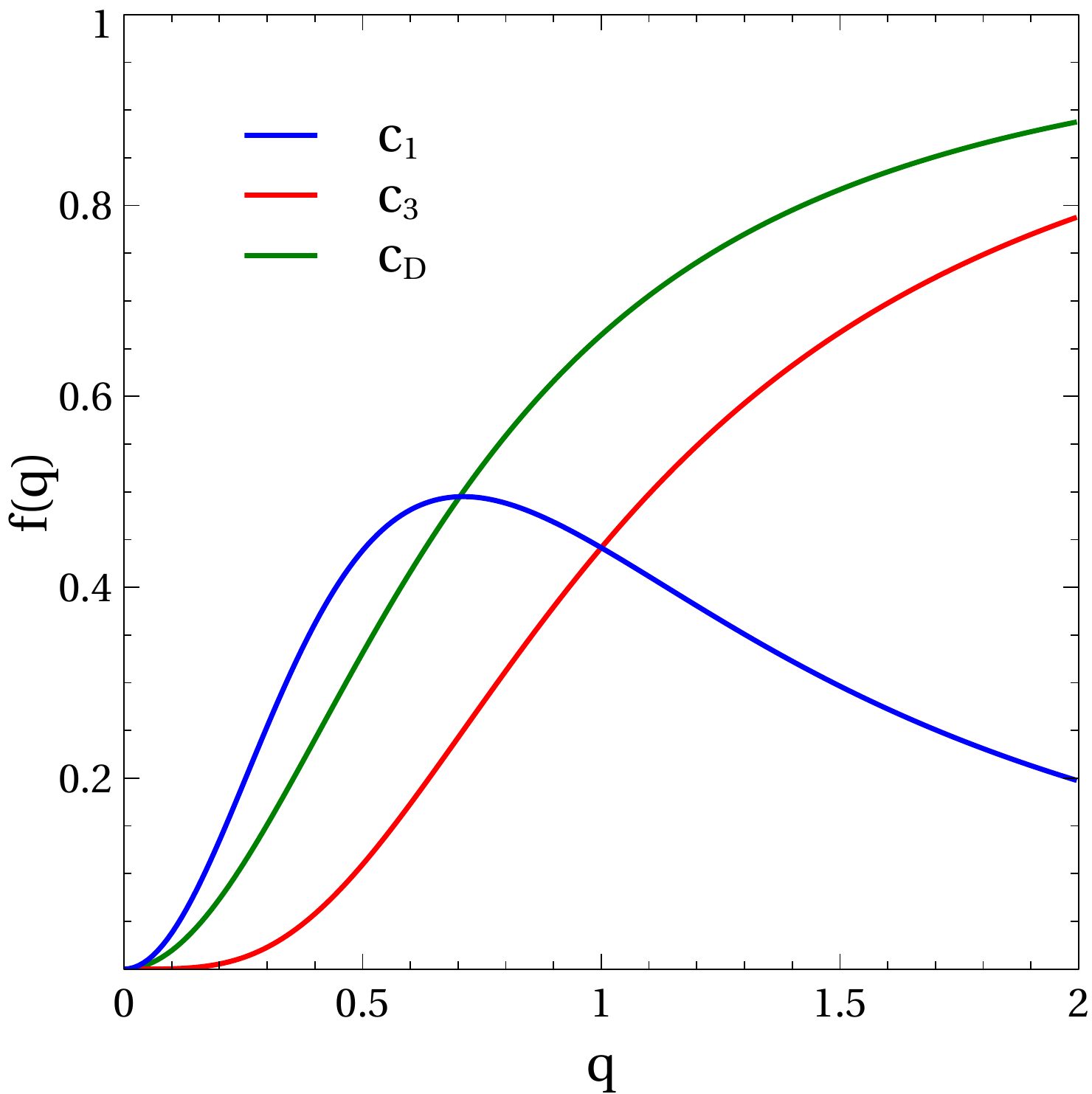}~~~

	\caption{Plot of the functions $f(q)$ in \eqref{eq:1d_ci} which are one-dimensional variants of the finite range 3N interactions ignoring spin-isospin.
	 The functions are plotted as a function of the 1-D momentum transfer $q$ variable.}
	\label{fig:ci_COMPARE}
\end{figure}

\end{document}